\documentclass[final,3p,times,fleqn,authoryear]{elsarticle} 

\usepackage{amssymb}
\usepackage{eqnarray,amsmath}
\usepackage{bm}
\usepackage[authoryear]{natbib}
\usepackage{multirow}
\usepackage{makecell}
\usepackage{subfigure}
\usepackage{float}
\usepackage{hyperref}
\usepackage{xcolor}
\usepackage{algorithm}  
\usepackage{algpseudocode}
\usepackage{lscape}
\usepackage{array}
\usepackage{comment}
\usepackage{rotating}
\usepackage{booktabs}
\usepackage{appendix}
\renewcommand{\algorithmicrequire}

\begin{document}
\begin{frontmatter}
\title{Capacity drop accounting for microscopic vehicle interaction effects: analytical model and validation with high-resolution trajectories}
\author[Yu]{Yu Han}
\author[Liu]{Pan Liu}
\author[Liu]{Zhiyuan Liu}
\author[LL]{Ludovic Leclercq}
\address[Yu]{Thrust of Intelligent Transportation, The Hong Kong University of Science and Technology (Guangzhou), 511400, Guangzhou, China}
\address[Liu]{School of Transportation, Southeast University, 211189, Nanjing, China.}
\address[LL]{LICIT-ECO7, Université Gustave Eiffel, ENTPE, F-69675, Lyon, France.}




\begin{abstract}
 
Capacity drop is a traffic phenomenon in which the discharge flow from a queue is lower than the theoretical infrastructure capacity. This paper proposes a generic analytical method to estimate the queue discharge flow of freeway traffic. Capacity drop is primarily attributed to hesitant vehicles, defined as vehicles that stochastically and temporarily enter an acceleration delay state and generate voids (i.e., extra gaps) in front of them. The proposed method estimates the expected total void length generated by all hesitant vehicles, based on the distributions of their spatial and temporal locations as well as the associated delays. It also accounts for interactions between the waves triggered by downstream hesitant vehicles and the voids generated by upstream ones. Our analysis reveals that this interaction is the key mechanism behind the differing extents of capacity drop observed between standing queues and jam waves in previous studies. The accuracy of the model is validated through both numerical simulations and real-world trajectories. Overall, the proposed method offers a deeper understanding of capacity drop, which can be leveraged in traffic flow modeling and control.

\end{abstract}


\end{frontmatter}

\section{Introduction}
\label{sec:introduction}


Capacity drop is a traffic phenomenon observed downstream of freeway bottlenecks, where the queue discharge flow (QDF) falls below the free-flow capacity. Accurately estimating the QDF is crucial for traffic design, management, and control. For example, a primary objective of freeway traffic control is to alleviate capacity drop and thereby increase the throughput of the freeway network \citep{papageorgiou1991alinea, cassidy2005increasing}.

Extensive empirical studies have shown that the extent of capacity drop varies and can be categorized into three aspects. First, capacity drop differs across various sites \citep{banks1991two, cassidy1999some, bertini2005empirical, oh2012estimation, srivastava2013empirical}, with reported reductions in QDFs ranging from 3\% to 18\% at active bottlenecks in different locations. Second, QDFs fluctuate over time at the same site. For instance, \cite{chung2007relation} observed QDF reductions ranging from 5\% to 18\% at a merge bottleneck, and \cite{kerner2002empirical} documented significant fluctuations in QDFs at multiple merge bottlenecks on German freeways. Third, QDFs vary by bottleneck type. \cite{yuan2017capacity} found that the QDFs of standing queues at fixed-location bottlenecks are significantly higher than those of jam waves, which are characterized by an upstream-moving head and tail.

The reasons behind variations in QDFs have been investigated from several perspectives. From a macroscopic viewpoint, \cite{oh2012estimation} found that the extent of capacity drop is inversely related to the number of lanes. Additionally, the presence of off-ramps downstream of merging bottlenecks was observed to reduce capacity drop, attributed to the smoothing effect that decreases disruptive lane changes in inner lanes \citep{cassidy2010smoothing}. \cite{srivastava2013empirical} discovered that capacity drop increases with higher merging flows, suggesting that greater merging flows lead to more severe traffic breakdowns and consequently larger capacity drop. Furthermore, \cite{chen2018capacity} examined capacity drop at weaving bottlenecks and found that a more balanced distribution of merging and diverging flows results in a lower capacity drop.

From a microscopic perspective, factors contributing to QDF variations can be summarized into two main aspects. The first is vehicle heterogeneity. \cite{leclercq2011capacity, leclercq2016capacity} analyzed how variations in initial speeds and acceleration capabilities of merging vehicles affect capacity drop. \citet{yuan2019geometric} argued that the stochastic nature of desired acceleration is a primary cause of capacity drop. The second aspect is the variation in driving behavior under different conditions. \citet{treiber2006understanding} examined how drivers adapt their desired time headway based on local speed variations, noting that drivers typically choose longer time headway during congestion compared to free-flow conditions. \citet{yeo2008asymmetric} linked capacity drop to asymmetric driving behavior, where deceleration and acceleration patterns differ. \citet{chen2012behavioral} observed distinct reaction patterns of timid drivers during deceleration and acceleration from trajectory data and validated their impact on capacity drop through numerical simulations. \cite{yuan2017microscopic} suggested that drivers' reaction times are negatively related to speed in congestion, leading to QDF variation.

While the mechanisms behind QDF variation have been extensively investigated, analytical models for quantifying QDFs are relatively scarce. \cite{leclercq2011capacity, leclercq2016capacity} proposed analytical formulas to estimate QDFs for merge bottlenecks with single and multiple lanes, respectively. Since the cause of capacity drop was attributed to the bounded acceleration capabilities of low-speed merging vehicles, the applicability of these models is limited to merge bottlenecks. \cite{jin2018kinematic} developed analytical derivations for lane drop and tunnel bottlenecks, assuming that capacity drop results from increasing time gaps during acceleration. \cite{yuan2017microscopic} analyzed the effects of acceleration spread and reaction time on QDFs from jam waves, finding that acceleration spread has a limited impact on capacity drop. Beyond these, most existing methods for quantifying QDFs rely on numerical simulations \citep{coifman2011extended, chen2014periodicity, jin2017first, yuan2019geometric, xu2020statistical}, which are challenging to integrate into traffic system optimization frameworks. Therefore, an analytical model that can quantify QDFs across different types of bottlenecks is still needed.

This paper proposes an analytical method to estimate QDFs. Building on previous empirical observations, stochastic acceleration delays may occur in some vehicles, referred to as hesitant vehicles, representing a temporary acceleration delay state often triggered by lane changes and sharp decelerations. These delays generate voids in front of affected vehicles and contribute to capacity drop \citep{han2025capacity}. When such delays occur at different locations within close temporal proximity, the void generated by an upstream vehicle may be diminished by the wave generated by a downstream vehicle. In this study, the wave–void interaction is formalized as a stochastic process that accounts for the spatial and temporal distributions of acceleration delay states and their associated response durations. The expected void generated by each event is derived by considering its probability of interacting with waves generated by other events. The contributions of this paper are twofold. First, we develop an analytical framework for estimating capacity drop that links microscopic stochastic triggering events to macroscopic traffic flow characteristics. The framework captures both inter-vehicle heterogeneity and intra-vehicle variability in driving behavior. Second, the proposed framework provides a unified explanation for the different extents of capacity drop observed in standing queues and moving jam waves, as reported in previous studies. The accuracy of the model is validated using both numerical simulations and real-world trajectory data.

The remainder of this paper is structured as follows. Section 2 reviews the existing literature on capacity drop. Section 3 illustrates the mechanism of capacity drop using field trajectory data as examples. Section 4 elaborates on the impact of wave-void interaction on QDF and presents the analytical formulation for estimating QDFs. Section 5 validates the proposed analytical model through numerical simulations and real-world data. Finally, Section 6 concludes the paper with a summary of findings and discussions.

\section{Literature review}

This section summarizes existing studies on capacity drop from two perspectives. Section 2.1 reviews past observations of capacity drop, while Section 2.2 introduces various mechanisms underlying capacity drop, based on different theories and assumptions.

\subsection{Observations of capacity drop}

The studies by \cite{hall1991freeway} and \cite{banks1991two} are among the earliest works identifying capacity drop, reporting that freeway capacity diminishes by 3-6\% following the formation of an upstream queue. \cite{cassidy1995methodology} proposed an efficient method to observe capacity drop by constructing a rescaled cumulative curve, where variations in the slope indicate changes in flow. Using this method, extensive empirical studies have documented capacity drops at various types of bottlenecks in different locations worldwide, mostly using aggregated traffic flow data from fixed-location sensors \citep{cassidy1999some, bertini2005empirical, yuan2017capacity}.

There are some important observations in these empirical studies. For instance, \cite{chung2007relation} found that the occurrence of capacity drop is closely related to the density in the bottleneck area; once the density exceeds a threshold, a reduction in QDF is observed. \cite{oh2012estimation} identified a negative relationship between capacity drop and the number of lanes, indicating that the extent of capacity drop diminishes as the number of lanes at the bottleneck increases. They also found that QDF reductions tend to be less pronounced when an off-ramp is located downstream of the bottleneck. \cite{srivastava2013empirical} showed that QDF variations are related to on-ramp flows, revealing that QDFs increase as on-ramp flow increases. \cite{yuan2017capacity} observed differences in QDFs between standing queues and jam waves on two different Dutch freeways, finding that the outflows of jam waves are significantly lower than those of standing queues.

While aggregated traffic flow data reveals certain phenomena related to capacity drop, trajectory data is crucial for elucidating the mechanisms of capacity drop and validating existing theories. In prior studies, NGSIM data has played a vital role in examining the correlation between vehicle behavior and capacity drop. For instance, \citet{chen2014periodicity} utilized NGSIM vehicle trajectory data to discern the differences in reactions between timid and aggressive drivers. Using the same dataset, \citet{oh2015impact} further illuminated that the extent of behavioral change among timid drivers depends on the severity of stop-and-go waves, which are primarily triggered by lane changes. NGSIM data predominantly capture congested traffic states, but analyzing capacity drop requires datasets covering a broader temporal dimension, spanning transitions between non-congested and congested states, and a wider spatial dimension, encompassing both the bottleneck area and its free-flow downstream section \citep{coifman2011extended}. Therefore, there remains a critical need for empirical research aimed at directly establishing the link between capacity drop and microscopic traffic behavior. In a recent study, \citet{han2025capacity} conducted an in-depth analysis of the causes of capacity drop using high-resolution trajectory data with extensive spatial and temporal coverage. They identified the primary contributor to capacity drop as stochastic acceleration response delays, during which vehicles enter an acceleration delay state and generate voids (i.e., extra gaps) in front of them. This empirical finding provides the foundation for the analytical method proposed in this paper.

\subsection{Mechanisms on capacity drop}

Many researchers believe that capacity drop is strongly related to lane-changing (LC) behavior. Some studies suggest that lane-changing vehicles momentarily occupy two lanes during the process, which temporarily reduces the link’s capacity \citep{coifman2006impact, jin2010kinematic}. One of the most widely recognized explanations is the bounded acceleration mechanism. When a vehicle changes lanes at a lower speed than the target lane, its limited acceleration capability prevents it from quickly matching the speed of the surrounding traffic. This creates a void in front of the vehicle and contributes to capacity drop \citep{laval2006lane, duret2010onset}. This mechanism is particularly relevant at merge bottlenecks, where there is a significant speed difference between the mainline traffic and on-ramp vehicles \citep{leclercq2011capacity, leclercq2016capacity}. However, when the speed of the inserting vehicle is comparable to that of the target lane, a temporary flow increment may occur. This phenomenon can be explained by the relaxation mechanism, where vehicles initially accept shorter headway following the insertion, then gradually return to their desired headway \citep{leclercq2007relaxation, laval2008microscopic}. \citet{kim2013driver} argued that capacity drop results from a reduction from supersaturated flow (beyond capacity) to saturated flow (the actual capacity), due to this relaxation process. \citet{chen2018capacity} analyzed capacity drop at multi-lane bottlenecks, considering spatially distributed LCs. They suggested that in heavy congestion, diverging traffic creates vacancies quickly filled by immediate followers, potentially mitigating capacity drop, particularly when the LC rate is high. As a result, capacity is lower when insertions occur downstream and desertions occur upstream, compared to the reverse scenario where desertions occur downstream and insertions upstream.

Beyond lane-changing behavior, numerous studies have explored capacity drop by analyzing vehicles' car-following (CF) behavior. The impact of heterogeneous drivers or inter-driver variability on capacity drop has been investigated in several studies. For instance, \citet{wong2002multi} replicated capacity drop using a multi-class traffic flow model, positing that different vehicle classes adhere to different equilibrium states and consequently, distinct flow-density relationships. More extensive research has been conducted to explain capacity drop by focusing on variations in driving behavior depending on traffic conditions, referred to as intra-driver variability. \citet{treiber2006understanding} examined the adaptation of desired time headway as a function of local speed variances, assuming that drivers choose longer time headway in congestion compared to free-flow scenarios. \citet{yeo2008asymmetric} linked capacity drop to asymmetric driving behavior, where drivers' deceleration and acceleration processes follow distinct curves. \citet{chen2014periodicity} identified distinct reaction patterns in aggressive and timid drivers when faced with disturbances. Their numerical simulation demonstrated that these varying driver reaction patterns can significantly impact bottleneck discharge flows. This observation received further validation through an empirical study conducted by \citet{oh2015impact}, which also found that the severity of stop-and-go waves can amplify the extent of behavioral changes. Building upon these findings, \citet{yuan2017microscopic} conducted an analytical derivation of the impact of drivers' extended reaction times on capacity drop in stop-and-go waves. \citet{jin2018kinematic} assumed a varying time gap within a tunnel or sag bottleneck and derived kinematic wave models to reproduce capacity drop for those bottlenecks. Random noise in vehicles' desired acceleration has also been considered a potential trigger for capacity drop. Several studies have developed microscopic simulations incorporating these acceleration errors to replicate capacity drop \citep{xu2020statistical}.

\section{The impact of acceleration delay on QDFs}
\label{empirical}

This section explains how the acceleration delays of vehicles impact QDFs, using trajectories collected from the field as illustration. Section \ref{subsec:description_data} describes the data used for analysis. Section \ref{subsec:data_delay} presents how voids are created from the acceleration delays and discusses the relationship between average acceleration delay and QDFs.

\subsection{Description of data}
\label{subsec:description_data}

The trajectory data were collected from two distinct sites in Nanjing, a provincial capital city in China, using high-resolution 4K cameras mounted on Unmanned Aerial Vehicles (UAVs). Each site required two UAVs to ensure sufficient spatial coverage. At each site, trajectories were extracted from individual videos and subsequently spliced using state-of-the-art trajectory processing algorithms. The topological features and traffic flow characteristics of these two sites are briefly introduced below.

\begin{figure}[!ht]
	\centering
	\includegraphics[width=0.98\textwidth]{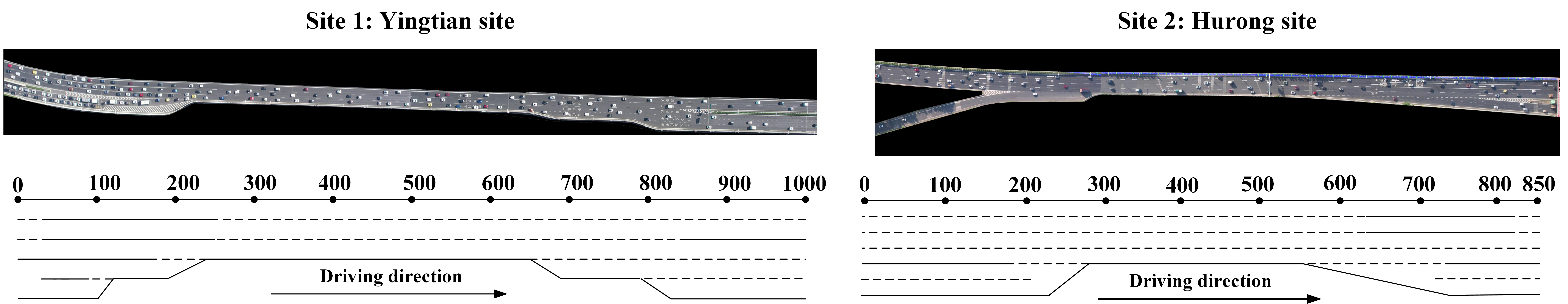}
	\caption{The geographical and topological map of the sites for data collection.}\label{fig:site}
\end{figure} 

\textbf{Site 1}: This site is a weaving section located on the Yingtian urban expressway with a speed limit of 80 km/h. The topological structure of the site is depicted in the left picture of Figure \ref{fig:site}. The site spans approximately 1 kilometer. The weaving area, which extends from 250 to 650 meters, is an active bottleneck. Downstream of this region, traffic diverges into two directions, with the number of lanes increasing from three to five. Data collection occurred during the morning peak hour (7:00-8:00) on three sunny workdays: July 7, July 8, and July 11, 2022. At this site, the traffic stream is predominantly composed of passenger cars. Vehicles longer than 7 m account for less than 5\% of the total traffic volume.

\begin{figure}[!ht]
	\centering
 	\includegraphics[width=0.75\textwidth]{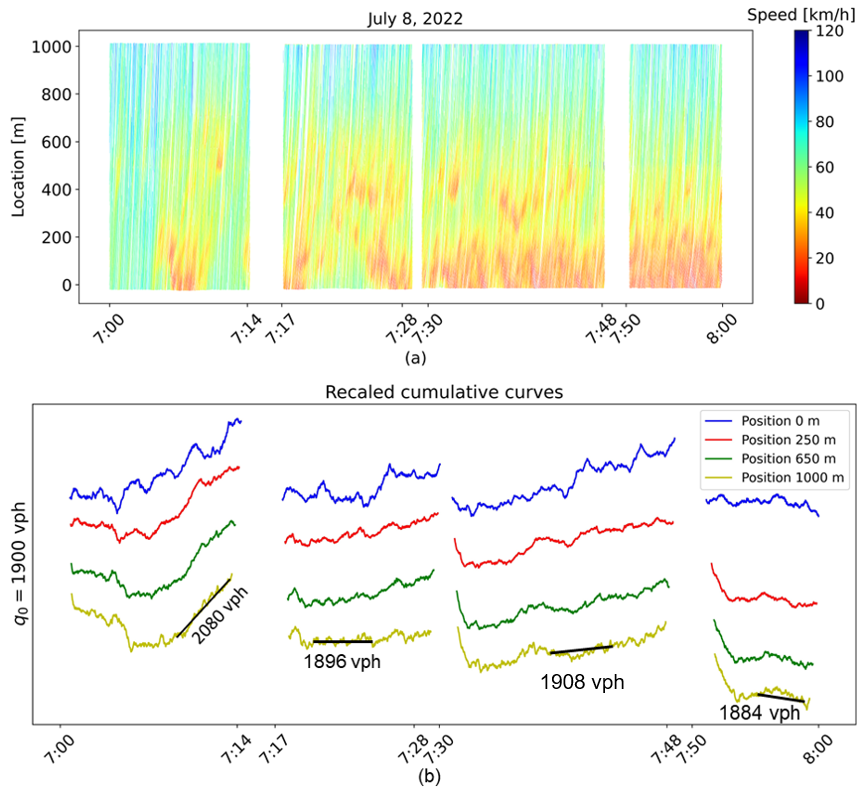}
	\caption{(a) Color-coded vehicle trajectories of the innermost lane of site 1, and (b) The rescaled cumulative curves.}
	\label{fig:site1}
\end{figure}

Figure \ref{fig:site1} (a) shows the vehicle trajectories on July 8, color-coded by speed, for the innermost lane of this site. Figure \ref{fig:site1} (b) presents the rescaled cumulative curves, with a background flow reduction of $q_{\text{0}}=1900$ veh/h/lane, at four different locations: the upstream end (0 m), the merge point (250 m), the diverge point (650 m), and the downstream end (1000 m), where the QDFs were measured. On that day, a traffic breakdown occurred at approximately 7:04. A post-breakdown recovery flow, sustained for 5 minutes, was recorded as the maximum flow of the site, at 2080 veh/h/lane. For model validation, data from this site were sampled at 12 different 5-minute intervals, as summarized in Table \ref{tab:the result of Yingtian}.

\textbf{Site 2}: This site is another weaving section located on the Hurong freeway. There are four mainstream lanes in this site, with a speed limit of 100 km/h. The topological structure is shown in the right panel of Figure \ref{fig:site}, and its geometric layout closely resembles that of Site 1. Data were collected during the morning peak hours on two sunny workdays, June 16 and 17, 2022.

\begin{figure}[!ht]
	\centering
 	\includegraphics[width=0.85\textwidth]{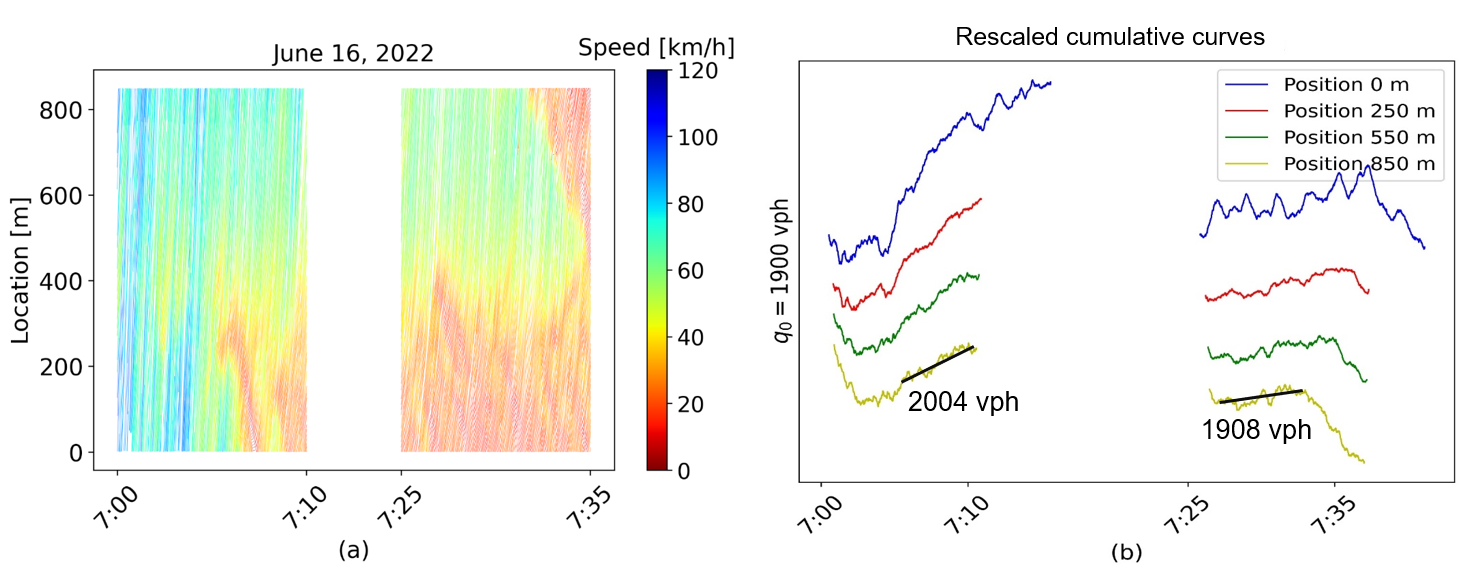}
	\caption{(a) Color-coded vehicle trajectories of the innermost lane of site 2, and (b) The rescaled cumulative curve.}
	\label{fig:site2}
\end{figure}

As depicted in Figure \ref{fig:site2} (a), a traffic breakdown occurred around 7:06 on July 16 at Site 2. Approximately 2.5 km downstream, another bottleneck was activated around the same time, causing congestion to propagate toward the downstream end of the study site. To analyze QDFs, free flow conditions should prevail downstream of the bottleneck. As a result, the effective data collection period at this site was relatively short. Data from this site were sampled across five 5-minute intervals, as summarized in Table \ref{tab:the result of Hurong}.

\subsection{Driver's acceleration delay and its impact on QDF}
\label{subsec:data_delay}

Acceleration response delay typically occurs when a vehicle decelerates and subsequently accelerates again. For ordinary passenger vehicles, this delay may be triggered by small disturbances, such as sudden braking of the preceding vehicle or aggressive lane changes. For trucks and other heavy vehicles, delayed acceleration may occur not due to a specific disturbance but as a consequence of their limited acceleration capabilities. It causes the following vehicle to deviate from its equilibrium state, creating a void in front of it. It should be noted that the term "hesitant" used in this study differs from timid drivers in the literature, which refers to drivers who generally maintain large following distances under any driving condition. In contrast, "hesitant" here describes a temporary and probabilistic behavioral state that may occur during the acceleration phase, rather than a fixed characteristic of specific drivers or vehicles. 

\begin{figure}[!h]
	\centering
	\includegraphics[width=0.9\textwidth]{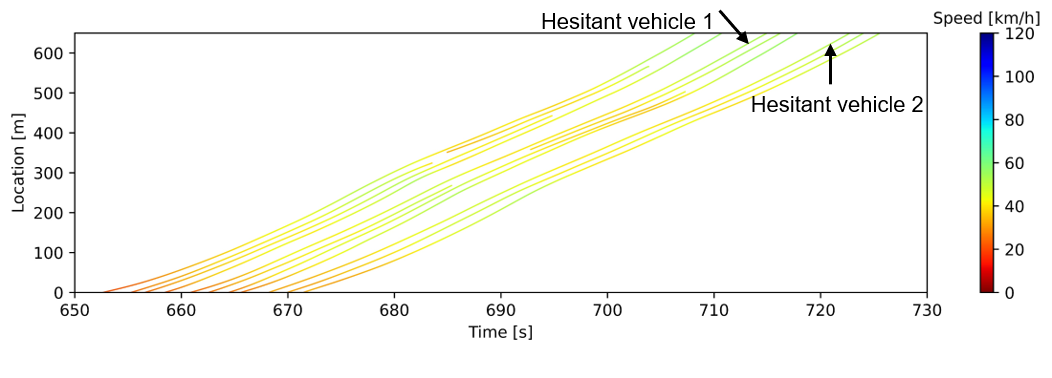}
	\caption{An example of a platoon where hesitant vehicles created voids.}
	\label{fig:platoon}
\end{figure}

\begin{figure}[!h]
	\centering
	\includegraphics[width=0.9\textwidth]{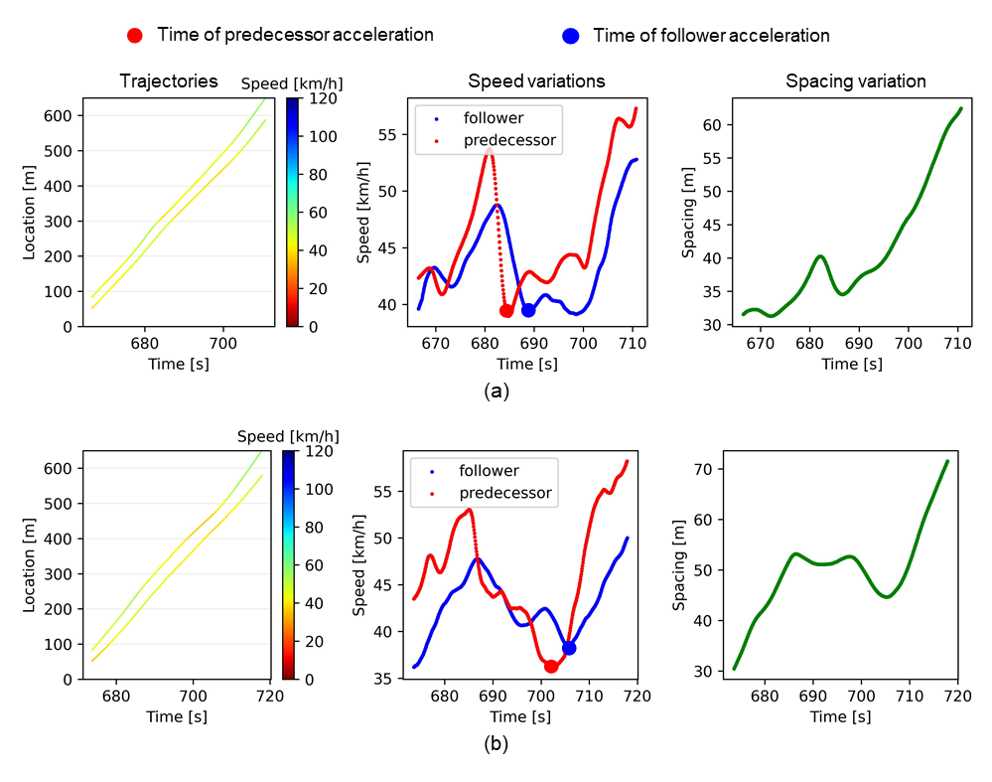}
	\caption{Acceleration response delays of vehicle 1 (a) and 2 (b) during deceleration-acceleration processes.}
	\label{fig:example}
\end{figure}

Figure \ref{fig:platoon} illustrates the creation of voids by hesitant vehicles in a platoon. Two hesitant vehicles were identified from the trajectory data, with each vehicle experiencing two deceleration-acceleration instances. During the second instance, the acceleration delays of the first and second hesitant vehicles were approximately 3.5 seconds and 4.2 seconds, respectively, as illustrated in Figure \ref{fig:example}. These delays led to the formation of large voids, resulting in a reduction of QDF.

\begin{figure}[!ht]
	\centering
	\includegraphics[width=1.0\textwidth]{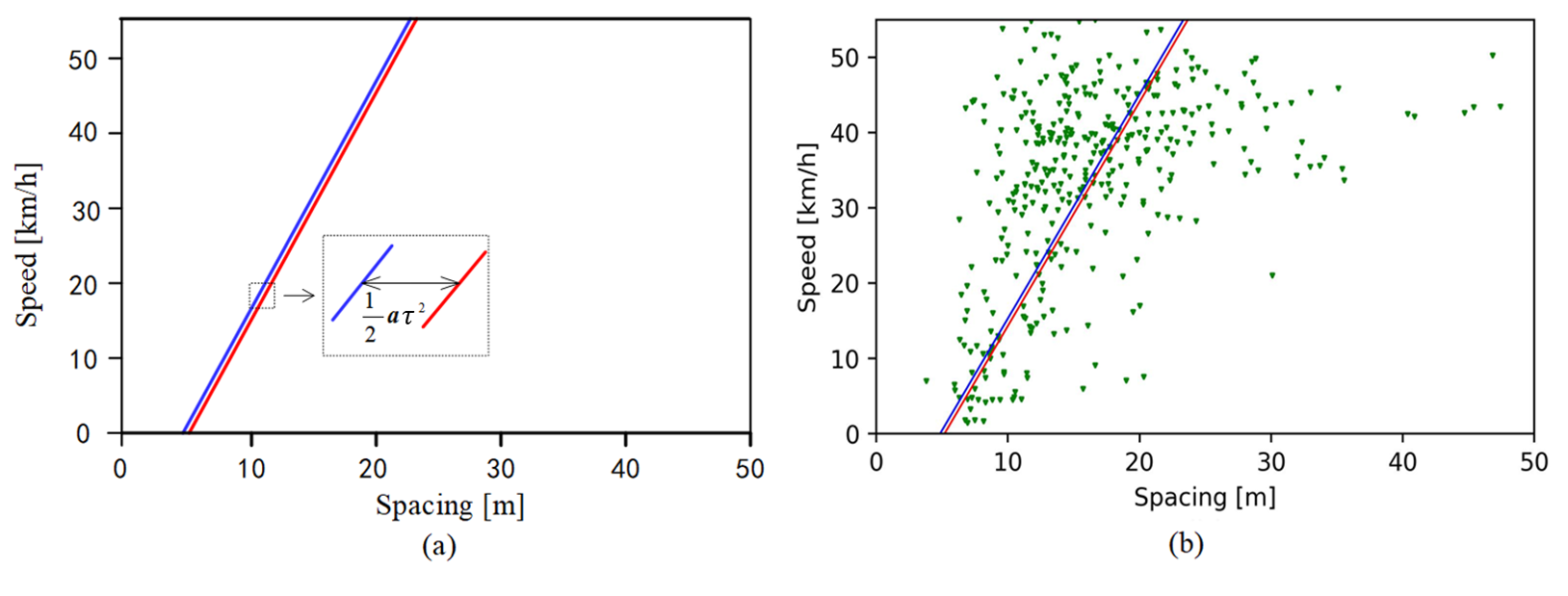}
	\caption{Comparison of the spacing-speed status of vehicles at the onset of acceleration with the equilibrium status.}
	\label{response_delay}
\end{figure}

The acceleration delay state of a vehicle is determined by comparing its spacing--speed condition at the onset of acceleration with the corresponding equilibrium condition described by Newell’s simplified car-following model, which assumes no acceleration delay \citep{newell2002simplified}. In Newell’s framework, the reaction time $\tau_{0}$ can be interpreted as the slope of the car-following branch in the spacing--speed diagram, as illustrated by the blue line in Fig.~\ref{response_delay}(a). This relationship is obtained by fitting the spacing--speed data from stable car-following states, where no deceleration--acceleration process occurs.

Assuming that the predecessor starts accelerating at speed $v_{0}$, the follower accelerates after a reaction time $\tau_{0}$. The spacing of the follower at the onset of acceleration, denoted by $s^{\text{ac}}$, is given by
\begin{align}
\label{sac}
s^{\text{ac}} = s^{\text{eq}} + \frac{1}{2} a \tau_{0}^{2},
\end{align}
where $s^{\text{eq}}$ is the equilibrium spacing corresponding to speed $v_{0}$, and $a$ is the assumed constant acceleration rate. Accordingly, the spacing--speed states at acceleration onset predicted by Newell’s model form a curve shifted to the right of the equilibrium car-following relation by $\frac{1}{2} a \tau_{0}^{2}$, shown as the red line in Fig.~\ref{response_delay}(a). Data points located to the right of this line are identified as exhibiting acceleration delay.

The actual acceleration response time is measured by comparing the timing of acceleration between a vehicle and its predecessor. If the follower’s speed at the onset of acceleration is equal to or greater than that of the predecessor, the response time is defined as the time difference between the moments when the predecessor and the follower begin to accelerate. Otherwise, the response time is defined as the time difference between the moment when the predecessor begins to accelerate and the moment when the follower’s speed reaches the speed at which the predecessor initiated acceleration. The measured response time is then compared with the Newell reaction time $\tau_{0}$. If the observed response time exceeds $\tau_{0}$, the vehicle is classified as being in the acceleration delay state, and the magnitude of the delay is defined as the difference between the observed response time and $\tau_{0}$. For a detailed presentation, readers are referred to \citep{han2025capacity}.

\begin{figure}[!ht]
	\centering
 	\includegraphics[width=0.9\textwidth]{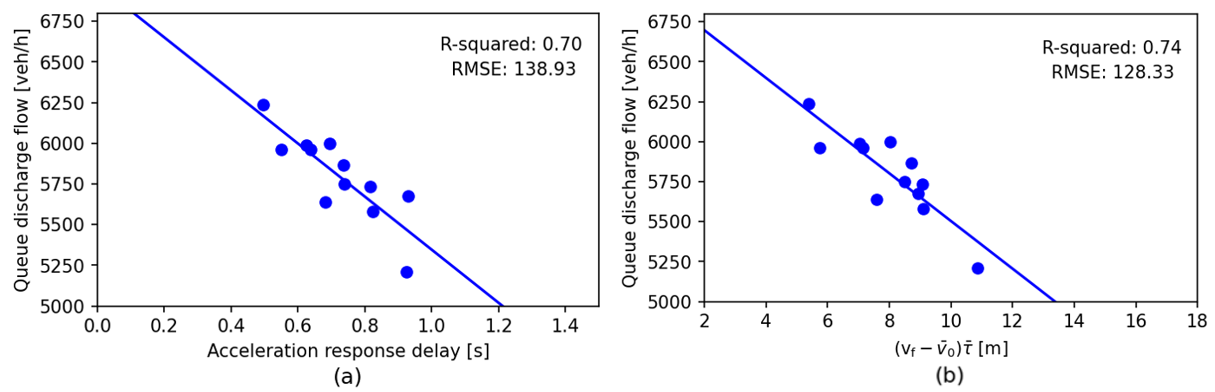}
	\caption{Relationships between QDF and (a) average acceleration delay, and (b) $(v_{\text{f}} - \bar{v}_{0}) \cdot \bar{\tau}$ for site 1.}
	\label{trend2}
\end{figure}

As an illustrative example, the relationship between average acceleration delay, $\bar{\tau}$, and QDFs for different time intervals at site 1 is depicted in Figure \ref{trend2} (a). A clear trend shows that QDF decreases as the average acceleration delay increases. This observation is intuitive: larger acceleration delays create larger voids, leading to a more substantial reduction in QDF. It is also straightforward to derive that the speed prior to acceleration, denoted as $v_0$, has a positive correlation with QDF. For a given response time $\tau'$, the spacing between the predecessor and the follower when both reach free-flow speed, $v_{\text{f}}$, is given by $s^{\text{eq}} + (v_{\text{f}} - v_0) \cdot \tau'$. This indicates that the impact of acceleration delay on QDF is more pronounced when the speed in congestion is lower. Figure \ref{trend2} (b) illustrates the relationship between $(v_{\text{f}} - \bar{v}_0) \cdot \bar{\tau}$ and QDF, where the regression line exhibits even better fitting performance than the one in Figure \ref{trend2} (a).

\section{The impact of wave-void interaction on QDF}

This section provides a detailed analysis of the impact of wave-void interaction on QDF. Section \ref{subsec: illustration} introduces the physical process of wave-void interaction, explaining how it occurs and its effects on traffic flow. Section \ref{subsec: analytical} derives an analytical formula to estimate QDF, incorporating the effects of wave-void interaction.

\subsection{Illustration of wave-void interaction}
\label{subsec: illustration}

\begin{figure}[!ht]
	\centering
 	\includegraphics[width=0.9\textwidth]{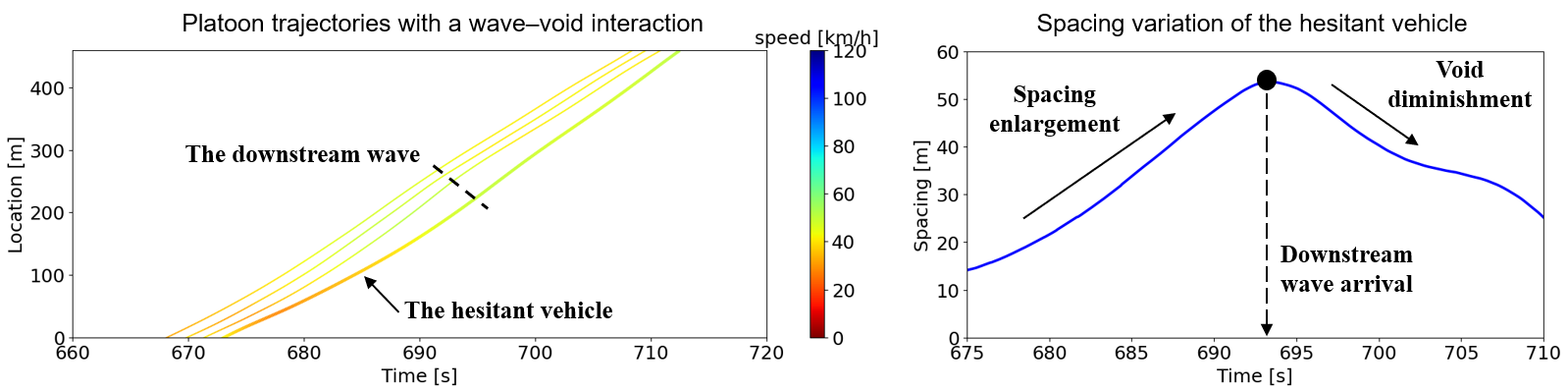}
	\caption{An example of wave-void interaction in real trajectories.}
	\label{interaction}
\end{figure}

As discussed, when a vehicle enters an acceleration delay state, it creates a void in front of it while simultaneously generating a wave that propagates upstream. When the void generated by an upstream vehicle intersects with the wave triggered by a downstream vehicle entering the same state, the void may be partially or fully diminished. Figure \ref{interaction} illustrates this mechanism. The left panel shows a platoon encountering two consecutive waves. During the first wave, one vehicle experiences a pronounced acceleration delay, resulting in an enlarged spacing. When the second, downstream wave reaches this location, the previously formed void is substantially reduced due to wave–void interaction, as shown in the right panel.

\begin{figure}[!ht]
	\centering
 	\includegraphics[width=0.95\textwidth]{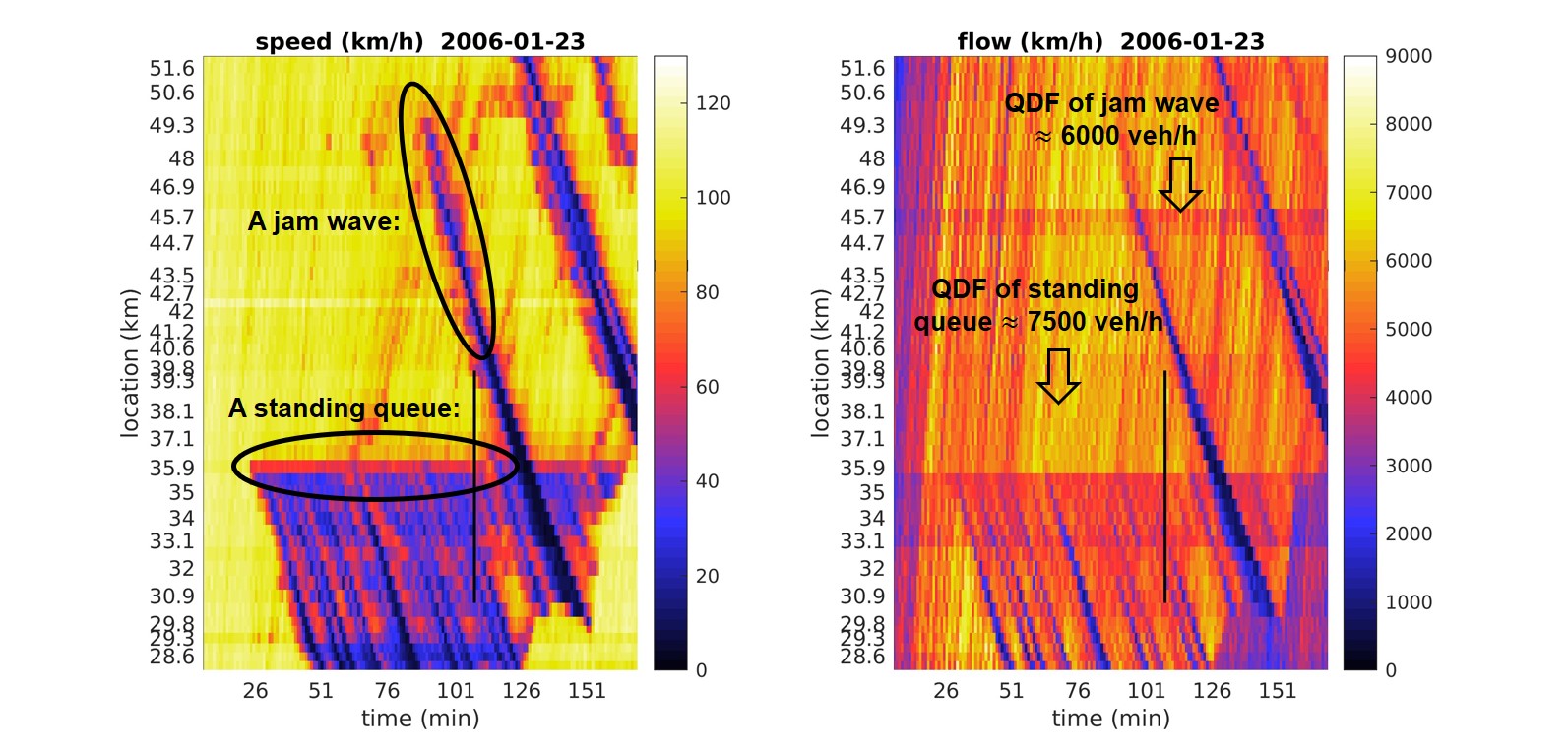}
	\caption{An example of a standing queue and jam wave observed in the field.}
	\label{fig:jams}
\end{figure}

Two types of traffic jams are usually identified on freeways \citep{hegyi2008specialist}. Traffic jams with the head fixed at the bottleneck are known as standing queues, and jams that have an upstream moving head and tail are known as jam
waves (also known as wide moving jams in some studies, e.g., \citet{kerner1996experimental}). Figure \ref{fig:jams} depicts two types of traffic jams: a standing queue originating from a fixed-location bottleneck and a jam wave, characterized by an upstream-moving jam head and tail, observed on a Dutch freeway. In this example, the QDF of the jam wave was significantly lower than that of the standing queue. A similar finding was reported in \cite{yuan2017capacity}. 

\begin{figure}[!ht]
	\centering
 	\includegraphics[width=0.95\textwidth]{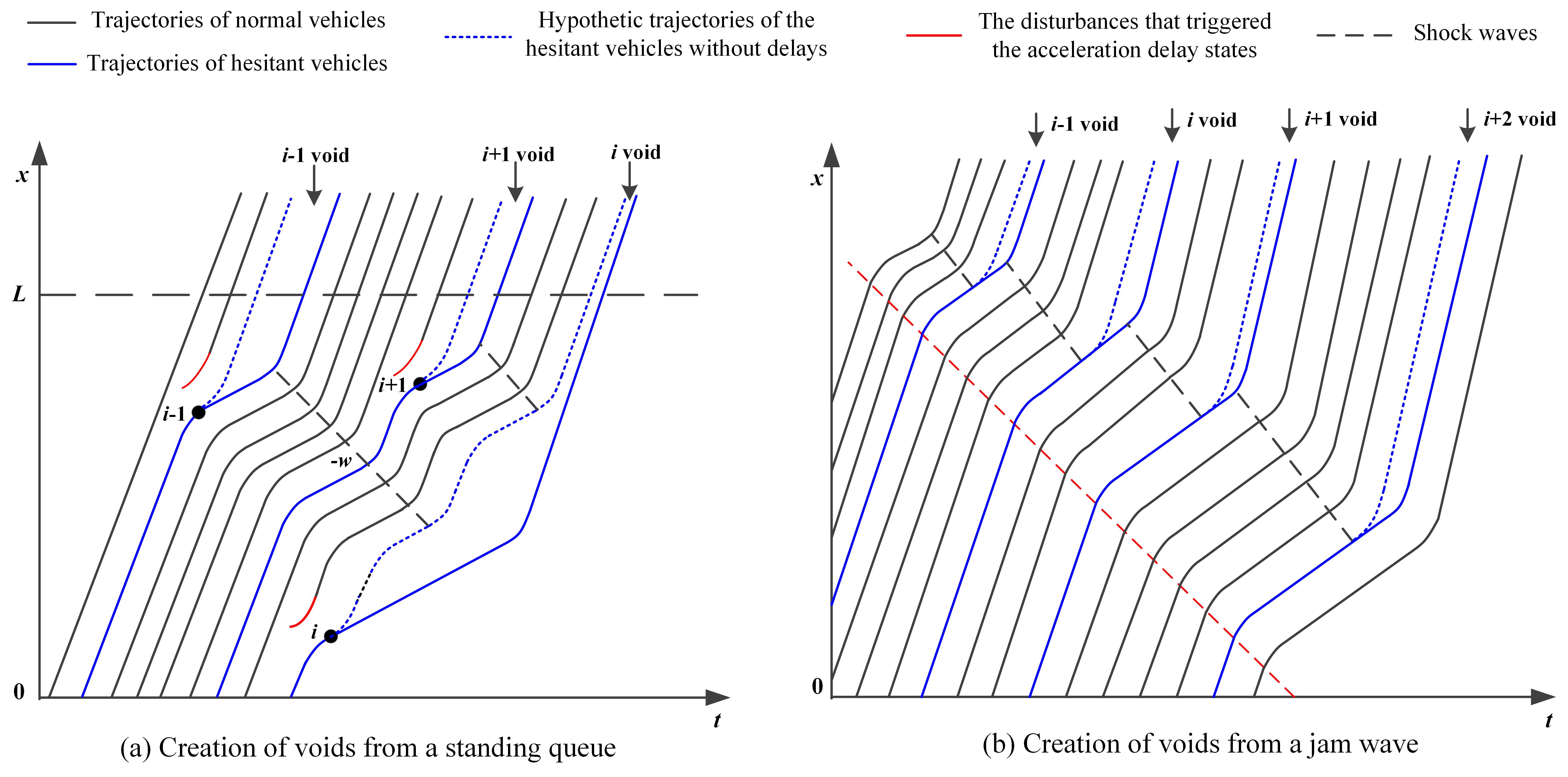}
	\caption{Creation of voids from (a) an active bottleneck and (b) a jam wave.}
	\label{fig:description}
\end{figure}

We attribute this difference to the role of wave–void interactions. Figure \ref{fig:description} (a) illustrates an example of wave-void interactions occurring in an active bottleneck. The positions of 0 and $L$ represent the upstream and downstream ends of the bottleneck area, respectively. Disturbances that trigger acceleration delays, shown as red curves in the figure, may occur at any position within the bottleneck. In this example, suppose three delays—$i-1$, $i$, and $i+1$—are triggered sequentially over time. Since the triggering positions of vehicles $i-1$ and $i+1$ are downstream of vehicle $i$, the waves they generate both intersect with the void created by vehicle $i$. Consequently, the initially large void created by vehicle $i$ diminishes after these wave-void interactions. This wave-void interaction mitigates the reduction in QDF, counteracting the capacity drop.

Figure \ref{fig:description} (b) illustrates how voids are created by acceleration delays in jam waves. Vehicles passing through a jam wave accelerate one after another, meaning that the position where vehicle $i$ is triggered is generally downstream of the position for vehicle $i+1$. As a result, the chances of wave-void interaction are significantly lower in jam waves compared to standing queues that form at active bottlenecks. Since wave-void interactions help mitigate the reduction in QDF, the fewer instances of wave-void interactions in jam waves may explain why the capacity drop is greater in jam waves compared to standing queues. The following section analytically derives the QDFs with and without wave-void interactions. 

\subsection{Analytical model to estimate QDF}
\label{subsec: analytical}

\begin{figure}[!ht]
	\centering
 	\includegraphics[width=0.75\textwidth]{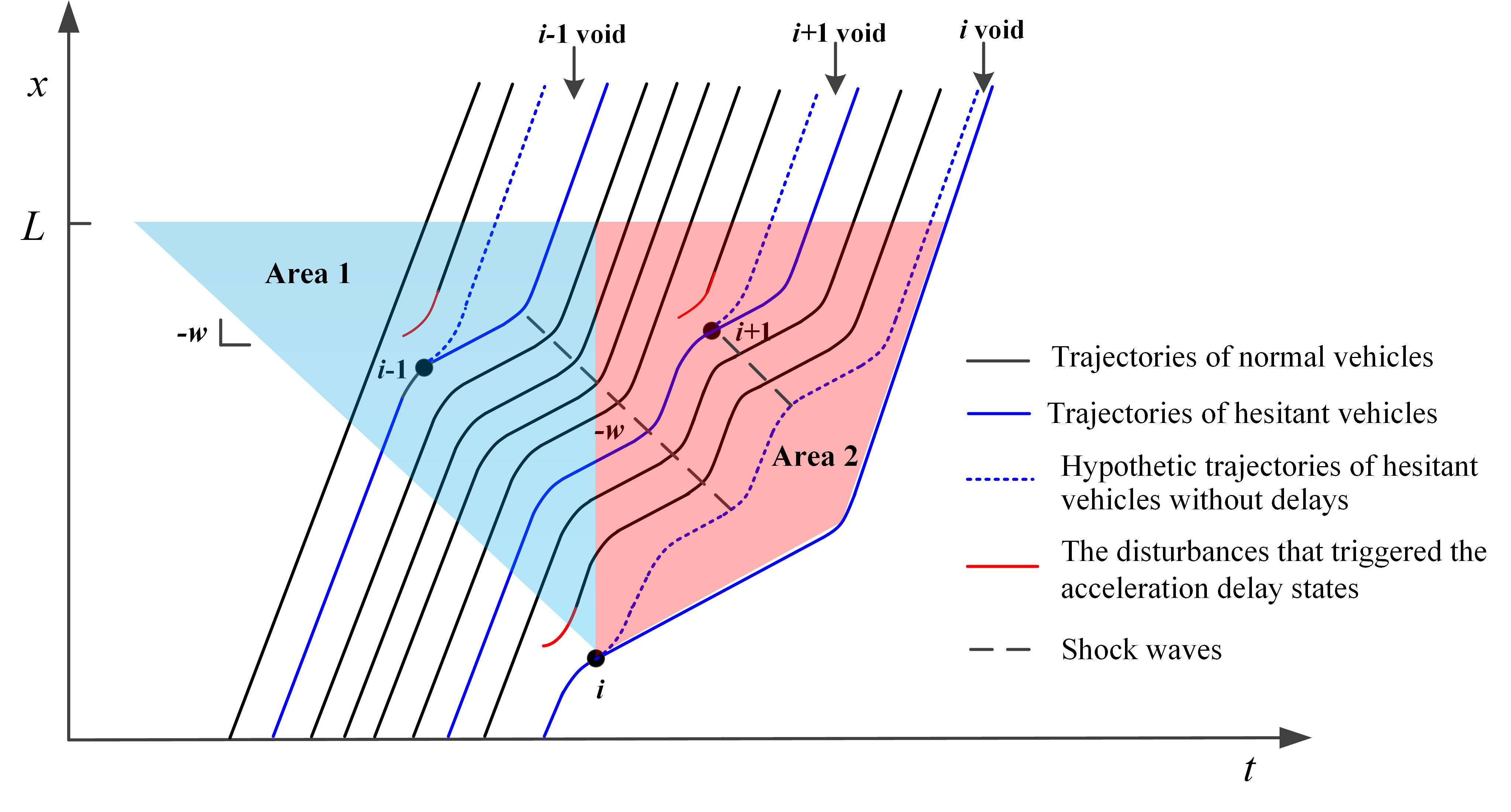}
	\caption{The probabilities of wave-void interactions.}
	\label{fig:probability}
\end{figure}

To analytically derive the formula for estimating the QDF, we begin by formulating the probability of wave-void interaction. Let $p_{\text{int}}^{i \rightarrow{i-1}}$ and $p_{\text{int}}^{i \rightarrow{i+1}}$ represent the probabilities that the waves induced by vehicles $i-1$ and $i+1$, respectively, affect the void created by vehicle $i$. Let $(t_i, x_i)$ denote the time and location where vehicle $i$ would begin to accelerate in the absence of any response delay. The void created by vehicle $i$ can potentially be influenced by waves generated by both vehicles $i-1$ and $i+1$, provided that $x_{i-1} > x_i$ and $x_{i+1} > x_i$. Given $(t_i, x_i)$, the probability that the wave induced by vehicle $i-1$ interacts with the void of vehicle $i$ is equivalent to the probability that $(t_{i-1}, x_{i-1})$ falls within area 1, as shown in Figure \ref{fig:probability}. Similarly, the probability that the wave induced by vehicle $i+1$ interacts with the void created by vehicle $i$ corresponds to the probability that $(t_{i+1}, x_{i+1})$ falls within area 2.

\begin{figure}[!ht]
	\centering
 	\includegraphics[width=0.48\textwidth]{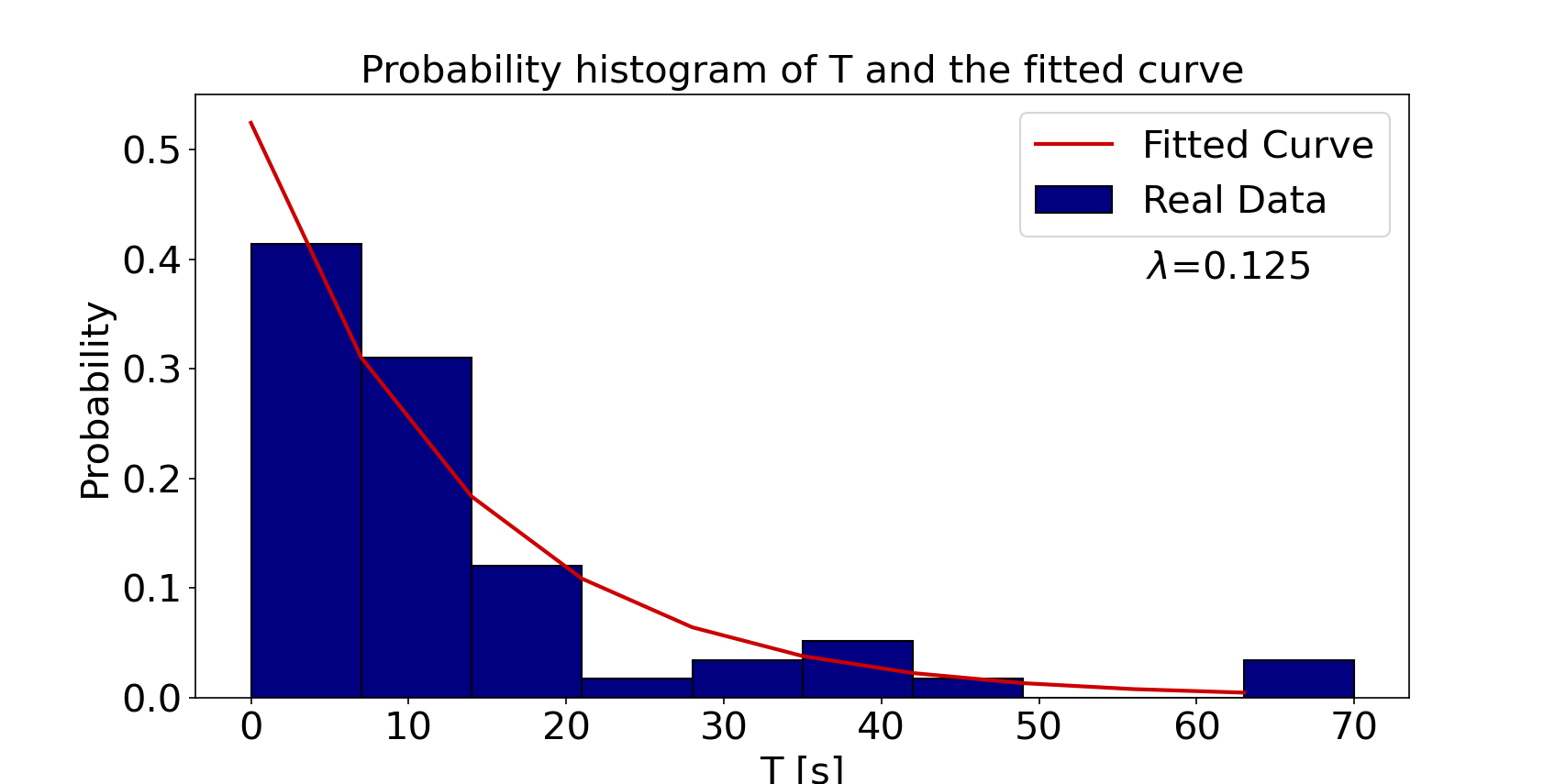}
 	\includegraphics[width=0.48\textwidth]{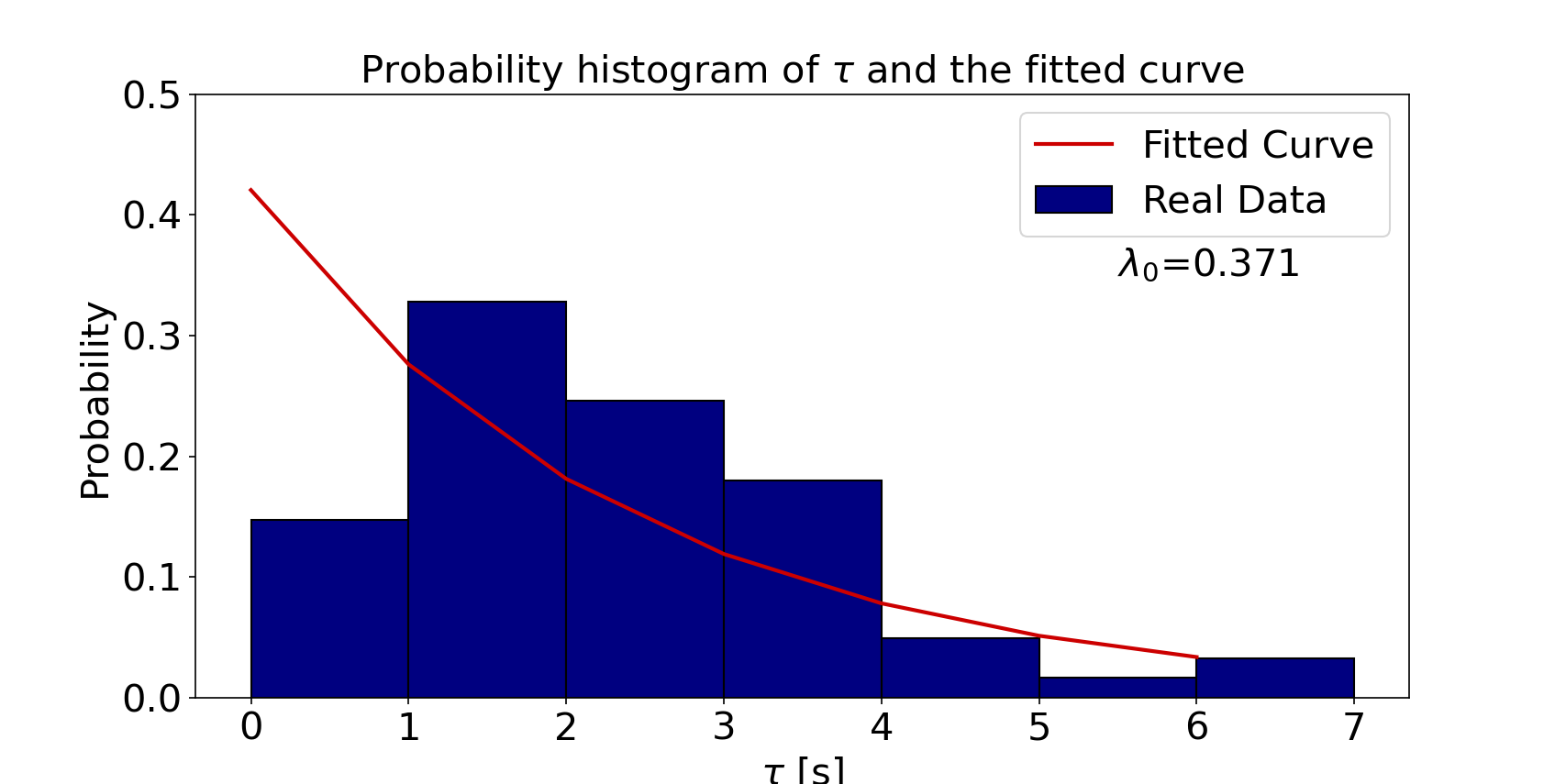}
	\caption{Fitting with real data for $T$ (left) and $\tau$ (right).}
	\label{fig:fitting}
\end{figure}

To derive the formulations of $p_{\text{int}}^{i \rightarrow i-1}$ and $p_{\text{int}}^{i \rightarrow i+1}$, the following assumptions are made regarding the spatial and temporal distributions of vehicles entering the acceleration delay state and the associated response delays:

\begin{enumerate} 

\item It is assumed that the triggering time and spatial location of the acceleration delay state are mutually independent.

\item The time intervals, $T$, between the triggering of consecutive delayed states follow a negative exponential distribution:
\begin{align}\label{time}
f(T; \lambda) = \lambda e^{-\lambda T},
\end{align}
where $\lambda$ is the rate parameter of the distribution for $T$. The left picture in Figure \ref{fig:fitting} shows the probability histogram of $T$ with a bin width of 5 second, along with the fitted curve based on data collected from site 1. 

\item The spatial location, $x$, where each delay is triggered follows a uniform distribution:
\begin{align}\label{location}
f(x; L) = \frac{1}{L},
\end{align}
where $L$ is the length of the bottleneck area.

\item The duration of the acceleration response delay, $\tau$, follows a negative exponential distribution:
\begin{align}\label{tau}
f(\tau; \lambda_{0}) = \lambda_{0} e^{-\lambda_{0} \tau},
\end{align}
where $\lambda_{0}$ is the rate parameter of the distribution for $\tau$. The right picture in Figure \ref{fig:fitting} shows the probability histogram of $\tau$ with a bin width of 1 second, along with the fitted curve based on data collected from site 1. It is worth noting that the empirical data exhibit a relatively low probability near zero compared to a standard exponential distribution. We selected the exponential distribution because it captures the overall trend of the observed delays while enabling a tractable analytical derivation. Alternative distributions such as the Gamma distribution could provide additional flexibility in fitting but would make the derivation intractable and reduce the analytical insights of the framework. More importantly, the accuracy of the proposed analytical model depends on the deviation between sampled and observed data. Although the exponential distribution underestimates the probability mass near zero, it reproduces the combined probability of the first two intervals with good accuracy. As a result, the overall deviation remains small, ensuring that the model outcomes are not significantly affected by this approximation.

\end{enumerate}

It is important to note that lane change is considered an endogenous factor triggering the response delay states. Hence, the proposed model does not explicitly account for lane-change behavior but incorporates its impact by assuming its spatial and temporal distributions. Based on the assumed distributions of $T$ and $x$, the probability of wave-void interaction $p_{\text{int}}^{i \rightarrow i-1}$ can be derived using the following equation:   

\begin{equation}
\begin{aligned}
\label{eq:interaction_left}
  p_{\operatorname{int}}^{i\to i-1} & =\frac{1}{L}\int_{0}^{L}{\int_{0}^{\frac{L-x}{w}}{\lambda }{{e}^{-\lambda t}}\cdot \frac{L-x-wt}{L}dt}dx \\ 
 & =-\frac{{{w}^{2}}}{{{\lambda }^{2}}{{L}^{2}}}{{e}^{-\lambda \frac{L}{w}}}+\frac{{{w}^{2}}}{{{\lambda }^{2}}{{L}^{2}}}-\frac{w}{\lambda L}+\frac{1}{2} \\ 
\end{aligned}
\end{equation}
where $w$ is the congestion wave speed. For brevity, the detailed derivation process has been moved to the appendix. Similarly, for a given $\tau_i$, the probability $p_{\text{int}}^{i \rightarrow i+1}$ can be derived using the following equations:

\begin{equation}
\begin{aligned}
\label{eq:interaction_right}
p_{\operatorname{int}}^{i\to i+1} & = \frac{1}{L}\int_{0}^{L} \left( \int_{0}^{\tau }{\lambda }{{e}^{-\lambda t}}\cdot \frac{L-x-{{v}_{0}}t}{L}dt+\int_{\tau }^{\frac{L-x-{{v}_{0}}\tau }{{{v}_\text{f}}}+\tau }{\lambda }{{e}^{-\lambda t}}\cdot (\frac{L-x-{{v}_{0}}\tau -{{v}_\text{f}}\left( t-\tau  \right)}{L})dt \right) dx \\ 
 & =\frac{1}{L}\left( \frac{{{v}_{0}}-{{v}_\text{f}}}{\lambda }{{e}^{-\lambda \tau }}+\frac{L\lambda -{{v}_{0}}}{\lambda }-\frac{L}{2}+\frac{{{v}_\text{f}}^{2}}{L{{\lambda }^{2}}}{{e}^{-\lambda \frac{-{{v}_{0}}\tau +{{v}_\text{f}}\tau }{{{v}_\text{f}}}}}-\frac{{{v}_\text{f}}^{2}}{L{{\lambda }^{2}}}{{e}^{-\lambda \frac{L-{{v}_{0}}\tau +{{v}_\text{f}}\tau }{{{v}_\text{f}}}}} \right) \\ 
\end{aligned}
\end{equation}

\begin{figure}[H]
	\centering
 	\includegraphics[width=0.75\textwidth]{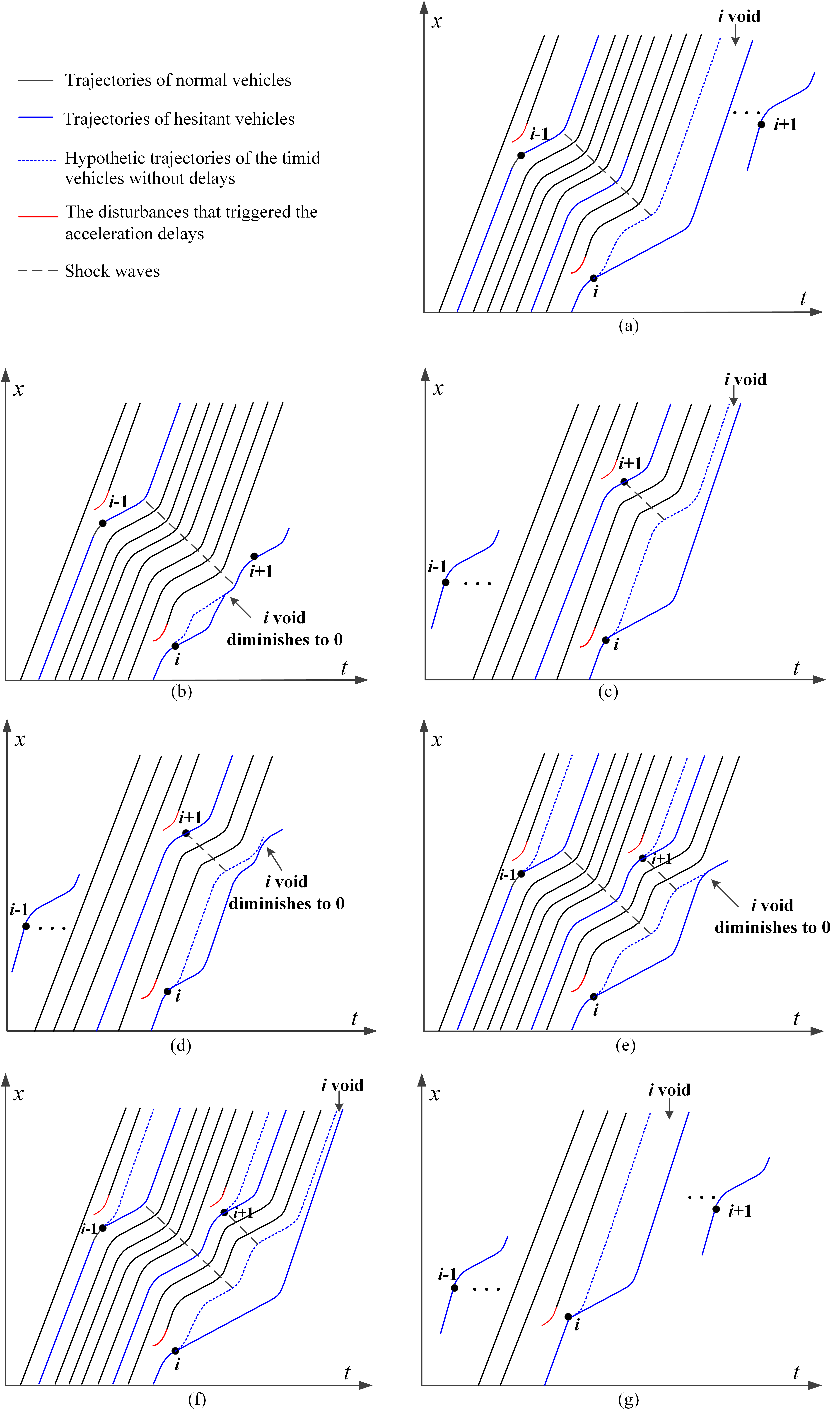}
	\caption{Illustration of the seven wave–void interaction scenarios.}
	\label{fig:seven}
\end{figure}

To derive the formulation for the expected void created by vehicle $i$, denoted as $\mathbb{E}[s_{\text{void}}^i]$, we first consider the expected void given that $\tau_{i}$ is known, i.e., $\mathbb{E}[s_{\text{void}}^i \mid \tau_{i}]$. The following conditions are considered to describe the different scenarios of wave-void interaction and the resulting void size created by vehicle $i$:

\begin{enumerate}
 
    \item As depicted in Figure \ref{fig:seven} (a), if the void of vehicle $i$ interacts only with the wave created by vehicle $i-1$, and the acceleration delay of vehicle $i$ ($\tau_i$) is greater than that of vehicle $i-1$ ($\tau_{i-1}$), then the size of the void created by vehicle $i$ is given by $s_{\text{void}}^{i} = (v_{\text{f}} - v_{\text{0}}) \cdot (\tau_i - \tau_{i-1})$. The probability of this scenario occurring is:

    \begin{equation}
    P_1 = p(\tau_{i-1} \leq \tau_i) \cdot p_{\text{int}}^{i \rightarrow i-1} \cdot (1 - p_{\text{int}}^{i \rightarrow i+1})
    \end{equation}
    where the probability that $\tau_{i-1} < \tau_i$ is given by $p(\tau_{i-1} < \tau_i) = 1 - e^{-\lambda_{0} \tau_{i}}$. 
    
    \item As depicted in Figure \ref{fig:seven} (b), if the void of vehicle $i$ interacts only with the wave created by vehicle $i-1$, and $\tau_i \leq \tau_{i-1}$, then the void diminishes to 0, and $s_{\text{void}}^{i} = 0$. In this case, the downstream wave serves as the next triggering event (i.e., $i+1$) for the acceleration delay state, and the void associated with that event is calculated independently from the current one.
    
    \item As depicted in Figure \ref{fig:seven} (c), if the void of vehicle $i$ interacts only with the wave created by vehicle $i+1$, and $\tau_i > \tau_{i+1}$, then the void size is given by $s_{\text{void}}^{i} = (v_{\text{f}} - v_{\text{0}}) \cdot (\tau_i - \tau_{i+1})$. The probability of this scenario is:
    \begin{equation}
    P_3 = p(\tau_{i+1} \leq \tau_i) \cdot (1 - p_{\text{int}}^{i \rightarrow i-1}) \cdot p_{\text{int}}^{i \rightarrow i+1},
    \end{equation}
    where $p(\tau_{i+1} \leq \tau_i) = 1 - e^{-\lambda_{0} \tau_{i}}$.
    
    \item As depicted in Figure \ref{fig:seven} (d), if the void of vehicle $i$ interacts only with the wave created by vehicle $i+1$, and $\tau_i \leq \tau_{i+1}$, then $s_{\text{void}}^{i} = 0$.
    
    \item As depicted in Figure \ref{fig:seven} (e), if the void of vehicle $i$ interacts with the waves of both vehicles $i-1$ and $i+1$, and $\tau_i \leq \tau_{i-1} + \tau_{i+1}$, then $s_{\text{void}}^{i} = 0$.
    
    \item As depicted in Figure \ref{fig:seven} (f), if the void of vehicle $i$ interacts with the waves created by both vehicles $i-1$ and $i+1$, and $\tau_i > \tau_{i-1} + \tau_{i+1}$, then $s_{\text{void}}^{i} = (v_{\text{f}} - v_{\text{0}}) \cdot (\tau_i - \tau_{i-1} - \tau_{i+1})$. The probability of this scenario is:
    \begin{equation}
    P_6 = P(z < \tau_i) \cdot p_{\text{int}}^{i \rightarrow i-1} \cdot p_{\text{int}}^{i \rightarrow i+1}, 
    \end{equation}
    where $z = \tau_{i-1} + \tau_{i+1}$ follows a gamma distribution with shape parameter 2 and rate parameter $\lambda_{0}$:
    \begin{equation}
    f(z) = \lambda_{0}^{2} z e^{-\lambda z}.
    \end{equation}
    The probability that $z < \tau_i$ is given by $p(z < \tau_i) = 1 - \left( \lambda _0 \tau_i + 1 \right) e^{-\lambda_0 \tau_i}$.
    
    \item As depicted in Figure \ref{fig:seven} (g), if the void of vehicle $i$ interacts with neither the wave created by vehicle $i-1$ nor $i+1$, then the void size is $s_{\text{void}}^{i} = (v_{\text{f}} - v_{\text{0}}) \cdot \tau_i$. The probability of this scenario is:
    \begin{equation}
    P_7 = (1 - p_{\text{int}}^{i \rightarrow i-1}) \cdot (1 - p_{\text{int}}^{i \rightarrow i+1}).
    \end{equation}
    
\end{enumerate}

Consequently, for a given $\tau_i$, the expected void created by vehicle $i$ is calculated as:

\begin{equation}
\label{eq:void_expectation}
\begin{aligned}
\mathbb{E}[s_{\text{void}}^i \mid \tau_i] = & (v_{\text{f}} - v_{\text{0}}) \cdot \big( \mathbb{E} [P_1 \cdot (\tau_i - \tau_{i-1}) \mid \tau_i] +  \mathbb{E} [P_3 \cdot (\tau_i - \tau_{i+1}) \mid \tau_i] \\
& + \mathbb{E} [P_6 \cdot (\tau_i - \tau_{i-1} - \tau_{i+1}) \mid \tau_i] + \mathbb{E} [P_7 \cdot \tau_{i} \mid \tau_i] \big). \\
\end{aligned}
\end{equation}

To derive the formulation of $\mathbb{E}[s_{\text{void}}^i \mid \tau_i]$, each term is calculated individually, starting with the term associated with scenario 1:

\begin{equation}
\label{eq:term 1}
\begin{aligned}
&\mathbb{E} [P_1 \cdot (\tau_{i} - \tau_{i-1}) \mid \tau_i] \\
&= P_1\cdot \left( \tau _i- \frac{\int_0^{\tau_{i}}{\tau_{i-1} f(\tau_{i-1}; \lambda_{0}) }d\tau_{i-1}}{\int_0^{\tau_{i}}f(\tau_{i-1}; \lambda_{0})d\tau _{i-1}} \right) \\
&=\frac{1}{\lambda _0}(-\frac{{{w}^{2}}}{{{\lambda }^{2}}{{L}^{2}}}{{e}^{-\lambda \frac{L}{w}}}+\frac{{{w}^{2}}}{{{\lambda }^{2}}{{L}^{2}}}-\frac{w}{\lambda L}+\frac{1}{2}) \cdot \left( \lambda _0\tau _i-1+e^{-\lambda _0\tau _i} \right) \\
&\; \; \; \cdot \left(1 - \frac{1}{L}\left( \frac{{{v}_{0}}-{{v}_\text{f}}}{\lambda }{{e}^{-\lambda \tau_{i} }}+\frac{L\lambda -{{v}_{0}}}{\lambda }-\frac{L}{2}+\frac{{{v}_\text{f}}^{2}}{L{{\lambda }^{2}}}{{e}^{-\lambda \frac{-{{v}_{0}}\tau_{i} +{{v}_\text{f}}\tau_{i} }{{{v}_\text{f}}}}}-\frac{{{v}_\text{f}}^{2}}{L{{\lambda }^{2}}}{{e}^{-\lambda \frac{L-{{v}_{0}}\tau_{i} +{{v}_\text{f}}\tau_{i} }{{{v}_\text{f}}}}} \right) \right) \\
\end{aligned}
\end{equation}

Similarly, the term associated with scenario 3 is calculated as follows:

\begin{equation}
\label{eq:term 3}
\begin{aligned}
&\mathbb{E} [P_3 \cdot (\tau_i - \tau_{i+1}) \mid \tau_i] \\
&= P_3 \cdot \left( \tau _i- \frac{\int_0^{\tau _i}{\tau_{i+1} f(\tau_{i+1}; \lambda_{0}) }d\tau_{i+1}}{\int_0^{\tau _i}f(\tau_{i+1}; \lambda_{0})d\tau _{i+1}} \right) \\
&=\frac{1}{\lambda _0}(\frac{{{w}^{2}}}{{{\lambda }^{2}}{{L}^{2}}}{{e}^{-\lambda \frac{L}{w}}}-\frac{{{w}^{2}}}{{{\lambda }^{2}}{{L}^{2}}}+\frac{w}{\lambda L}+\frac{1}{2}) \cdot \left( \lambda _0\tau _i-1+e^{-\lambda _0\tau _i} \right) \\
&\; \; \; \cdot \frac{1}{L}\left( \frac{{{v}_{0}}-{{v}_\text{f}}}{\lambda }{{e}^{-\lambda \tau_{i} }}+\frac{L\lambda -{{v}_{0}}}{\lambda }-\frac{L}{2}+\frac{{{v}_\text{f}}^{2}}{L{{\lambda }^{2}}}{{e}^{-\lambda \frac{-{{v}_{0}}\tau_{i} +{{v}_\text{f}}\tau_{i} }{{{v}_\text{f}}}}}-\frac{{{v}_\text{f}}^{2}}{L{{\lambda }^{2}}}{{e}^{-\lambda \frac{L-{{v}_{0}}\tau_{i} +{{v}_\text{f}}\tau_{i} }{{{v}_\text{f}}}}} \right) \\
\end{aligned}
\end{equation}

For the term associated with scenario 6:

\begin{equation}
\label{eq:term 6}
\begin{aligned}
&\mathbb{E} [P_6 \cdot (\tau_i - z) \mid \tau_i] \\
&= P_6 \cdot \left( \tau _i- \frac{\int_0^{\tau _i}{z f(z) }d\tau_{i+1}}{\int_0^{\tau _i}f(z)dz} \right) \\
&=\frac{1}{\lambda _0}(\frac{{{w}^{2}}}{{{\lambda }^{2}}{{L}^{2}}}{{e}^{-\lambda \frac{L}{w}}}-\frac{{{w}^{2}}}{{{\lambda }^{2}}{{L}^{2}}}+\frac{w}{\lambda L}+\frac{1}{2}) \cdot \left( \left( \lambda _0\tau _i-2 \right) +\left( \lambda _0\tau _i+2 \right) e^{-\lambda _0\tau _i} \right) \\
&\; \; \; \cdot\left(1 - \frac{1}{L}\left( \frac{{{v}_{0}}-{{v}_\text{f}}}{\lambda }{{e}^{-\lambda \tau_{i} }}+\frac{L\lambda -{{v}_{0}}}{\lambda }-\frac{L}{2}+\frac{{{v}_\text{f}}^{2}}{L{{\lambda }^{2}}}{{e}^{-\lambda \frac{-{{v}_{0}}\tau_{i} +{{v}_\text{f}}\tau_{i} }{{{v}_\text{f}}}}}-\frac{{{v}_\text{f}}^{2}}{L{{\lambda }^{2}}}{{e}^{-\lambda \frac{L-{{v}_{0}}\tau_{i} +{{v}_\text{f}}\tau_{i} }{{{v}_\text{f}}}}} \right) \right) \\
\end{aligned}
\end{equation}

For the term associated with scenario 7:

\begin{equation}
\begin{aligned}
\label{eq:term 7}
&\mathbb{E} [P_7 \cdot \tau_i \mid \tau_i] \\
& = (\frac{{{w}^{2}}}{{{\lambda }^{2}}{{L}^{2}}}{{e}^{-\lambda \frac{L}{w}}}-\frac{{{w}^{2}}}{{{\lambda }^{2}}{{L}^{2}}}+\frac{w}{\lambda L}+\frac{1}{2}) \cdot \tau_i\\
&\; \; \; \cdot \left(1 - \frac{1}{L}\left( \frac{{{v}_{0}}-{{v}_\text{f}}}{\lambda }{{e}^{-\lambda \tau_{i} }}+\frac{L\lambda -{{v}_{0}}}{\lambda }-\frac{L}{2}+\frac{{{v}_\text{f}}^{2}}{L{{\lambda }^{2}}}{{e}^{-\lambda \frac{-{{v}_{0}}\tau_{i} +{{v}_\text{f}}\tau_{i} }{{{v}_\text{f}}}}}-\frac{{{v}_\text{f}}^{2}}{L{{\lambda }^{2}}}{{e}^{-\lambda \frac{L-{{v}_{0}}\tau_{i} +{{v}_\text{f}}\tau_{i} }{{{v}_\text{f}}}}} \right) \right) \\
\end{aligned}
\end{equation}

After deriving $\mathbb{E}[s_{\text{void}}^{i} \mid \tau_i]$, the expected void size for all delayed states, $\mathbb{E}[s_{\text{void}}]$, is computed using the law of total probability:

\begin{equation}
\label{eq:void_expectation2}
\mathbb{E}[s_{\text{void}}] = \int_{0}^{\infty} f(\tau_i; \lambda_{0}) \cdot
\mathbb{E}[s_{\text{void}}^{i} \mid \tau_i] d\tau_{i}
\end{equation}

The derivation of $\mathbb{E}[s_{\text{void}}]$ is also conducted on a term-by-term basis. The expected void size created by the vehicles in scenario 1, denoted as $\mathbb{E}[s_{\text{void-term 1}}]$, is calculated as:

\begin{equation}
\begin{aligned}
\label{eq:toal_probability_term1}
\mathbb{E}[s_{\text{void-term 1}}^{i}] & = \int_0^{\infty} (v_{\text{f}} - v_{\text{0}}) \cdot f(\tau_i; \lambda_{0}) \cdot \mathbb{E} [P_1 \cdot (\tau_{i} - \tau_{i-1}) \mid \tau_i] d\tau _i \\
\end{aligned}
\end{equation}

Incorporating the results of Equation \eqref{eq:term 1} into \eqref{eq:toal_probability_term1}, the final expression of $\mathbb{E}[s_{\text{void-term 1}}]$ is derived as, 

\begin{equation}
\begin{aligned}
\label{eq:term 1_x}
\mathbb{E}[s_{\text{void-term 1}}] &=\left( v_{\text{f}}-v_{\text{0}} \right) \left( -\frac{w^2}{\lambda ^2L^2}e^{-\lambda \frac{L}{w}}+\frac{w^2}{\lambda ^2L^2}-\frac{w}{\lambda L}+\frac{1}{2} \right) \\
&\; \; \; \cdot \left( \frac{1}{2\lambda _0}-\frac{1}{L}\left( \begin{array}{c}	\frac{v_{\text{0}}-v_{\text{f}}}{\lambda}\cdot \left( \frac{1}{\lambda +2\lambda _0}-\frac{\lambda}{\left( \lambda +\lambda _0 \right) ^2} \right) +\left( \frac{L\lambda -v_{\text{0}}}{\lambda}-\frac{L}{2} \right) \cdot \frac{1}{2\lambda _0}\\	+\frac{{v_{\text{f}}}^2}{L\lambda ^2}\cdot \left( 1-e^{-\frac{\lambda L}{v_{\text{f}}}} \right) \cdot \left( \frac{\lambda v_{\text{f}}\left( v_{\text{0}}-v_{\text{f}} \right)}{\left( \lambda v_{\text{0}}-\lambda v_{\text{f}}-\lambda _0v_{\text{f}} \right) ^2}-\frac{v_{\text{f}}}{\lambda v_{\text{0}}-\lambda v_{\text{f}}-2\lambda _0v_{\text{f}}} \right)\\\end{array} \right) \right) \\
\end{aligned}
\end{equation}

The expected void size created by the vehicles in scenario 3, denoted as $\mathbb{E}[s_{\text{void-term 3}}]$, is calculated as:

\begin{equation}
\begin{aligned}
\label{eq:term 3_x}
\mathbb{E}[s_{\text{void-term 3}}^{i}] & = \int_0^{\infty} (v_{\text{f}} - v_{\text{0}}) \cdot f(\tau_i; \lambda_{0}) \cdot \mathbb{E} [P_3 \cdot (\tau_i - \tau_{i+1}) \mid \tau_i] d\tau _i \\
&=\left( v_{\text{f}} - v_{\text{0}} \right) \left( \frac{w^2}{\lambda ^2L^2}e^{-\lambda \frac{L}{w}}-\frac{w^2}{\lambda ^2L^2}+\frac{w}{\lambda L}+\frac{1}{2} \right) \\
&\cdot \frac{1}{L}\left( \begin{array}{c}	\frac{v_{\text{0}}-v_{\text{f}}}{\lambda}\cdot \left( \frac{1}{\lambda +2\lambda _0}-\frac{\lambda}{\left( \lambda +\lambda _0 \right) ^2} \right) +\left( \frac{L\lambda -v_{\text{0}}}{\lambda}-\frac{L}{2} \right) \cdot \frac{1}{2\lambda _0}\\	+\frac{{v_{\text{f}}}^2}{L\lambda ^2}\cdot \left( 1-e^{-\frac{\lambda L}{v_{\text{f}}}} \right) \cdot \left( \frac{\lambda v_{\text{f}}\left( v_{\text{0}}-v_{\text{f}} \right)}{\left( \lambda v_{\text{0}}-\lambda v_{\text{f}}-\lambda _0v_{\text{f}} \right) ^2}-\frac{v_{\text{f}}}{\lambda v_{\text{0}}-\lambda v_{\text{f}}-2\lambda _0v_{\text{f}}}. \right)\\\end{array} \right) \\
\end{aligned}
\end{equation}

The expected void size created by the vehicles in scenario 6, denoted as $\mathbb{E}[s_{\text{void-term 6}}]$, is calculated as:

\begin{equation}
\begin{aligned}
\label{eq:term 6_x}
\mathbb{E}[s_{\text{void-term 6}}^{i}] & = \int_0^{\infty} (v_{\text{f}} - v_{\text{0}}) \cdot f(\tau_i; \lambda_{0}) \cdot \mathbb{E} [P_6 \cdot (\tau_i - z) \mid \tau_i] d\tau _i \\
&={\lambda \lambda _0}^2\left( v_{\text{f}}-v_{\text{0}} \right)(\frac{{{w}^{2}}}{{{\lambda }^{2}}{{L}^{2}}}{{e}^{-\lambda \frac{L}{w}}}-\frac{{{w}^{2}}}{{{\lambda }^{2}}{{L}^{2}}}+\frac{w}{\lambda L}+\frac{1}{2}) \\
&\; \; \; \cdot\left( \begin{array}{c}	\left( 1-\frac{1}{L}\left( \frac{L\lambda -v_{\text{0}}}{\lambda}-\frac{L}{2} \right) \right) \left( \frac{2\lambda _0+3\lambda}{\left( \lambda +\lambda _0 \right) ^2}+\frac{1}{\lambda} \right) -\frac{1}{L}\frac{v_{\text{0}}-v_{\text{f}}}{\lambda ^4}\left( \frac{2\lambda _0+5\lambda}{\left( 2\lambda +\lambda _0 \right) ^2}-\frac{\lambda +2\lambda _0}{\left( \lambda +\lambda _0 \right) ^2} \right)\\	+\frac{1}{L}\left( 1-e^{-\frac{L\lambda}{v_{\text{f}}}} \right) \left( \frac{v_{\text{f}}\left( 2\lambda v_{\text{0}}-5\lambda v_{\text{f}}-2\lambda_0 v_{\text{f}} \right)}{\left( \lambda v_{\text{0}}-2\lambda v_{\text{f}}-\lambda _0v_{\text{f}} \right) ^2}+\frac{v_{\text{f}}\left( 2\lambda v_{\text{0}}-3\lambda v_{\text{f}}-2\lambda _0v_{\text{f}} \right)}{\left( \lambda v_{\text{0}}-\lambda v_{\text{f}}-\lambda _0v_{\text{f}} \right) ^2} \right)\\\end{array} \right) \\
\end{aligned}
\end{equation}

$\mathbb{E}[s_{\text{void-term 7}}]$ is calculated as:

\begin{equation}
\begin{aligned}
\label{eq:term 7_x}
\mathbb{E}[s_{\text{void-term 7}}^{i}] & = \int_0^{\infty} (v_{\text{f}} - v_{\text{0}}) \cdot f(\tau_i; \lambda_{0}) \cdot \mathbb{E} [P_7 \cdot \tau_i \mid \tau_i] d\tau _i \\
&= \lambda_{0} \left( v_{\text{f}}-v_{\text{0}} \right) \cdot (\frac{{{w}^{2}}}{{{\lambda }^{2}}{{L}^{2}}}{{e}^{-\lambda \frac{L}{w}}}-\frac{{{w}^{2}}}{{{\lambda }^{2}}{{L}^{2}}}+\frac{w}{\lambda L}+\frac{1}{2}) \\
&\; \; \; \cdot \left( \frac{1}{{\lambda _{0}}^2}-\frac{1}{L}\left( \begin{array}{c}	\frac{v_0-v_{\text{f}}}{\lambda}\cdot \frac{1}{\left( \lambda +\lambda _{0} \right) ^2}+\left( \frac{L\lambda -v_0}{\lambda}-\frac{L}{2} \right) \cdot \frac{1}{{\lambda _{0}}^2}\\	+\frac{{v_{\text{f}}}^2}{L\lambda ^2}\left( 1-e^{-\frac{L\lambda}{v_{\text{f}}}} \right) \left( \frac{{v_{\text{f}}}^2}{\left( \lambda v_0-\lambda v_{\text{f}}-\lambda _{0}v_{\text{f}} \right) ^2} \right)\\\end{array} \right) \right)\\
\end{aligned}
\end{equation}

Summing these four terms gives the final expression for $\mathbb{E}[s_{\text{void}}]$:

\begin{equation}
\mathbb{E}[s_{\text{void}}] = \mathbb{E}[s_{\text{void-term 1}}^{i}] + \mathbb{E}[s_{\text{void-term 3}}^{i}] + \mathbb{E}[s_{\text{void-term 6}}^{i}] + \mathbb{E}[s_{\text{void-term 7}}^{i}]
\end{equation}

Finally, the QDF, denoted by $Q_{\text{d}}$, is calculated as:

\begin{equation}
\label{eq:qdf_estimation}
Q_{\text{d}} = \frac{v_{\text{f}}} {s_{\text{cri}} + \alpha \mathbb{E}[s_{\text{void}}]}
\end{equation}
where $s_{\text{cri}}$ is the critical spacing at equilibrium, and $\alpha$ represents the expected probability of a vehicle entering an acceleration delay state, which is estimated from trajectory data as the proportion of vehicles exhibiting this state.

The proposed model assumes that one void intersects with the wave generated by only one downstream hesitant vehicle. It is possible, however, for the void to intersect with multiple downstream hesitant vehicles. To simplify the analytical derivation, interactions with additional downstream hesitant vehicles are not considered. Nevertheless, the same analytical framework can be extended to account for such interactions. In the following section, we demonstrate that the impact of considering multiple downstream hesitant vehicles is minor.

\section{Model validation}
\label{sec:validation}

This section evaluates the accuracy of the proposed analytical formula for estimating QDF. In Section \ref{subsec:comparison}, the estimated QDFs for two distinct types of traffic jams—jam waves and standing queues—are compared. Section \ref{subsec:numerical} contrasts the analytical estimates with results obtained from numerical simulations. Finally, Section \ref{subsec:data} validates the proposed formula using field data.

\subsection{QDF comparison between standing queues and jam waves}
\label{subsec:comparison}

This section explores the relationship between wave-void interaction and the previously observed phenomenon that QDFs from standing queues are significantly lower than those from jam waves. We assume that in a jam wave scenario, wave-void interactions do not occur because vehicles accelerate sequentially along the jam wave. As a result, the void created by a hesitant vehicle will not intersect with a downstream wave, as illustrated in Figure \ref{fig:description} (b). In this scenario, for a given $\tau_i$, the void size is expressed as: $s_{\text{void}}^{i} = (v_{\text{f}} - v_{\text{0}}) \cdot \tau_i$. Assuming $\tau$ follows an exponential distribution with parameter $\lambda_{0}$, the expected void created by a jam wave, denoted as $\mathbb{E}[s_{\text{void}}^{\text{jw}}]$, is calculated as:

\begin{equation}
\label{eq:void_jw}
\begin{aligned}
&\mathbb{E}[s_{\text{void}}^{\text{jw}}] = (v_{\text{f}} - v_{\text{0}}) \cdot \bar{\tau},
\end{aligned}
\end{equation}
where $\bar{\tau}$ is the mean response delay. By incorporating $\mathbb{E}[s_{\text{void}}^{\text{jw}}]$ into Eq. \eqref{eq:qdf_estimation}, the QDF for a jam wave scenario can be determined.

We calculate the QDFs for both standing queue and jam wave scenarios using the following settings. The equilibrium traffic state is modeled with a triangular fundamental diagram of density and flow. The parameters are as follows: the free-flow speed, $v_{\text{f}}$, is set to 20 m/s, and the critical spacing, $s_{\text{cri}}$, is set to 36 m, yielding a free-flow capacity of 2000 vehicles per hour per lane. The wave speed, $w$, is set to -5 m/s. The proportion of vehicles exhibiting the acceleration delay state, $\alpha$, is set to $\frac{1}{3}$. For standing queue scenarios, the length of the bottleneck area, $L$, is set to 400 m. The parameter $\lambda$, which characterizes the time interval between two consecutive hesitant vehicles being triggered, is set to $\frac{1}{6}$, representing an average interval of 6 seconds. The parameter $\lambda_{0}$, which is the reciprocal of the average response delay $\bar{\tau}$, varies from 0.1 to 2, corresponding to a mean response delay ranging from 10 to 0.5 seconds. When $\lambda_{0}$ changes, $v_{0}$, the speed before acceleration, is fixed at 10 m/s. Alternatively, when $v_{0}$ varies between 0 and 20 m/s, $\lambda_{0}$ is kept constant at 1 s. 

\begin{figure}[!h]
	\centering
 	\includegraphics[width=0.48\textwidth]{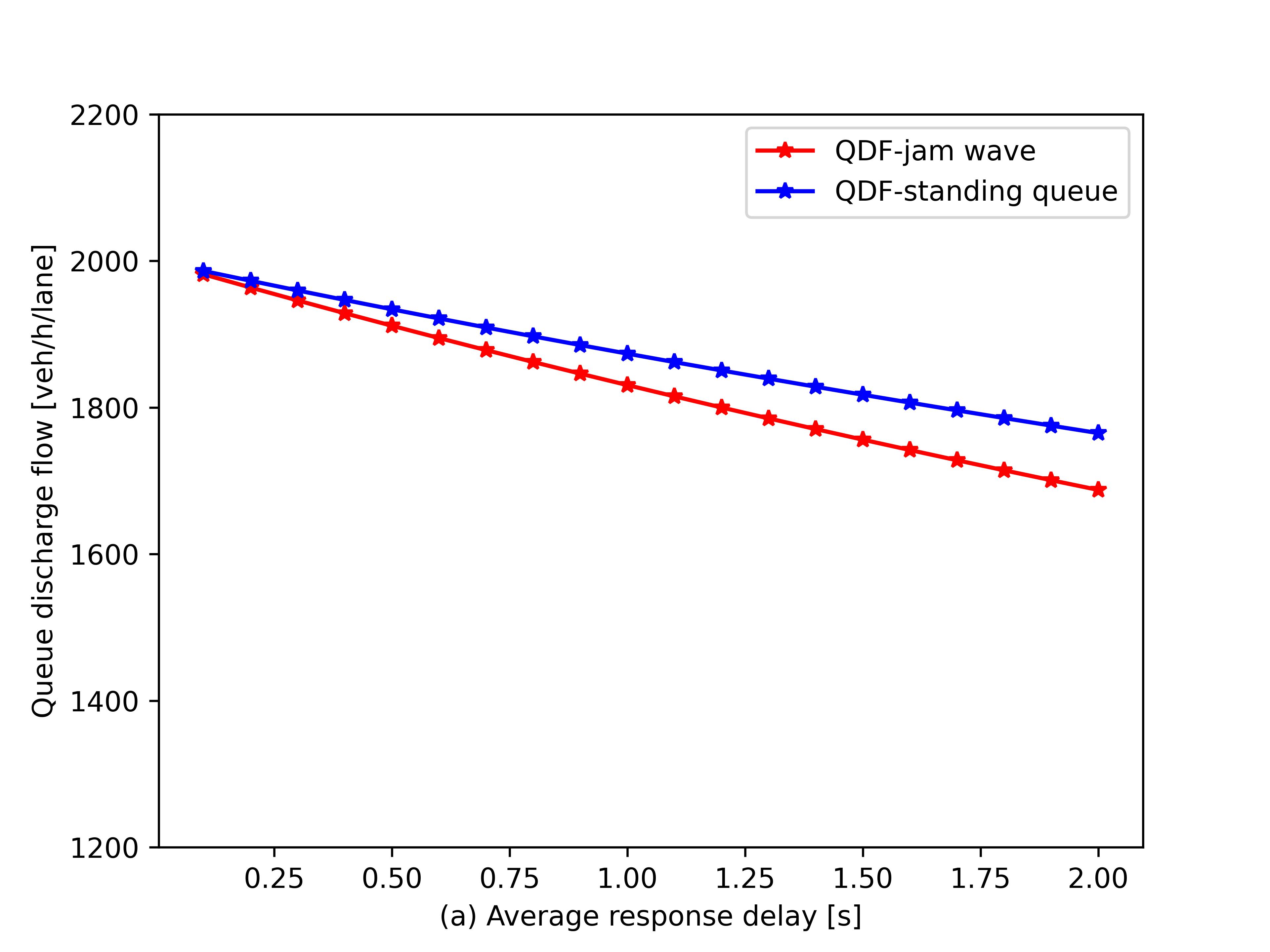}
 	\includegraphics[width=0.48\textwidth]{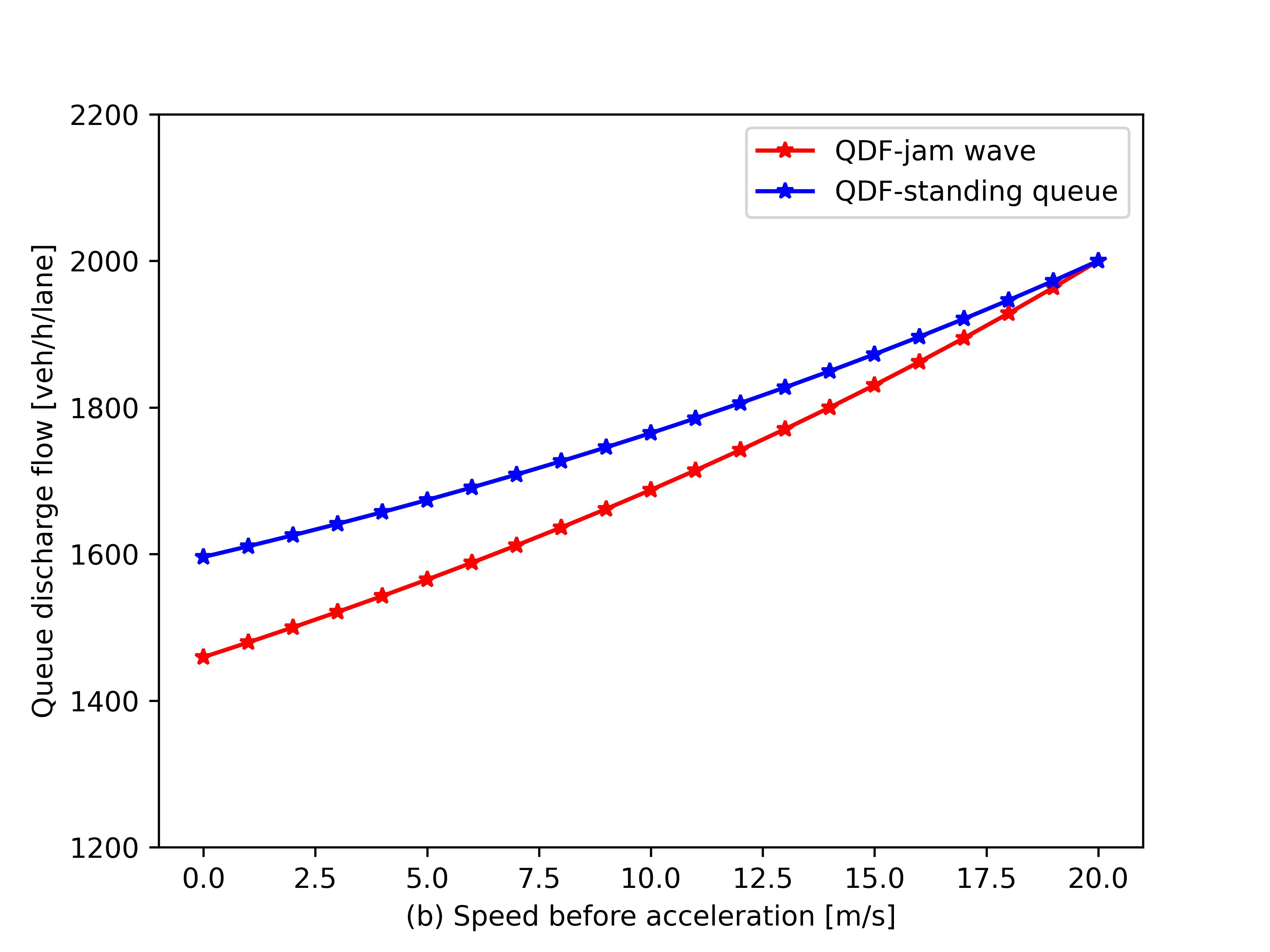}
	\caption{QDF comparison between jam wave and standing queue scenarios with the variation of $\bar{\tau}$ and $v_{\text{0}}$.}
	\label{fig:sq_jw}
\end{figure}

Figure \ref{fig:sq_jw} (a) and (b) present the estimated QDFs using the proposed analytical formula for jam waves and standing queues, with variations in $\bar{\tau}$ and $v_{0}$, respectively. It is clear that QDFs in jam waves are significantly lower than those in standing queues, emphasizing the substantial impact of wave-void interactions on capacity drop. For instance, when $v_{0}$ is as low as 0 m/s, the QDF reduction in a jam wave can reach up to 30\% with an average response delay of 1 second for hesitant vehicles. This finding aligns with the empirical observations in \cite{kerner2002empirical} and \cite{yuan2017capacity}. Two factors may contribute to this phenomenon: First, vehicles in a jam wave tend to travel at lower speeds compared to those in a standing queue, as also noted in \cite{yuan2017capacity}. Second, wave-void interactions occur less frequently in jam waves than in standing queues.

\subsection{Validation through Numerical Simulations}
\label{subsec:numerical}

This section evaluates the accuracy of the proposed analytical formula for estimating the QDF by comparing it with results obtained from numerical simulations. Unlike conventional traffic simulators, which model comprehensive traffic dynamics, the numerical simulation used here directly applies a Monte Carlo approach to replicate the physical processes outlined in Section \ref{subsec: illustration}. In this method, a large number of samples are generated, each representing a specific instance of interaction between vehicle $i$ and its neighboring vehicles, $i-1$ and $i+1$. For each sample, the time interval between vehicle triggers $T$, the triggering position $x$, and the acceleration delay $\tau$ for three consecutive hesitant vehicles—$i-1$, $i$, and $i+1$—are randomly selected from their respective distributions. The actual void created by vehicle $i$ is then calculated based on the scenario each sample represents, as described in Section \ref{subsec: analytical}.

QDFs estimated by the analytical formula were compared with those obtained from numerical simulations under several scenarios involving variations of parameters $\lambda_{0}$, $L$, and $v_{\text{0}}$. The settings for other parameters, including those defining the equilibrium traffic state, i.e., $v_\text{f}$, $s_\text{cri}$, and $w$, as well as the ratio and temporal distribution of vehicles in the acceleration delay state, $\alpha$ and $\lambda$, remain consistent with those described in Section \ref{subsec:comparison}.

\begin{enumerate}

    \item \textbf{QDFs under varying $\lambda_{0}$}: In this scenario, $\lambda_{0}$ varies from 0.1 to 2.0 with a 0.1 interval, representing a mean response delay ranging from 2 to 0.1 seconds. $v_{0}$ is fixed at 10 \( \text{m/s} \) for each vehicle, and the bottleneck length, $L$, is set to 400 m. For each $\lambda_{0}$, 10,000 samples (interaction instances) were randomly generated. In each sample, the time interval between vehicle $i$ and $i-1$, as well as $i$ and $i+1$, was randomly generated based on the exponential distribution with parameter $\lambda$. Triggering positions of vehicles were randomly selected from a uniform distribution within $L$, and response delays were sampled from an exponential distribution with parameter $\lambda_{0}$. QDFs from numerical simulations were computed based on the average void created in each sample.

    \item \textbf{QDFs under varying $L$}: Here, $v_{0}$ remains fixed at 10 \( \text{m/s} \), and $\lambda_{0}$ is set to 0.5, corresponding to a mean response delay of 2 seconds for hesitant vehicles. The bottleneck length, $L$, varies from 200 to 1000 m in 50 m increments. For each value of $L$, 10,000 samples were generated, using the same spatial and temporal distributions of vehicles in the acceleration delay state and their response delays as described previously.

    \item \textbf{QDFs under varying $v_{0}$}: In this scenario, the bottleneck length, $L$, is fixed at 400 m, and $\lambda_{0}$ is set to 0.5. The initial speed, $v_{0}$, varies from 0 to 20 m/s in 1 m/s increments. For each value of $v_{0}$, 10,000 samples were generated, again using the same distributions for vehicle spacing, triggering positions, and response delays.
    
\end{enumerate}

\begin{figure}[H]
	\centering
 	\includegraphics[width=0.48\textwidth]{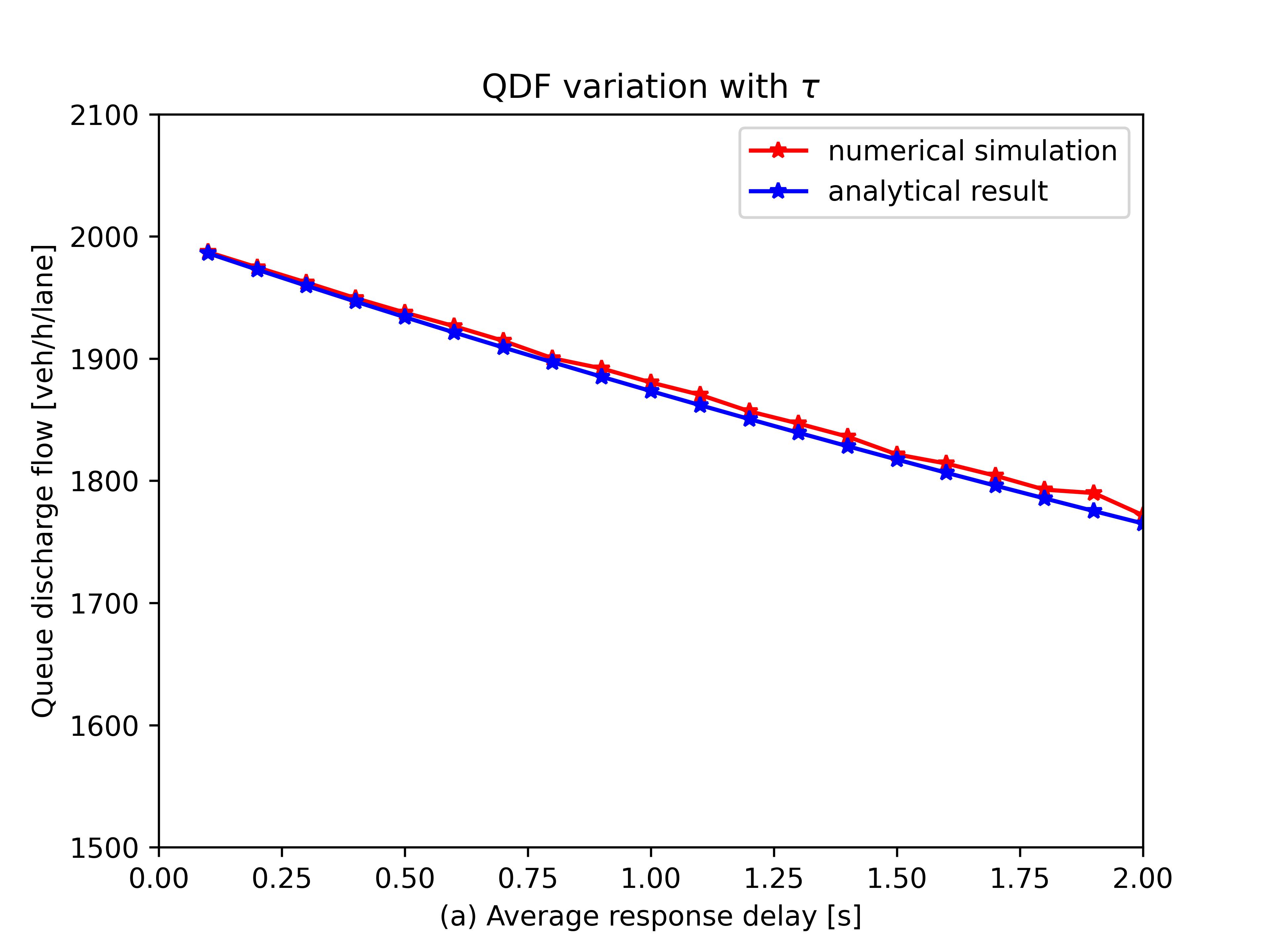}
 	\includegraphics[width=0.48\textwidth]{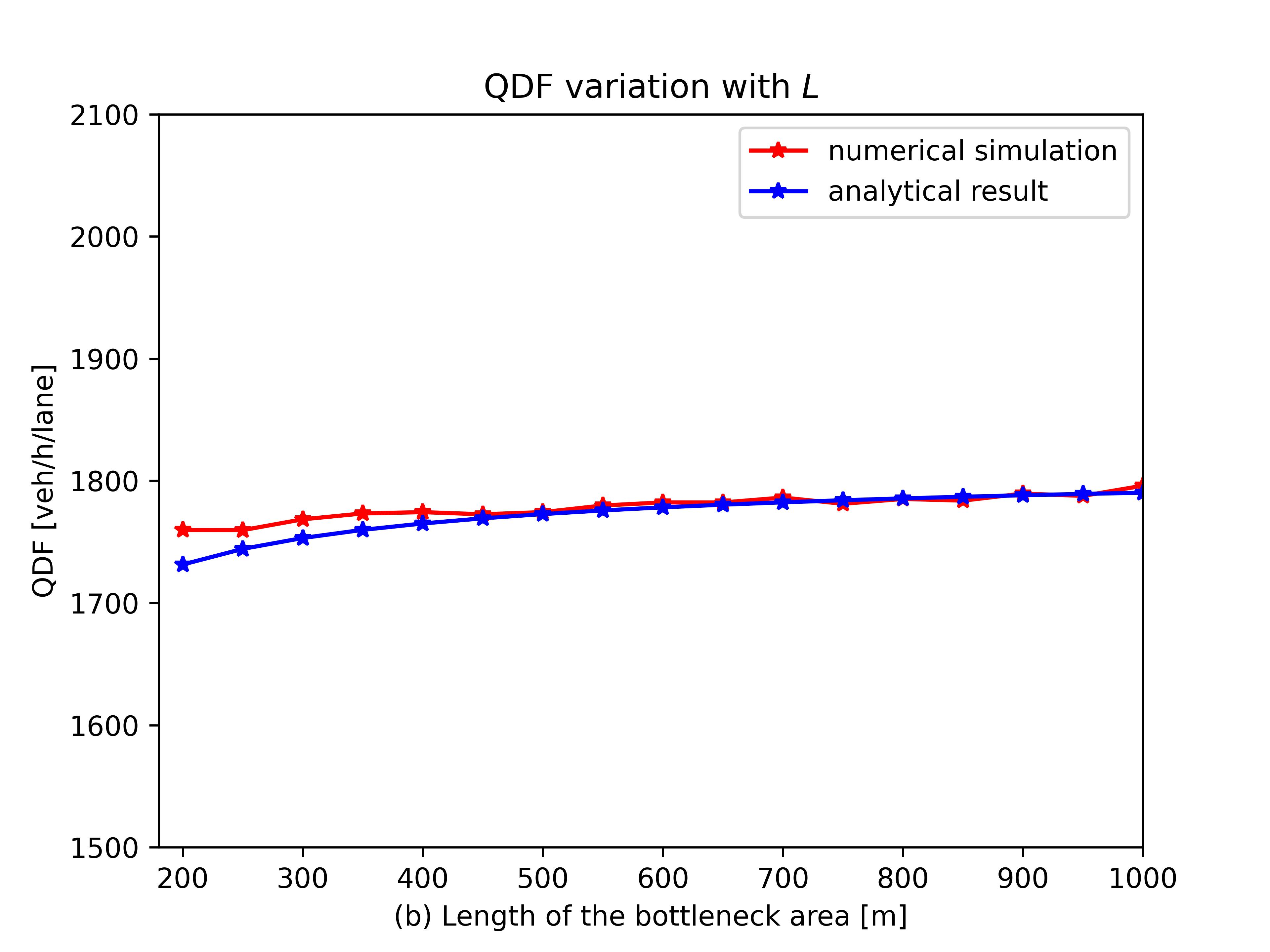}
 	\includegraphics[width=0.48\textwidth]{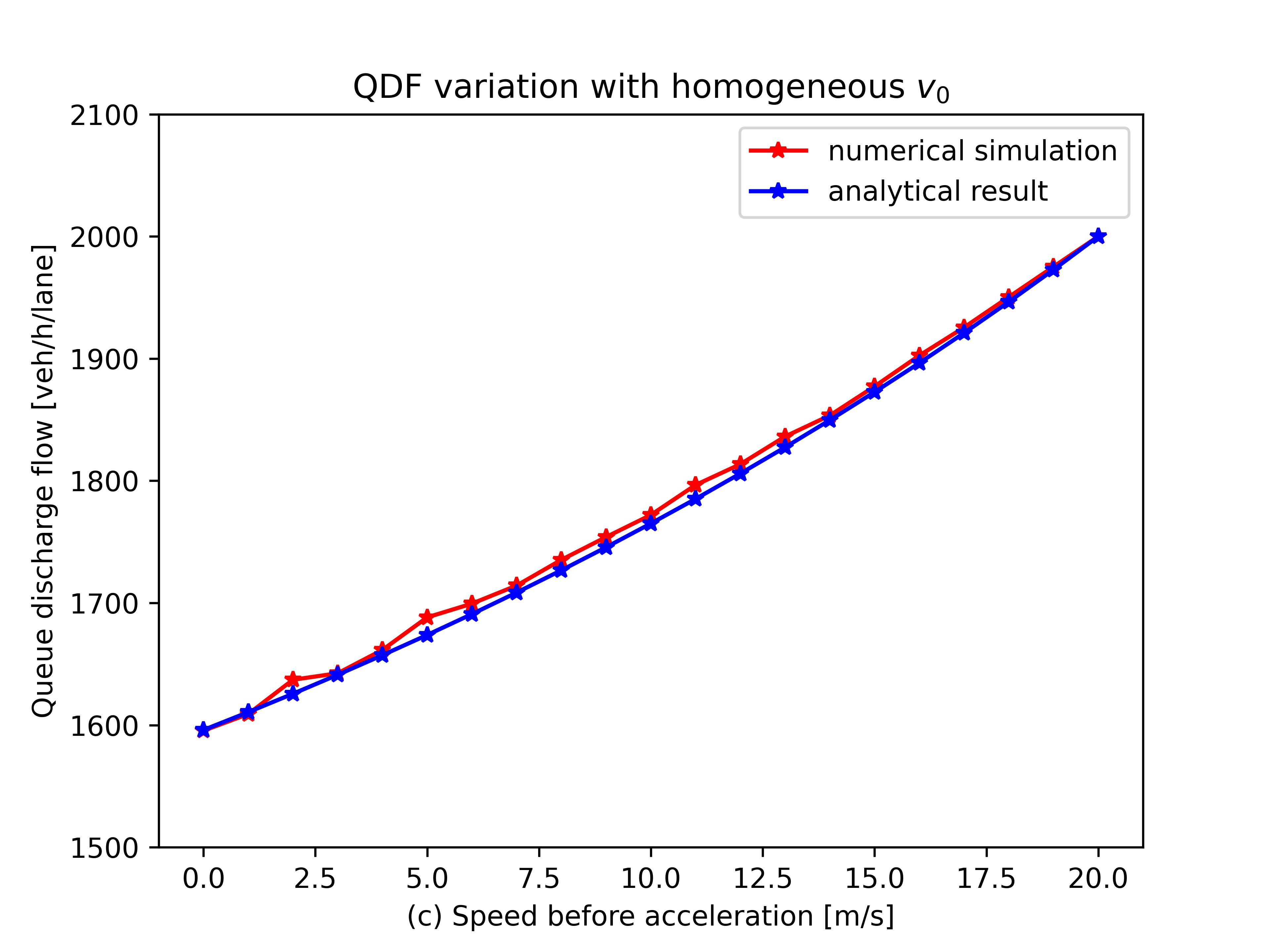}
   	\includegraphics[width=0.48\textwidth]{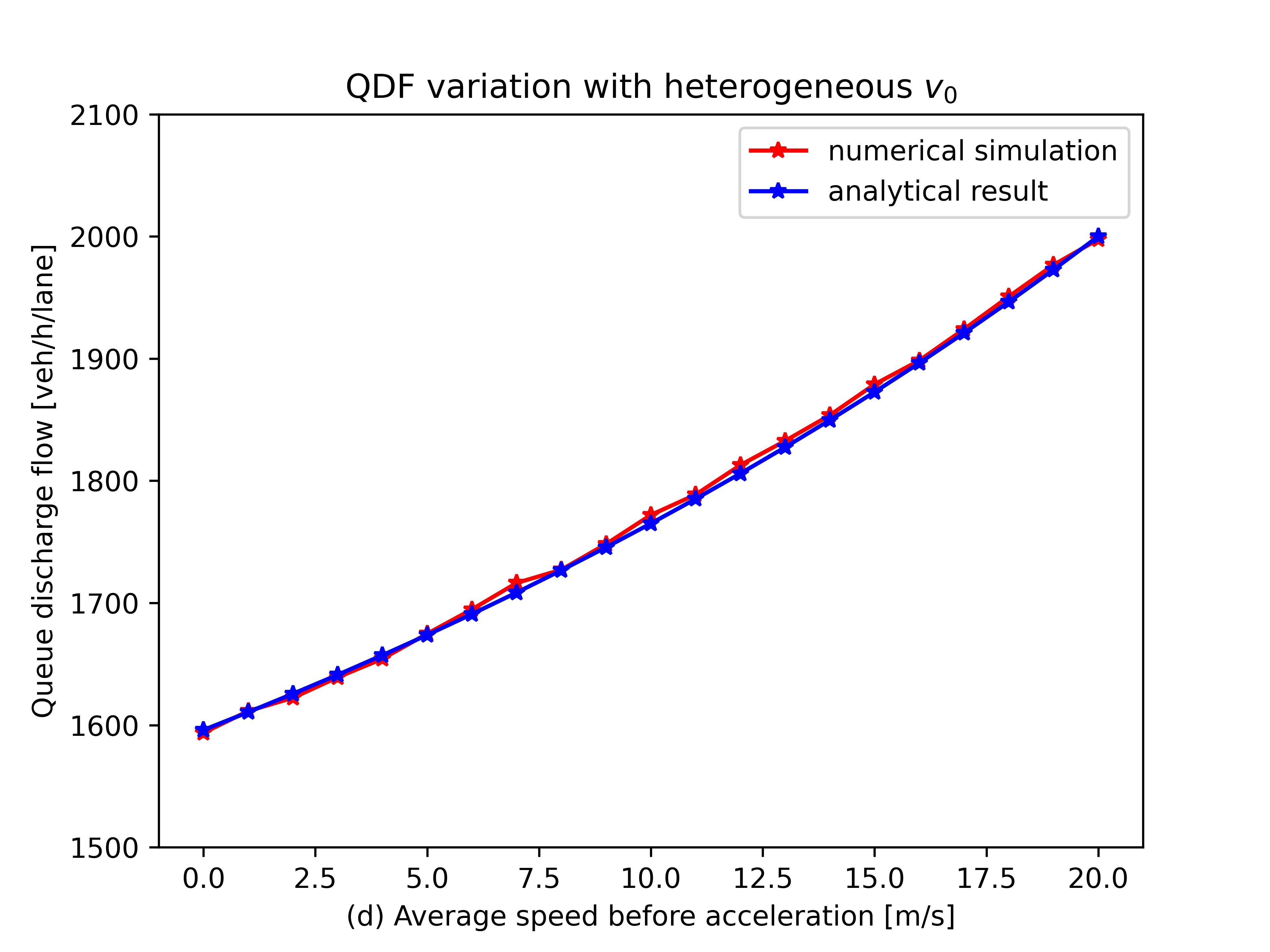}
	\caption{Comparison of QDFs between numerical simulations and analytical results for varying (a) $\tau$, (b) $L$, (c) homogeneous $v_{\text{0}}$, and (d) heterogeneous $v_{\text{0}}$.}
	\label{fig:numerical}
\end{figure}

The results of the three scenarios are depicted in Figure \ref{fig:numerical} (a-c). In all cases, the maximum deviation between the numerical and analytical results remains below 1\%, demonstrating that the analytical formula accurately estimates the average void created by hesitant vehicles, as governed by the physical process described in Section \ref{subsec: illustration}. The proposed analytical method assumes homogeneous initial speeds for all hesitant vehicles. However, in real traffic, the speed before acceleration may vary among vehicles. To further test the performance of the analytical formula under heterogeneous $v_{\text{0}}$, we modified Scenario 3 by introducing variability in $v_{\text{0}}$. Instead of using a fixed $v_{\text{0}}$ for each vehicle in the numerical simulation, we assumed that the initial speeds of vehicles exhibiting acceleration delay follow a normal distribution with a mean of $v_{\text{0}}$ and a standard deviation of $0.2 \cdot v_{\text{0}}$. The results, shown in Figure \ref{fig:numerical} (d), indicate that the errors in QDFs estimated by the analytical formula remain below 1\%. Therefore, we conclude that assuming homogeneity for the initial speed does not significantly affect the accuracy of the formula.
         
\subsection{Sensitivity analysis}
\label{subsec:sensitivity}

\begin{figure}[H]
	\centering
  	\includegraphics[width=0.48\textwidth]{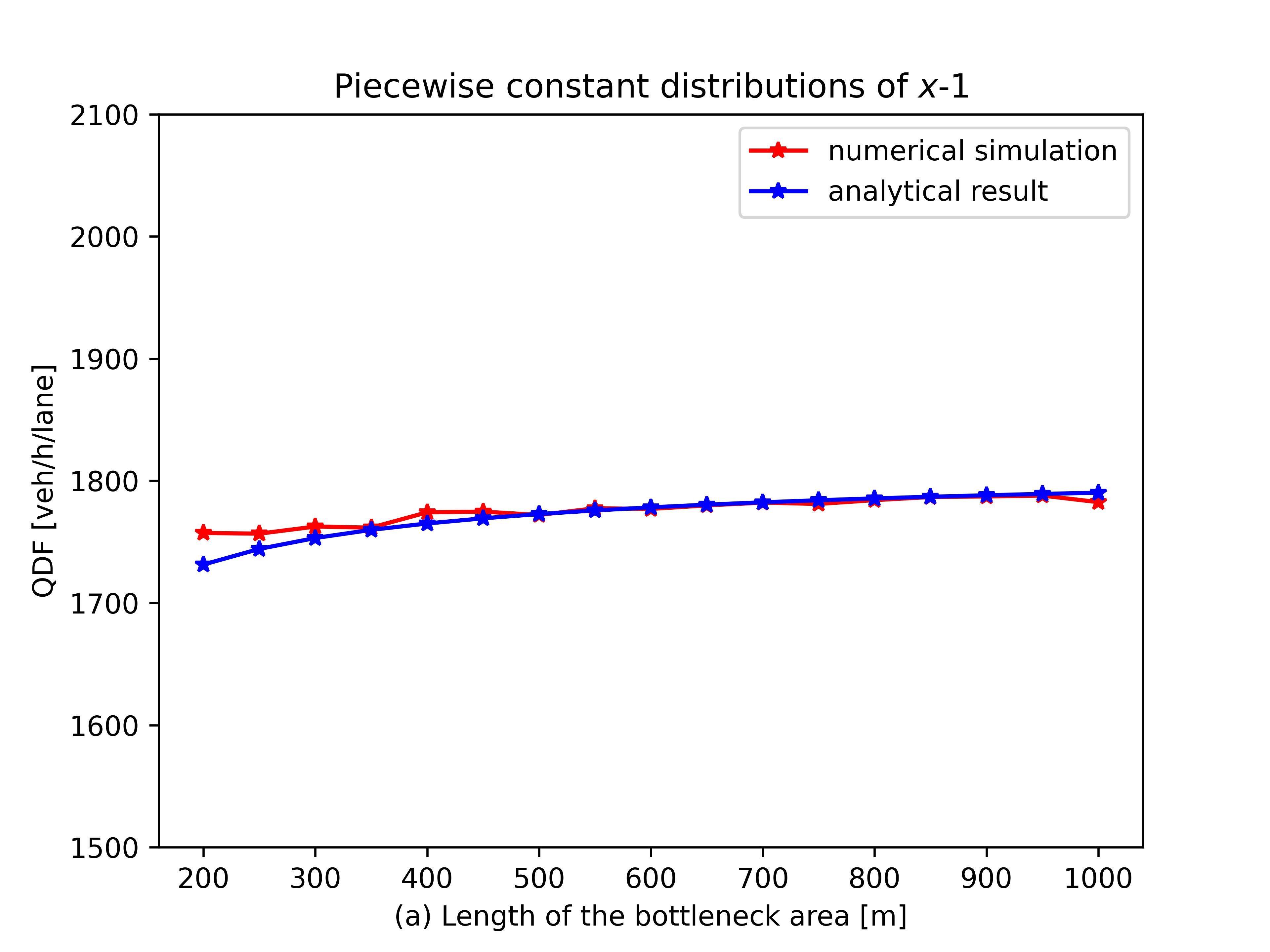}   
        \includegraphics[width=0.48\textwidth]{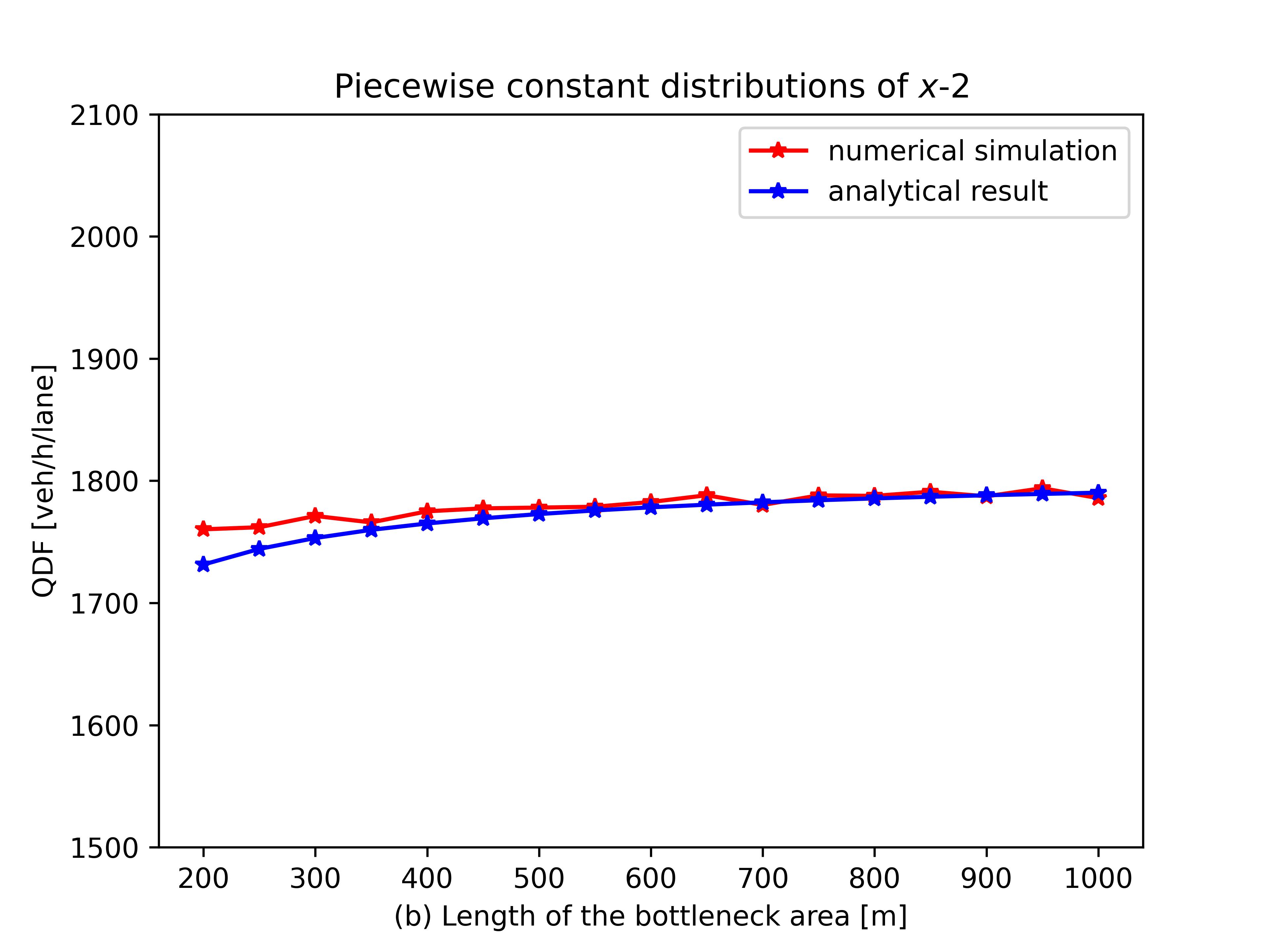}   
	\caption{Comparison of QDFs between numerical simulations and analytical results for two piecewise constant distributions of $x$.}
	\label{fig:heterogeneous}
\end{figure}

The assumption that $x$ follows a uniform distribution may not entirely align with practical traffic conditions. Therefore, we tested the performance of the proposed formula when $x$ does not follow a uniform distribution. We used two piecewise constant probability distributions to represent $x$. In the first distribution, the probabilities that $x$ falls into the ranges [0, $\frac{1}{4}L$], [$\frac{1}{4}L$, $\frac{1}{2}L$], [$\frac{1}{2}L$, $\frac{3}{4}L$], and [$\frac{3}{4}L$, $L$] are 0.1, 0.2, 0.3, and 0.4, respectively. In the second distribution, the probabilities that $x$ falls into the ranges [0, $\frac{1}{4}L$], [$\frac{1}{4}L$, $\frac{1}{2}L$], [$\frac{1}{2}L$, $\frac{3}{4}L$], and [$\frac{3}{4}L$, $L$] are 0.3, 0.2, 0.2, and 0.3. The results are depicted in Figure \ref{fig:heterogeneous} (a) and (b), respectively. The accuracy of the analytical formula in these scenarios are comparable to that in Scenario 2, which employed a uniform distribution for $x$ to generate random samples. Therefore, assuming a uniform distribution for $x$ does not compromise the accuracy of the formula.

\begin{figure}[H]
	\centering
  	\includegraphics[width=0.48\textwidth]{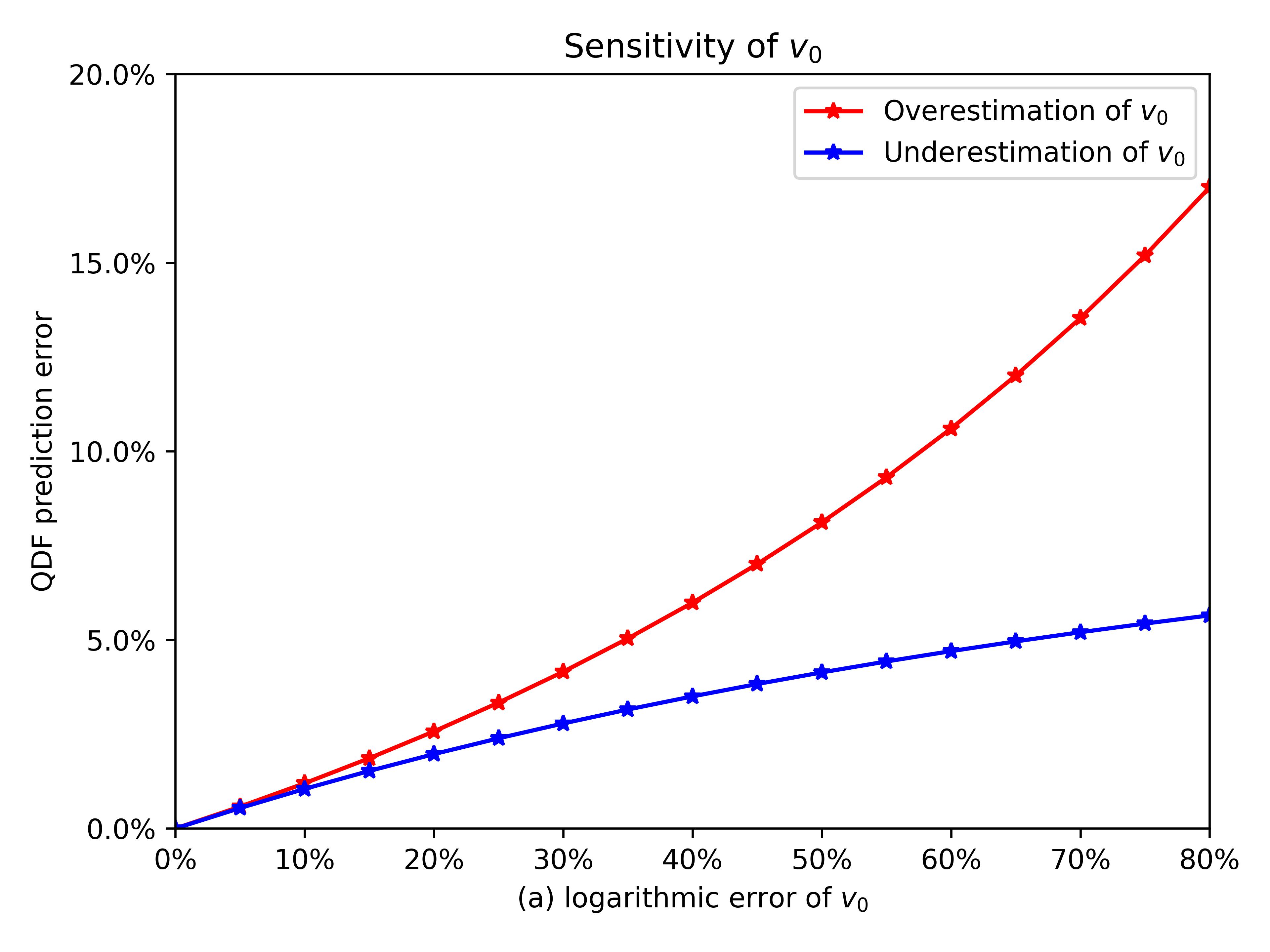}   
        \includegraphics[width=0.48\textwidth]{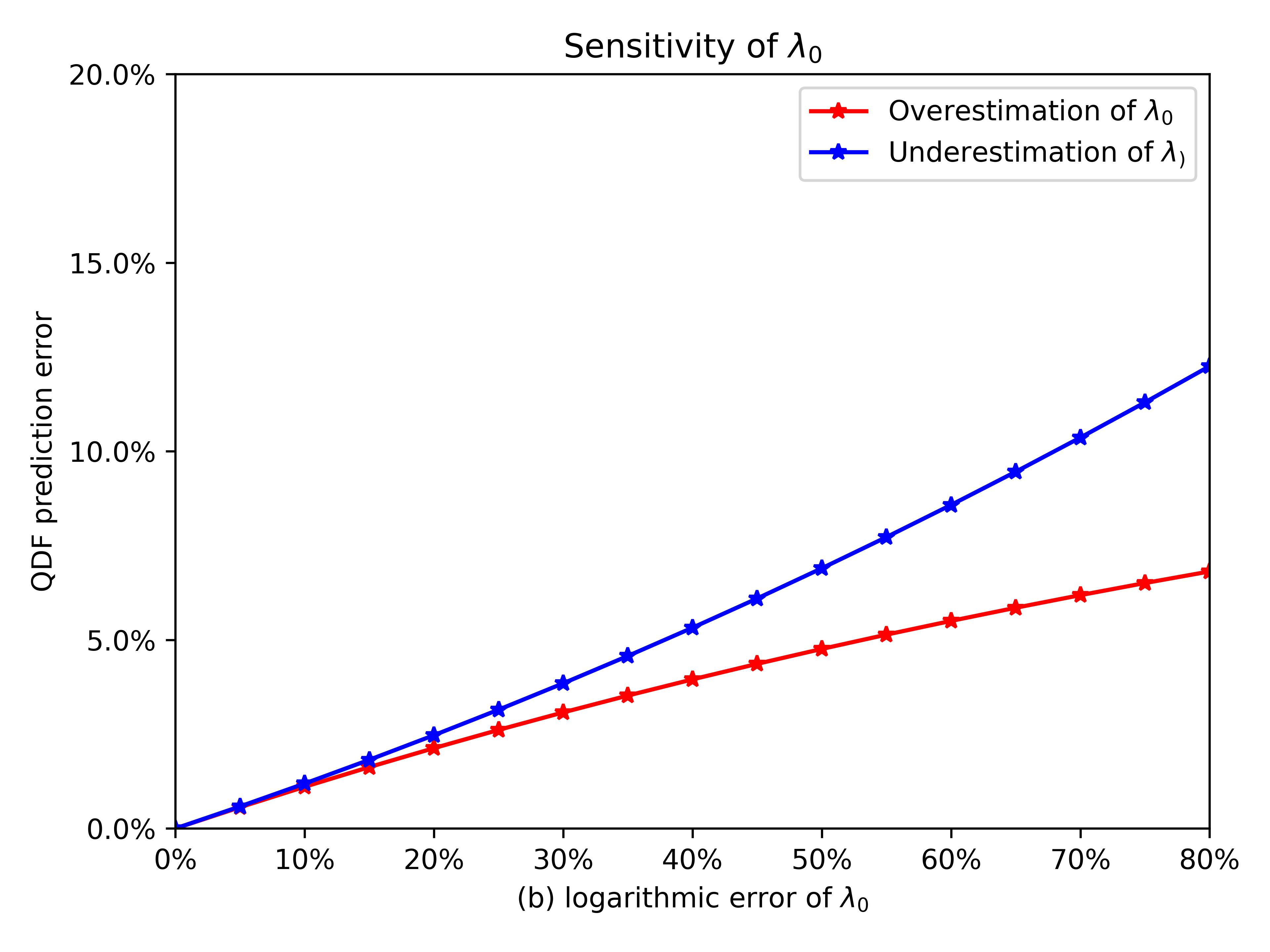}
        \includegraphics[width=0.48\textwidth]{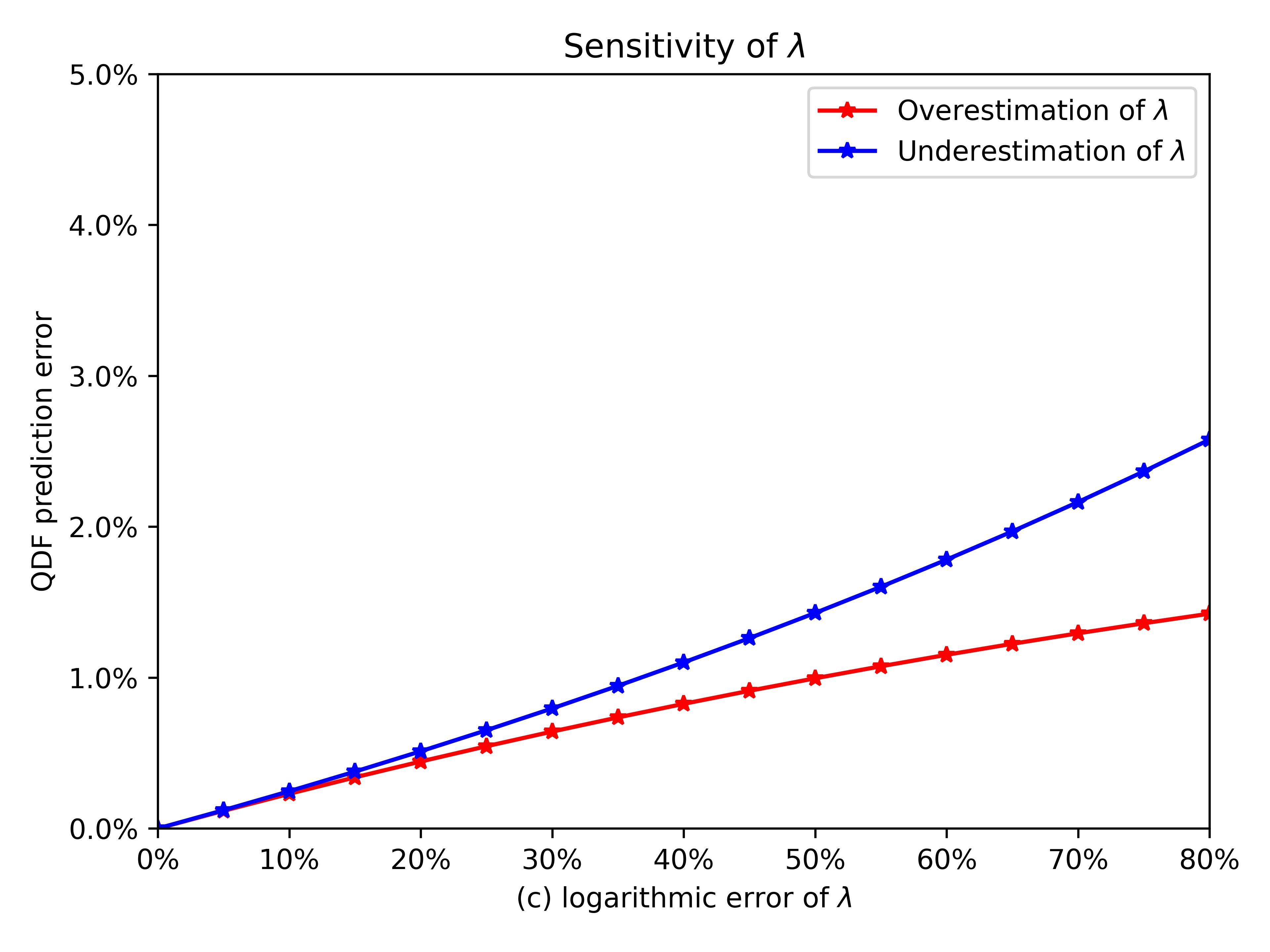}
        \includegraphics[width=0.48\textwidth]{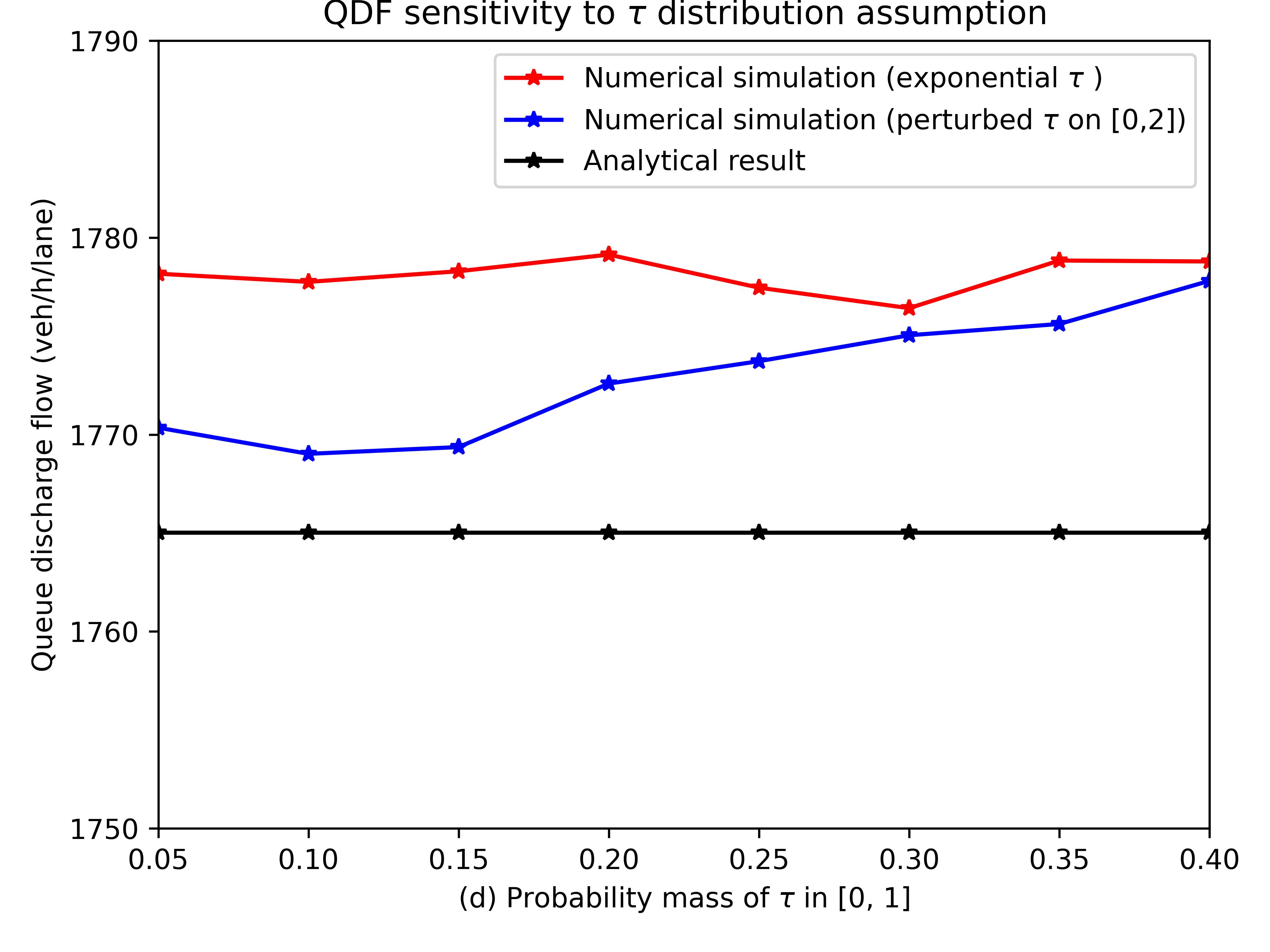}
	\caption{QDF prediction errors of the proposed model with different parameter estimation errors.}
	\label{fig:sensitivity}
\end{figure}

The prediction errors of the model were evaluated under different levels of estimation errors for parameters $v_0$, $\lambda_0$, and $\lambda$. For each parameter, we varied its logarithmic error from 0 to 80\% in both overestimation and underestimation scenarios. The benchmark values were set as follows: $v_0 = 10 m/s$, $\lambda_0 = 0.5$, and $\lambda = \frac{1}{6}$. Figure \ref{fig:sensitivity} (a-c) present the sensitivity analysis results. For $v_0$, underestimation leads to prediction errors below 5\% even when the estimation error reaches 80\%. In contrast, overestimation causes the prediction error to increase sharply with error magnitude, as a larger estimated $v_0$ implies that the pre-acceleration speed is closer to free-flow speed, thereby inflating the predicted QDF toward free-flow capacity. For $\lambda_0$, underestimation also results in large prediction errors, reflecting that acceleration delays are overestimated and the predicted QDF is consequently lower than the actual value. By comparison, the model is less sensitive to $\lambda$: for both underestimation and overestimation, prediction errors remain below 5\% when the estimation error is within 80\%.

The assumption that $\tau$ follows a negative exponential distribution may not fully reflect empirical traffic conditions. Specifically, the empirical data indicate that the probability of $\tau$ falling within the interval $[0,1]$ is lower than that within $[1,2]$, whereas the exponential distribution exhibits the opposite pattern. Nevertheless, the exponential distribution reproduces the combined probability mass over the interval $[0,2]$ with good accuracy. We therefore hypothesize that, although the probability mass is misallocated between the two sub-intervals, this approximation does not significantly affect the analytical estimation of the QDF.

To examine this hypothesis, we conducted a sensitivity analysis using numerical simulation. The total probability of $\tau$ over $[0,2]$ was fixed at 0.63, corresponding to a negative exponential distribution with parameter 0.5. Under this baseline distribution, the probability of $\tau \in [0,1]$ is 0.39. In the sensitivity analysis, this probability was varied from 0.05 to 0.40 in increments of 0.05, while maintaining the total probability over $[0,2]$ unchanged. The results are presented in Fig.~\ref{fig:sensitivity}(d). The black line denotes the analytical result. The red line corresponds to numerical simulation with $\tau$ sampled directly from the negative exponential distribution. The blue line represents numerical simulation results when the probability mass of $\tau$ is deliberately perturbed within $[0,2]$. In the baseline case, the QDF obtained from numerical simulation is 1778, whereas the analytical model yields 1765, resulting in an estimation error of approximately 0.7\%. When the distribution of $\tau$ is perturbed within $[0,2]$, the deviation between the analytical and numerical results remains within this range. These results indicate that the analytical model is robust to moderate deviations in the assumed $\tau$ distribution, particularly with respect to the allocation of probability mass within $[0,2]$.

We also evaluated the impact of assuming that a void interacts with only one downstream hesitant vehicle, as introduced in the previous section, on QDF estimation. Specifically, the differences in triggering time, positions, and delay durations of two downstream hesitant vehicles were randomly generated based on the previously introduced distributions. The results show that while the probability of a second downstream hesitant vehicle interacting with the upstream void is not negligible (approximately 10\%), the resulting bias in the calculated remaining void is small (less than 3\%), as most of the void is typically diminished after the first interaction. Consequently, the bias in QDF estimation is minimal—only 0.2\%. These findings suggest that explicitly modeling a second hesitant vehicle has a negligible impact on estimation accuracy.

\subsection{Validation using real data}
\label{subsec:data}

In this section, we validate the proposed analytical model using field data from the two sites described in Section 2. The model parameters were estimated and tested separately for each site (cross-validation was not performed) due to differences in characteristics such as speed limits. We begin by estimating the parameters for equilibrium traffic states \citep{wang2021model}, specifically $v_\text{f}$, $s_\text{cri}$, and $w$. For model validation, 12 five-minute intervals were sampled from site 1 and 5 intervals from site 2. Each interval contained approximately 440 to 520 vehicles, yielding a total of about 5000 trajectories for site 1 and 2400 for site 2.

To obtain the macroscopic equilibrium traffic state, we first extract microscopic equilibrium traffic states from vehicle platoons. These platoons consist of vehicles exhibiting stable driving behavior, characterized by consistent speed and spacing while following their predecessor. Several conditions should be satisfied to sample such states effectively. For example, data sampling should exclude lane change maneuvers for both the subject vehicle and its predecessor. Additionally, the variations in speed and spacing of the vehicle should be below a specified threshold, and the acceleration/deceleration rates of both the predecessor and the follower should not exceed another threshold. The state should be maintained for a sufficiently long time period. The following threshold values were chosen to balance the number of available data samples with their quality: the maximum variations in speed and spacing were set to 10\%, the maximum acceleration and deceleration rates were set to $\pm$ 0.2 $m/s^2$, and the minimum time interval was set to 4 seconds. 

\begin{figure}[!h]
	\centering
 	\includegraphics[width=0.48\textwidth]{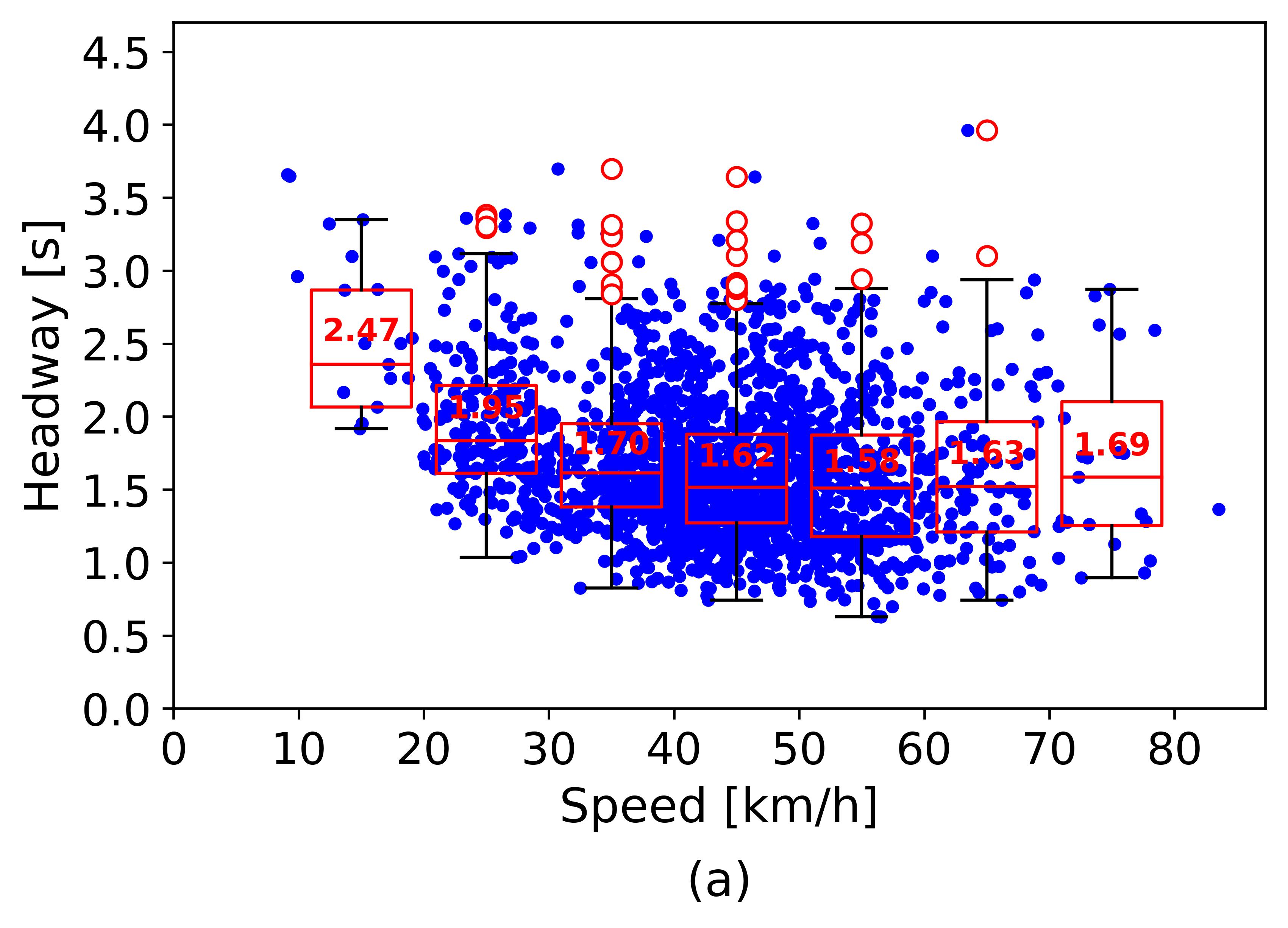}
 	\includegraphics[width=0.48\textwidth]{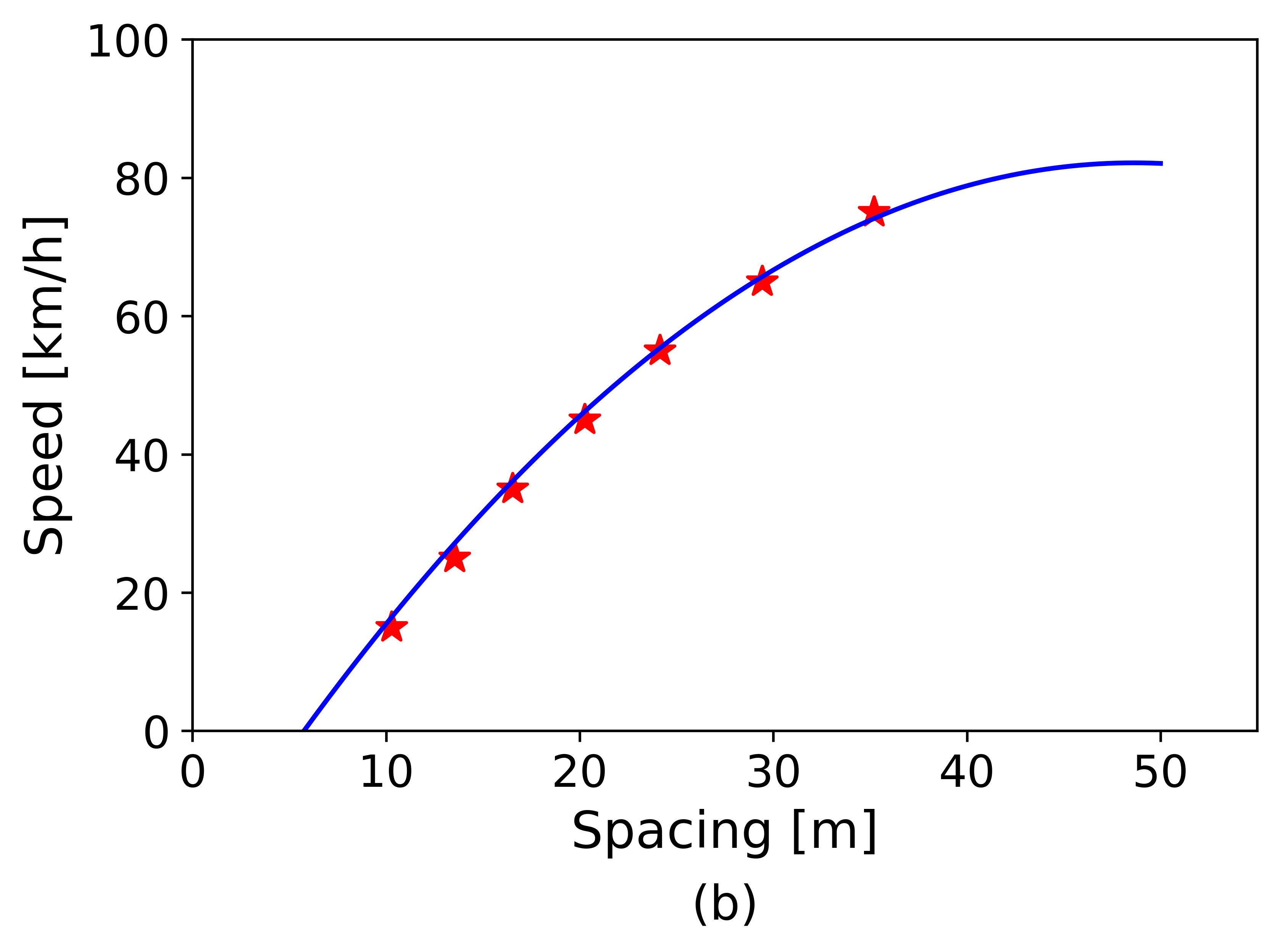}
	\caption{(a) Time headway distribution across different speed intervals, and (b) the corresponding spacing-speed relationship for site 1.}
	\label{theoretical_capacity}
\end{figure}

Figure \ref{theoretical_capacity} (a) illustrates the average time headway for groups of vehicles, categorized by speed intervals of 10 km/h, based on sampled data from Site 1. For each speed interval, an equilibrium data point is derived for the average spacing-speed relationship. Figure \ref{theoretical_capacity} (b) shows the spacing-speed fundamental diagram, with a line fitted to the equilibrium data points. Assuming equilibrium traffic states follow a triangular fundamental diagram for density and flow, the parameters for the equilibrium traffic states are estimated as follows: $s_\text{cri} = 36.0$ m and $w = -4.1$ m/s. The free-flow speed, $v_\text{f}$, is set to the site's speed limit of 80 km/h. We applied the same procedure to Site 2 and obtained the corresponding parameters as $s_\text{cri}$ = 43.0 m and $w$ = -5.0 m/s. Additionally, $v_\text{f}$ is set to 100 km/h.

The parameters of the analytical formula, including $\lambda$, $v_{0}$, $\lambda_{0}$, and $\alpha$, were estimated for each 5-minute interval using data from all identified vehicles exhibiting the acceleration delay state, including their triggering times, initial speeds, and delay durations. Specifically, all deceleration–acceleration events were first identified for every vehicle. By comparing the spacing–speed state at the onset of acceleration with the equilibrium state, vehicles exhibiting the acceleration delay state were extracted. For those states, the triggering time, initial speed before acceleration, and delay duration were recorded. Within each interval, $v_0$ was calculated as the mean of all sampled initial speeds. The time gap between consecutive triggering events, denoted as $T$, was obtained from the triggering times, and $\lambda$ was estimated by fitting $T$ to a negative exponential distribution. Similarly, $\lambda_0$ was derived by fitting the sampled delay durations to a negative exponential distribution. The number of vehicles entering the acceleration delay state in each interval used for parameter estimation is given by the total number of vehicles in that interval multiplied by $\alpha$. 

By incorporating the estimated parameters into the analytical formula, the QDFs for all intervals were calculated and are summarized in Tables \ref{tab:the result of Yingtian} and \ref{tab:the result of Hurong} for Site 1 and Site 2, respectively. The mean absolute estimation errors for the two sites are 2.1 \% and 1.5 \%, respectively, indicating that the proposed model can accurately predict the QDF. Parameter estimation was not conducted on a lane-by-lane basis because the number of acceleration delay samples within each 5-minute interval per lane was limited, and in some cases only a few observations were available. Such sparse samples could lead to biased estimates of $\lambda$ and $\lambda_{0}$. Furthermore, since inter-lane interactions are not explicitly modeled, estimating parameters using data from all lanes combined does not compromise the accuracy of the overall QDF estimation. The estimated parameters across the two sites are highly consistent, likely due to their comparable geometric configurations (300–400 m weaving sections) and similar aggregate driver behavior.

In the proposed analytical model, lane-changing is considered only as an endogenous trigger of acceleration delay and not as a mechanism that offsets capacity drop through gap filling. As a result, the model may slightly overestimate the magnitude of the capacity drop because it does not account for void utilization from adjacent lanes. However, in the validation sites used in this study, the bottlenecks are weaving bottlenecks where insertions typically occur upstream of desertions. For this type of configuration, as shown in \cite{chen2018capacity}, the effect of void utilization by lane-changing vehicles is limited because desertions are concentrated downstream, where vacancies tend to remain unfilled. Therefore, the omission of explicit inter-lane gap-filling behavior is not expected to significantly affect the validation results presented in this paper.

In practical applications, the proposed model offers several important advantages. First, as an analytical formulation, it can be computed very efficiently, requiring only the estimation of a few parameters—all of which have clear physical interpretations. As demonstrated in the numerical validation (Section 5.3), the model is relatively more sensitive to estimation errors in parameters $v_0$ and $\lambda_0$. In practice, $v_0$ is straightforward to estimate because speeds in congested conditions tend to vary little at recurrent bottlenecks. In our data, $v_0$ remained stable across different intervals at both sites, typically between 30 and 40 km/h. Parameter $\lambda_0$ is more challenging to estimate, as accurate delay measurements require complete vehicle trajectories. However, it can be calibrated using data from one site with complete trajectories and then transferred to another site with a similar driver population (e.g., within the same city), as demonstrated in Tables \ref{tab:the result of Yingtian} and \ref{tab:the result of Hurong}, where Site 1 and Site 2 belong to the same city. Another key advantage of the proposed model is its ability to capture the general trend of QDF variation under different congestion levels, infrastructure geometries, and driver population compositions. This feature is particularly valuable for developing efficient traffic management and control strategies, as most existing approaches treat the extent of capacity drop as a fixed constant.

\begin{table}[!ht]
    \centering
    \begin{tabular}{cccccccc}
    \toprule[1pt]
    Index&  \makecell[c]{Real QDFs\\veh/h/lane}&  \makecell[c]{Analytical QDFs \\veh/h/lane}&  \makecell[c]{$\lambda _0$\\}&  $\lambda$&  \makecell[c]{$v_{\text{0}}$\\m/s}&  $\alpha$&  Absolute errors \\
    \midrule[1pt]
    1&  2080&  2025&  0.435&  0.125&  11.14& 0.216& $2.64\%$\\
    2&  1896&  1906&  0.367&  0.157&  10.88& 0.299& $0.53\%$\\
    3&  1908&  1911&  0.311&  0.122&  10.50& 0.230& $0.16\%$\\
    4&  1884&  1908&  0.320&  0.156&  12.39& 0.298& $1.27\%$\\
    5&  1972&  1966&  0.381&  0.131&  10.72& 0.238& $0.30\%$\\
    6&  1756&  1855&  0.389&  0.170&  10.25& 0.359& $5.64\%$\\
    7&  1952&  1997&  0.322&  0.096&  11.53& 0.177& $2.31\%$\\
    8&  1860&  1951&  0.413&  0.145&  10.85& 0.281& $4.89\%$\\
    9&  1984&  1937&  0.382&  0.146&  10.42& 0.265& $2.37\%$\\
    10&  1848&  1903&  0.360&  0.152&  10.93& 0.296& $2.98\%$\\
    11&  1964&  1963&  0.383&  0.133&  10.79& 0.244& $0.05\%$\\
    12&  1976&  1918&  0.379&  0.153&  10.17& 0.279& $2.94\%$\\
    \bottomrule[1pt]
    \end{tabular}
    \caption{Estimated parameters, QDF, and observed QDF for each time interval at Site 1.}
    \label{tab:the result of Yingtian}
\end{table}

\begin{table}[!ht]
    \centering
    \begin{tabular}{cccccccc}
    \toprule[1pt]
    Index&  \makecell[c]{Real QDFs\\veh/h/lane}&  \makecell[c]{Analytical QDFs\\veh/h/lane}&  \makecell[c]{$\lambda _0$\\}&  $\lambda$&  \makecell[c]{$v_{\text{0}}$\\m/s}&  $\alpha$&  Absolute errors\\
    \midrule[1pt]
    1&  2004&  2045&  0.512&  0.108&  9.60& 0.212& $0.05\%$\\
    2&  1908&  1948&  0.518&  0.118&  6.98& 0.287& $1.42\%$\\
    3&  2031&  1989&  0.456&  0.113&  10.09& 0.262& $4.19\%$\\
    4&  1866&  1890&  0.394&  0.104&  7.66& 0.270& $1.88\%$\\
    5&  1911&  1907&  0.436&  0.126&  8.54& 0.302& $3.14\%$\\
    \bottomrule[1pt]
    \end{tabular}
    \caption{Estimated parameters, QDF, and observed QDF for each time interval at Site 2.}
    \label{tab:the result of Hurong}
\end{table}

\section{Conclusions and Discussions}

This paper proposes an analytical method to estimate the queue discharge flow (QDF) at freeway bottlenecks by explicitly incorporating wave–void interaction. Building on established findings, acceleration delay states generate voids in front of vehicles and contribute to local capacity reduction. At active bottlenecks, where the influence area typically extends several hundred meters, wave–void interactions may arise when such delay states are triggered at different locations within close temporal proximity. In these situations, the void created by an upstream vehicle may intersect with the wave generated by a downstream vehicle, thereby reducing the void size and its impact on QDF. In such cases, the void created by an upstream hesitant vehicle can intersect with the wave generated by a downstream hesitant vehicle, diminishing the void size and thereby reducing its impact on QDF. Beyond the micro-level, this wave–void interaction process may also shed light on macroscopic density heterogeneity. It helps explain why different QDF values can be observed under the same average density. Without wave–void interactions, vehicles experiencing response delays generate large voids ahead of them, which increases the variance of inter-vehicle spacing and results in greater spatial heterogeneity. By contrast, wave–void interactions diminish these enlarged gaps and lead to a more homogeneous vehicle distribution at the bottleneck. This mechanism provides a behavioral explanation for why similar density levels can yield different QDF values depending on the degree of spatial heterogeneity at the aggregate level.

This interaction and its effect on QDF are modeled as a stochastic process, where the spatial and temporal distributions of acceleration delay states and their associated delay durations are characterized based on real-world observations. In principle, the proposed analytical framework can be applied to different types of bottlenecks, provided that the corresponding distributional parameters of acceleration delay states are available. Furthermore, the framework is flexible in accommodating different capacity drop mechanisms. Although this study focuses on acceleration delay states, other void-inducing mechanisms, such as the bounded acceleration of merging vehicles, can be incorporated by treating these vehicles as being in an acceleration delay state. It should be noted that the assumed forms of the underlying distributions, including the exponential distribution for delay duration and the uniform distribution for trigger location, are primarily introduced for analytical tractability. Their use is supported by robustness checks rather than exact empirical fits. In practical applications, these distributions and their parameters may vary with bottleneck type, number of lanes, road geometry, and driver population characteristics, and should therefore be calibrated accordingly for different scenarios.

The accuracy of the analytical formula was first evaluated by comparing it with numerical simulations that replicate the physical process of wave–void interactions at active bottlenecks. The results show that the analytical model produces outcomes that are highly consistent with the simulation results. In addition, the proposed framework reproduces the observed phenomenon that QDFs in jam waves are lower than those at active bottlenecks, which is attributed to the reduced frequency of wave–void interactions in jam waves. However, it should be noted that this explanation is supported primarily by analytical reasoning and numerical simulation. Due to the lack of field trajectory data capturing complete moving jam dynamics, direct empirical validation for jam wave scenarios was not conducted. Instead, real-world validation was limited to active bottlenecks, where the model demonstrates good predictive accuracy, with an average error of approximately 2\%.

In future research, we plan to extend the proposed framework to explicitly account for multi-lane interactions. Although the present study validated the model using field data from multi-lane active bottlenecks, the results represent lane-averaged conditions. Lane-changing was modeled solely as an endogenous factor triggering acceleration delay, and inter-lane interactions were not explicitly incorporated. In practice, a void generated by a vehicles exhibiting acceleration delay may be filled by a lane-changing vehicle from an adjacent lane, partially mitigating the reduction in QDF. This effect is particularly relevant in merge or diverge bottlenecks. Future work will therefore focus on extending the analytical model to capture inter-lane gap-filling dynamics across different bottleneck types. Additionally, future research will explore dynamic traffic control strategies aimed at enhancing bottleneck QDF based on the mechanisms identified in this study. By leveraging connected and autonomous vehicle technologies, it may be possible to reduce the triggers of acceleration delay states and optimize their spatial and temporal distribution, thereby mitigating void formation and increasing bottleneck throughput.

\section*{Acknowledgment}

This research is supported by the National Natural Science Foundation of China (No.52525204, No.52232012, No.52131203).

\clearpage

\begin{appendix} 
\section{Supplementary derivation process of equations}

The complete derivation process for Equation \eqref{eq:interaction_left} is provided below: 

\begin{equation}
\begin{aligned}
\label{eq:interaction_left_appendix}
  & p_{\operatorname{int}}^{i\to i-1}=\frac{1}{L}\int_{0}^{L}{\int_{0}^{\frac{L-x}{w}}{\lambda }{{e}^{-\lambda t}}\cdot \frac{L-x-wt}{L}dt}dx \\ 
 & =\frac{1}{L}\int_{0}^{L}{\left( -\left. \frac{L-x}{L}{{e}^{-\lambda t}} \right|_{0}^{\frac{L-x}{w}}+\left. \frac{w}{L}(t{{e}^{-\lambda t}}+\frac{1}{\lambda }{{e}^{-\lambda t}}) \right|_{0}^{\frac{L-x}{w}} \right)}dx \\ 
 & =\frac{1}{L}\int_{0}^{L}{\left( \frac{w}{\lambda L}{{e}^{-\lambda \frac{L-x}{w}}}-\frac{w}{\lambda L}+\frac{L-x}{L} \right)}dx \\ 
 & =\frac{1}{L}\left( \left. \frac{w}{\lambda L}(\frac{w}{\lambda }{{e}^{-\lambda \frac{L-x}{w}}}) \right|_{0}^{L}+\left. (x-\frac{wx}{\lambda L}-\frac{{{x}^{2}}}{2L}) \right|_{0}^{L} \right) \\ 
 & =\frac{1}{L}\left( -\frac{{{w}^{2}}}{{{\lambda }^{2}}L}{{e}^{-\lambda \frac{L}{w}}}+\frac{{{w}^{2}}}{{{\lambda }^{2}}L}-\frac{w}{\lambda }+\frac{L}{2} \right) \\ 
 & =-\frac{{{w}^{2}}}{{{\lambda }^{2}}{{L}^{2}}}{{e}^{-\lambda \frac{L}{w}}}+\frac{{{w}^{2}}}{{{\lambda }^{2}}{{L}^{2}}}-\frac{w}{\lambda L}+\frac{1}{2}. \\ 
\end{aligned}
\end{equation}

The derivation process for Equation \eqref{eq:interaction_right} involves two steps. First, we perform the temporal integration over $t$:

\begin{equation}
\begin{aligned}
\label{eq:interaction_right_appendix_temporal}
& \int_{0}^{\tau }{\lambda }{{e}^{-\lambda t}}\cdot \frac{L-x-{{v}_{0}}t}{L}dt+\int_{\tau }^{\frac{L-x-{{v}_{0}}\tau }{{{v}_\text{f}}}+\tau }{\lambda }{{e}^{-\lambda t}}\cdot (\frac{L-x-{{v}_{0}}\tau -{{v}_\text{f}}\left( t-\tau  \right)}{L})dt \\
& =-\left. \frac{L-x}{L}{{e}^{-\lambda t}} \right|_{0}^{\tau }+\left. \frac{{{v}_{0}}}{L}(t{{e}^{-\lambda t}}+\frac{1}{\lambda }{{e}^{-\lambda t}}) \right|_{0}^{\tau }+\int_{\tau }^{\frac{L-x-{{v}_{0}}\tau }{{{v}_\text{f}}}+\tau }{\lambda }{{e}^{-\lambda t}}\frac{L-x-{{v}_{0}}\tau +{{v}_\text{f}}\tau }{L}dt +\int_{\tau }^{\frac{L-x-{{v}_{0}}\tau }{{{v}_\text{f}}}+\tau }{\lambda }{{e}^{-\lambda t}}\frac{-{{v}_\text{f}}t}{L}dt \\ 
& =-\frac{L-x}{L}{{e}^{-\lambda \tau }}+\frac{L-x}{L}+\frac{{{v}_{0}}}{L}(\tau {{e}^{-\lambda \tau }}+\frac{1}{\lambda }{{e}^{-\lambda \tau }})-\frac{{{v}_{0}}}{L\lambda }+\frac{L-x-{{v}_{0}}\tau +{{v}_\text{f}}\tau }{L}\left( -{{e}^{-\lambda \frac{L-x-{{v}_{0}}\tau }{{{v}_\text{f}}}-\lambda \tau }}+{{e}^{-\lambda \tau }} \right) \\ 
& \; \; \; +\frac{{{v}_\text{f}}}{L}\left( \frac{L-x-{{v}_{0}}\tau +{{v}_\text{f}}\tau }{{{v}_\text{f}}}{{e}^{-\lambda \frac{L-x-{{v}_{0}}\tau +{{v}_\text{f}}\tau }{{{v}_\text{f}}}}}+\frac{1}{\lambda }{{e}^{-\lambda \frac{L-x-{{v}_{0}}\tau +{{v}_\text{f}}\tau }{{{v}_\text{f}}}}}-\tau {{e}^{-\lambda \tau }}-\frac{1}{\lambda }{{e}^{-\lambda \tau }} \right) \\ 
& =\frac{{{v}_{0}}-{{v}_\text{f}}}{L\lambda }{{e}^{-\lambda \tau }}+\frac{{{v}_\text{f}}}{L\lambda }{{e}^{-\lambda \frac{L-x-{{v}_{0}}\tau +{{v}_\text{f}}\tau }{{{v}_\text{f}}}}}+\frac{L\lambda -\lambda x-{{v}_{0}}}{L\lambda }. \\ 
\end{aligned}
\end{equation}

Next, we perform the spatial integration over $x$:

\begin{equation}
\begin{aligned}
  & p_{\operatorname{int}}^{i\to i+1}=\frac{1}{L}\int_{0}^{L}{\frac{{{v}_{0}}-{{v}_\text{f}}}{L\lambda }{{e}^{-\lambda \tau }}+\frac{{{v}_\text{f}}}{L\lambda }{{e}^{-\lambda \frac{L-x-{{v}_{0}}\tau +{{v}_\text{f}}\tau }{{{v}_\text{f}}}}}+\frac{L\lambda -\lambda x-{{v}_{0}}}{L\lambda }}dx \\ 
 & =\frac{1}{L}\left( \left. x\left( \frac{{{v}_{0}}-{{v}_\text{f}}}{L\lambda }{{e}^{-\lambda \tau }}+\frac{L\lambda -{{v}_{0}}}{L\lambda } \right) \right|_{0}^{L}-\left. \frac{{{x}^{2}}}{2L} \right|_{0}^{L}+\left. \frac{{{v}_\text{f}}^{2}}{L{{\lambda }^{2}}}{{e}^{-\lambda \frac{L-x-{{v}_{0}}\tau +{{v}_\text{f}}\tau }{{{v}_\text{f}}}}} \right|_{0}^{L} \right) \\ 
 & =\frac{1}{L}\left( \frac{{{v}_{0}}-{{v}_\text{f}}}{\lambda }{{e}^{-\lambda \tau }}+\frac{L\lambda -{{v}_{0}}}{\lambda }-\frac{L}{2}+\frac{{{v}_\text{f}}^{2}}{L{{\lambda }^{2}}}{{e}^{-\lambda \frac{-{{v}_{0}}\tau +{{v}_\text{f}}\tau }{{{v}_\text{f}}}}}-\frac{{{v}_\text{f}}^{2}}{L{{\lambda }^{2}}}{{e}^{-\lambda \frac{L-{{v}_{0}}\tau +{{v}_\text{f}}\tau }{{{v}_\text{f}}}}} \right). \\ 
\end{aligned}
\end{equation}

The complete derivation process of Equation \eqref{eq:term 1} is as follows: 

\begin{equation}
\label{eq:term 1_appendix}
\begin{aligned}
&P_1\cdot \left( \tau _i-\frac{\int_0^{\tau _i}{\tau _{i-1}\lambda _0e^{-\lambda _0\tau _{i-1}}}d\tau _{i-1}}{\int_0^{\tau _i}{\lambda _0e^{-\lambda _0\tau _{i-1}}}d\tau _{i-1}} \right) = (1 - {e}^{-\lambda_{0} \tau_{i}}) (-\frac{{{w}^{2}}}{{{\lambda }^{2}}{{L}^{2}}}{{e}^{-\lambda \frac{L}{w}}}+\frac{{{w}^{2}}}{{{\lambda }^{2}}{{L}^{2}}}-\frac{w}{\lambda L}+\frac{1}{2}) \\
&\; \; \; \cdot \left(1 - \frac{1}{L}\left( \frac{{{v}_{0}}-{{v}_\text{f}}}{\lambda }{{e}^{-\lambda \tau_{i} }}+\frac{L\lambda -{{v}_{0}}}{\lambda }-\frac{L}{2}+\frac{{{v}_\text{f}}^{2}}{L{{\lambda }^{2}}}{{e}^{-\lambda \frac{-{{v}_{0}}\tau_{i} +{{v}_\text{f}}\tau_{i} }{{{v}_\text{f}}}}}-\frac{{{v}_\text{f}}^{2}}{L{{\lambda }^{2}}}{{e}^{-\lambda \frac{L-{{v}_{0}}\tau_{i} +{{v}_\text{f}}\tau_{i} }{{{v}_\text{f}}}}} \right) \right) \cdot \left( \tau _i-\frac{-\frac{e^{-\lambda _0\tau _{i-1}}\left( \lambda _0\tau _{i-1}+1 \right)}{{\lambda _0}^2}\mid_{0}^{\tau _i}}{-\frac{e^{-\lambda _0\tau _{i-1}}}{\lambda _0}\mid_{0}^{\tau _i}} \right) \\
&= (1 - {e}^{-\lambda_{0} \tau_{i}})(-\frac{{{w}^{2}}}{{{\lambda }^{2}}{{L}^{2}}}{{e}^{-\lambda \frac{L}{w}}}+\frac{{{w}^{2}}}{{{\lambda }^{2}}{{L}^{2}}}-\frac{w}{\lambda L}+\frac{1}{2}) \\
&\; \; \; \cdot \left(1 - \frac{1}{L}\left( \frac{{{v}_{0}}-{{v}_\text{f}}}{\lambda }{{e}^{-\lambda \tau_{i} }}+\frac{L\lambda -{{v}_{0}}}{\lambda }-\frac{L}{2}+\frac{{{v}_\text{f}}^{2}}{L{{\lambda }^{2}}}{{e}^{-\lambda \frac{-{{v}_{0}}\tau_{i} +{{v}_\text{f}}\tau_{i} }{{{v}_\text{f}}}}}-\frac{{{v}_\text{f}}^{2}}{L{{\lambda }^{2}}}{{e}^{-\lambda \frac{L-{{v}_{0}}\tau_{i} +{{v}_\text{f}}\tau_{i} }{{{v}_\text{f}}}}} \right) \right) \cdot \left( \frac{\lambda _0\tau _i-1+e^{-\lambda _0\tau _i}}{\lambda _0\left( 1-e^{-\lambda _0\tau _i} \right)} \right) \\
&=\frac{1}{\lambda _0}(-\frac{{{w}^{2}}}{{{\lambda }^{2}}{{L}^{2}}}{{e}^{-\lambda \frac{L}{w}}}+\frac{{{w}^{2}}}{{{\lambda }^{2}}{{L}^{2}}}-\frac{w}{\lambda L}+\frac{1}{2}) \\
&\; \; \; \cdot \left(1 - \frac{1}{L}\left( \frac{{{v}_{0}}-{{v}_\text{f}}}{\lambda }{{e}^{-\lambda \tau_{i} }}+\frac{L\lambda -{{v}_{0}}}{\lambda }-\frac{L}{2}+\frac{{{v}_\text{f}}^{2}}{L{{\lambda }^{2}}}{{e}^{-\lambda \frac{-{{v}_{0}}\tau_{i} +{{v}_\text{f}}\tau_{i} }{{{v}_\text{f}}}}}-\frac{{{v}_\text{f}}^{2}}{L{{\lambda }^{2}}}{{e}^{-\lambda \frac{L-{{v}_{0}}\tau_{i} +{{v}_\text{f}}\tau_{i} }{{{v}_\text{f}}}}} \right) \right) \cdot \left( \lambda _0\tau _i-1+e^{-\lambda _0\tau _i} \right). \\
\end{aligned}
\end{equation}

The complete derivation process of Equation \eqref{eq:term 3}: 

\begin{equation}
\label{eq:term 3_appendix}
\begin{aligned}
&P_3\cdot \left( \tau _i-\frac{\int_0^{\tau _i}{\tau _{i+1}\lambda _0e^{-\lambda _0\tau _{i+1}}}d\tau _{i+1}}{\int_0^{\tau _i}{\lambda _0e^{-\lambda _0\tau _{i+1}}}d\tau _{i+1}} \right) = (1 - {e}^{-\lambda_{0} \tau_{i}}) (\frac{{{w}^{2}}}{{{\lambda }^{2}}{{L}^{2}}}{{e}^{-\lambda \frac{L}{w}}}-\frac{{{w}^{2}}}{{{\lambda }^{2}}{{L}^{2}}}+\frac{w}{\lambda L}+\frac{1}{2}) \\
&\; \; \; \cdot \frac{1}{L}\left( \frac{{{v}_{0}}-{{v}_\text{f}}}{\lambda }{{e}^{-\lambda \tau_{i} }}+\frac{L\lambda -{{v}_{0}}}{\lambda }-\frac{L}{2}+\frac{{{v}_\text{f}}^{2}}{L{{\lambda }^{2}}}{{e}^{-\lambda \frac{-{{v}_{0}}\tau_{i} +{{v}_\text{f}}\tau_{i} }{{{v}_\text{f}}}}}-\frac{{{v}_\text{f}}^{2}}{L{{\lambda }^{2}}}{{e}^{-\lambda \frac{L-{{v}_{0}}\tau_{i} +{{v}_\text{f}}\tau_{i} }{{{v}_\text{f}}}}} \right)  \cdot \left( \tau _i-\frac{-\frac{e^{-\lambda _0\tau _{i+1}}\left( \lambda _0\tau _{i+1}+1 \right)}{{\lambda _0}^2}\mid_{0}^{\tau _i}}{-\frac{e^{-\lambda _0\tau _{i+1}}}{\lambda _0}\mid_{0}^{\tau _i}} \right) \\
&= (1 - {e}^{-\lambda_{0} \tau_{i}})(\frac{{{w}^{2}}}{{{\lambda }^{2}}{{L}^{2}}}{{e}^{-\lambda \frac{L}{w}}}-\frac{{{w}^{2}}}{{{\lambda }^{2}}{{L}^{2}}}+\frac{w}{\lambda L}+\frac{1}{2}) \\
&\; \; \; \cdot \frac{1}{L}\left( \frac{{{v}_{0}}-{{v}_\text{f}}}{\lambda }{{e}^{-\lambda \tau_{i} }}+\frac{L\lambda -{{v}_{0}}}{\lambda }-\frac{L}{2}+\frac{{{v}_\text{f}}^{2}}{L{{\lambda }^{2}}}{{e}^{-\lambda \frac{-{{v}_{0}}\tau_{i} +{{v}_\text{f}}\tau_{i} }{{{v}_\text{f}}}}}-\frac{{{v}_\text{f}}^{2}}{L{{\lambda }^{2}}}{{e}^{-\lambda \frac{L-{{v}_{0}}\tau_{i} +{{v}_\text{f}}\tau_{i} }{{{v}_\text{f}}}}} \right) \cdot \left( \frac{\lambda _0\tau _i-1+e^{-\lambda _0\tau _i}}{\lambda _0\left( 1-e^{-\lambda _0\tau _i} \right)} \right) \\
&= \frac{1}{\lambda _0}(\frac{{{w}^{2}}}{{{\lambda }^{2}}{{L}^{2}}}{{e}^{-\lambda \frac{L}{w}}}-\frac{{{w}^{2}}}{{{\lambda }^{2}}{{L}^{2}}}+\frac{w}{\lambda L}+\frac{1}{2}) \\
&\; \; \; \cdot \frac{1}{L}\left( \frac{{{v}_{0}}-{{v}_\text{f}}}{\lambda }{{e}^{-\lambda \tau_{i} }}+\frac{L\lambda -{{v}_{0}}}{\lambda }-\frac{L}{2}+\frac{{{v}_\text{f}}^{2}}{L{{\lambda }^{2}}}{{e}^{-\lambda \frac{-{{v}_{0}}\tau_{i} +{{v}_\text{f}}\tau_{i} }{{{v}_\text{f}}}}}-\frac{{{v}_\text{f}}^{2}}{L{{\lambda }^{2}}}{{e}^{-\lambda \frac{L-{{v}_{0}}\tau_{i} +{{v}_\text{f}}\tau_{i} }{{{v}_\text{f}}}}} \right) \cdot \left( \lambda _0\tau _i-1+e^{-\lambda _0\tau _i} \right) \\
\end{aligned}
\end{equation}

The complete derivation process of Equation \eqref{eq:term 6}: 

\begin{equation}
\label{eq:term 6_appendix}
\begin{aligned}
&P_6\cdot \left( \tau _i-\frac{\int_0^{\tau _i}{z^2{\lambda _0}^2e^{-\lambda _0z}}dz}{\int_0^{\tau _i}{{\lambda _0}^2ze^{-\lambda _0z}}dz} \right) = \left( 1-\left( \lambda _0\tau _i+1 \right) e^{-\lambda _0\tau _i} \right)(\frac{{{w}^{2}}}{{{\lambda }^{2}}{{L}^{2}}}{{e}^{-\lambda \frac{L}{w}}}-\frac{{{w}^{2}}}{{{\lambda }^{2}}{{L}^{2}}}+\frac{w}{\lambda L}+\frac{1}{2}) \\
&\; \; \; \cdot \left(1 - \frac{1}{L}\left( \frac{{{v}_{0}}-{{v}_\text{f}}}{\lambda }{{e}^{-\lambda \tau_{i} }}+\frac{L\lambda -{{v}_{0}}}{\lambda }-\frac{L}{2}+\frac{{{v}_\text{f}}^{2}}{L{{\lambda }^{2}}}{{e}^{-\lambda \frac{-{{v}_{0}}\tau_{i} +{{v}_\text{f}}\tau_{i} }{{{v}_\text{f}}}}}-\frac{{{v}_\text{f}}^{2}}{L{{\lambda }^{2}}}{{e}^{-\lambda \frac{L-{{v}_{0}}\tau_{i} +{{v}_\text{f}}\tau_{i} }{{{v}_\text{f}}}}} \right) \right) \cdot \left( \tau _i-\frac{-\frac{e^{-\lambda _0z}\left( {\lambda _0}^2+2\lambda _0z+2 \right)}{{\lambda _0}^3}\mid_{0}^{\tau _i}}{-\frac{e^{-\lambda _0z}\left( \lambda _0z+1 \right)}{{\lambda _0}^2}\mid_{0}^{\tau _i}} \right)\\
&=\frac{1}{\lambda _0}(\frac{{{w}^{2}}}{{{\lambda }^{2}}{{L}^{2}}}{{e}^{-\lambda \frac{L}{w}}}-\frac{{{w}^{2}}}{{{\lambda }^{2}}{{L}^{2}}}+\frac{w}{\lambda L}+\frac{1}{2}) \\
&\; \; \; \cdot \left( \frac{1}{{\lambda _{0}}^2}-\frac{1}{L}\left( \begin{array}{c}	\frac{v_0-v_\text{f}}{\lambda}\cdot \frac{1}{\left( \lambda +\lambda _{0} \right) ^2}+\left( \frac{L\lambda -v_0}{\lambda}-\frac{L}{2} \right) \cdot \frac{1}{{\lambda _{0}}^2}\\	+\frac{{v_\text{f}}^2}{L\lambda ^2}\left( 1-e^{-\frac{L\lambda}{v_\text{f}}} \right) \left( \frac{{v_\text{f}}^2}{\left( \lambda v_0-\lambda v_\text{f}-\lambda _{0}v_\text{f} \right) ^2} \right)\\\end{array} \right) \right)\\
\end{aligned}
\end{equation}

The complete derivation process of Equation \eqref{eq:term 7}: 

\begin{equation}
\label{eq:term 7_appendix}
\begin{aligned}
&P_7\cdot \tau _i = (\frac{{{w}^{2}}}{{{\lambda }^{2}}{{L}^{2}}}{{e}^{-\lambda \frac{L}{w}}}-\frac{{{w}^{2}}}{{{\lambda }^{2}}{{L}^{2}}}+\frac{w}{\lambda L}+\frac{1}{2}) \\
&\; \; \; \cdot \left(1 - \frac{1}{L}\left( \frac{{{v}_{0}}-{{v}_\text{f}}}{\lambda }{{e}^{-\lambda \tau_{i} }}+\frac{L\lambda -{{v}_{0}}}{\lambda }-\frac{L}{2}+\frac{{{v}_\text{f}}^{2}}{L{{\lambda }^{2}}}{{e}^{-\lambda \frac{-{{v}_{0}}\tau_{i} +{{v}_\text{f}}\tau_{i} }{{{v}_\text{f}}}}}-\frac{{{v}_\text{f}}^{2}}{L{{\lambda }^{2}}}{{e}^{-\lambda \frac{L-{{v}_{0}}\tau_{i} +{{v}_\text{f}}\tau_{i} }{{{v}_\text{f}}}}} \right) \right) \cdot \tau _i\\
\end{aligned}
\end{equation}

The complete derivation process of Equation \eqref{eq:term 1_x}: 

\begin{equation}
\label{eq:term 1_x_appendix}
\begin{aligned}
&\int_0^{\infty}{\lambda _0e^{-\lambda _0\tau _i}\cdot \left( v_{\text{f}}-v_{\text{0}} \right) \cdot}P_1\cdot \left( \tau _i-\frac{\int_0^{\tau _i}{\tau _{i-1}\lambda _0e^{-\lambda _0\tau _{i-1}}}d\tau _{i-1}}{\int_0^{\tau _i}{\lambda _0e^{-\lambda _0\tau _{i-1}}}d\tau _{i-1}} \right) d\tau _i\\
&=\left( v_{\text{f}}-v_{\text{0}} \right) \left( -\frac{w^2}{\lambda ^2L^2}e^{-\lambda \frac{L}{w}}+\frac{w^2}{\lambda ^2L^2}-\frac{w}{\lambda L}+\frac{1}{2} \right) \\
&\; \; \; \cdot \left(1 - \frac{1}{L}\left( \frac{{{v}_{0}}-{{v}_\text{f}}}{\lambda }{{e}^{-\lambda \tau_{i} }}+\frac{L\lambda -{{v}_{0}}}{\lambda }-\frac{L}{2}+\frac{{{v}_\text{f}}^{2}}{L{{\lambda }^{2}}}{{e}^{-\lambda \frac{-{{v}_{0}}\tau_{i} +{{v}_\text{f}}\tau_{i} }{{{v}_\text{f}}}}}-\frac{{{v}_\text{f}}^{2}}{L{{\lambda }^{2}}}{{e}^{-\lambda \frac{L-{{v}_{0}}\tau_{i} +{{v}_\text{f}}\tau_{i} }{{{v}_\text{f}}}}} \right) \right) \cdot \left( \left( \lambda _0\tau _i-1 \right) e^{-\lambda _0\tau _i}+e^{-2\lambda _0\tau _i} \right) \\
&=\left( v_{\text{f}}-v_{\text{0}} \right) \left( -\frac{w^2}{\lambda ^2L^2}e^{-\lambda \frac{L}{w}}+\frac{w^2}{\lambda ^2L^2}-\frac{w}{\lambda L}+\frac{1}{2} \right) \\
&\; \; \; \cdot \left( -\frac{e^{-2\lambda _0\tau _i}}{2\lambda _0}\mid_{0}^{\infty}-\frac{1}{L}\left( \begin{array}{c}	\frac{v_{\text{0}}-v_{\text{f}}}{\lambda}\cdot \left( \frac{e^{-\left( \lambda +\lambda _0 \right) \tau _i}\left( \lambda -\lambda _0\tau _i\left( \lambda +\lambda _0 \right) \right)}{\left( \lambda +\lambda _0 \right) ^2}\mid_{0}^{\infty}-\frac{e^{-\left( \lambda +2\lambda _0 \right) \tau _i}}{\lambda +2\lambda _0}\mid_{0}^{\infty} \right) +\left( \frac{L\lambda -v_{\text{0}}}{\lambda}-\frac{L}{2} \right) \cdot \left( -\frac{e^{-2\lambda _0\tau _i}}{2\lambda _0}\mid_{0}^{\infty} \right)\\	+\frac{{v_{\text{f}}}^2}{L\lambda ^2}\cdot \left( 1-e^{-\frac{\lambda L}{v_{\text{f}}}} \right) \cdot \left( \frac{e^{\frac{\lambda v_{\text{0}}-\lambda v_{\text{f}}-\lambda _0v_{\text{f}}}{v_{\text{f}}}\tau _i}\left( \lambda _0\left( \frac{\lambda v_{\text{0}}-\lambda v_{\text{f}}-\lambda _0v_{\text{f}}}{v_{\text{f}}}\tau _i-1 \right) -\frac{\lambda v_{\text{0}}-\lambda v_{\text{f}}-\lambda _0v_{\text{f}}}{v_{\text{f}}} \right)}{\left( \frac{\lambda v_{\text{0}}-\lambda v_{\text{f}}-\lambda _0v_{\text{f}}}{v_{\text{f}}} \right) ^2}\mid_{0}^{\infty}+\frac{e^{\frac{\lambda v_{\text{0}}-\lambda v_{\text{f}}-2\lambda _0v_{\text{f}}}{v_{\text{f}}}\tau _i}}{\frac{\lambda v_{\text{0}}-\lambda v_{\text{f}}-2\lambda _0v_{\text{f}}}{v_{\text{f}}}}\mid_{0}^{\infty} \right)\\\end{array} \right) \right) \\
&=\left( v_{\text{f}}-v_{\text{0}} \right) \left( -\frac{w^2}{\lambda ^2L^2}e^{-\lambda \frac{L}{w}}+\frac{w^2}{\lambda ^2L^2}-\frac{w}{\lambda L}+\frac{1}{2} \right) \\
&\; \; \; \cdot \left( \frac{1}{2\lambda _0}-\frac{1}{L}\left( \begin{array}{c}	\frac{v_{\text{0}}-v_{\text{f}}}{\lambda}\cdot \left( \frac{1}{\lambda +2\lambda _0}-\frac{\lambda}{\left( \lambda +\lambda _0 \right) ^2} \right) +\left( \frac{L\lambda -v_{\text{0}}}{\lambda}-\frac{L}{2} \right) \cdot \frac{1}{2\lambda _0}\\	+\frac{{v_{\text{f}}}^2}{L\lambda ^2}\cdot \left( 1-e^{-\frac{\lambda L}{v_{\text{f}}}} \right) \cdot \left( \frac{\lambda v_{\text{f}}\left( v_{\text{0}}-v_{\text{f}} \right)}{\left( \lambda v_{\text{0}}-\lambda v_{\text{f}}-\lambda _0v_{\text{f}} \right) ^2}-\frac{v_{\text{f}}}{\lambda v_{\text{0}}-\lambda v_{\text{f}}-2\lambda _0v_{\text{f}}} \right)\\\end{array} \right) \right) \\
\end{aligned}
\end{equation}

The complete derivation process of Equation \eqref{eq:term 3_x}: 

\begin{equation}
\label{eq:term 3_x_appendix}
\begin{aligned}
&\int_0^{\infty}{\lambda _0e^{-\lambda _0\tau _i}\cdot \left( v_{\text{f}}-v_{\text{0}} \right) \cdot}P_3\cdot \left( \tau _i-\frac{\int_0^{\tau _i}{\tau _{i+1}\lambda _0e^{-\lambda _0\tau _{i+1}}}d\tau _{i+1}}{\int_0^{\tau _i}{\lambda _0e^{-\lambda _0\tau _{i+1}}}d\tau _{i+1}} \right) d\tau _i\\
&=\left( v_{\text{f}}-v_{\text{0}} \right) \left( \frac{w^2}{\lambda ^2L^2}e^{-\lambda \frac{L}{w}}-\frac{w^2}{\lambda ^2L^2}+\frac{w}{\lambda L}+\frac{1}{2} \right) \\
&\; \; \; \cdot \frac{1}{L}\left( \frac{v_{\text{0}}-v_{\text{f}}}{\lambda}e^{-\lambda \tau _i}+\frac{L\lambda -v_{\text{0}}}{\lambda}-\frac{L}{2}+\frac{{v_{\text{f}}}^2}{L\lambda ^2}e^{-\lambda \frac{-v_{\text{0}}\tau _i+v_{\text{f}}\tau _i}{v_{\text{f}}}}-\frac{{v_{\text{f}}}^2}{L\lambda ^2}e^{-\lambda \frac{L-v_{\text{0}}\tau _i+v_{\text{f}}\tau _i}{v_{\text{f}}}} \right) \cdot \left( \left( \lambda _0\tau _i-1 \right) e^{-\lambda _0\tau _i}+e^{-2\lambda _0\tau _i} \right) \\
&=\left( v_{\text{f}}-v_{\text{0}} \right) \left( \frac{w^2}{\lambda ^2L^2}e^{-\lambda \frac{L}{w}}-\frac{w^2}{\lambda ^2L^2}+\frac{w}{\lambda L}+\frac{1}{2} \right) \\
&\; \; \; \cdot \frac{1}{L}\left( \begin{array}{c}	\frac{v_{\text{0}}-v_{\text{f}}}{\lambda}\cdot \left( \frac{e^{-\left( \lambda +\lambda _0 \right) \tau _i}\left( \lambda -\lambda _0\tau _i\left( \lambda +\lambda _0 \right) \right)}{\left( \lambda +\lambda _0 \right) ^2}\mid_{0}^{\infty}-\frac{e^{-\left( \lambda +2\lambda _0 \right) \tau _i}}{\lambda +2\lambda _0}\mid_{0}^{\infty} \right) +\left( \frac{L\lambda -v_{\text{0}}}{\lambda}-\frac{L}{2} \right) \cdot \left( -\frac{e^{-2\lambda _0\tau _i}}{2\lambda _0}\mid_{0}^{\infty} \right)\\	+\frac{{v_{\text{f}}}^2}{L\lambda ^2}\cdot \left( 1-e^{-\frac{\lambda L}{v_{\text{f}}}} \right) \cdot \left( \frac{e^{\frac{\lambda v_{\text{0}}-\lambda v_{\text{f}}-\lambda _0v_{\text{f}}}{v_{\text{f}}}\tau _i}\left( \lambda _0\left( \frac{\lambda v_{\text{0}}-\lambda v_{\text{f}}-\lambda _0v_{\text{f}}}{v_{\text{f}}}\tau _i-1 \right) -\frac{\lambda v_{\text{0}}-\lambda v_{\text{f}}-\lambda _0v_{\text{f}}}{v_{\text{f}}} \right)}{\left( \frac{\lambda v_{\text{0}}-\lambda v_{\text{f}}-\lambda _0v_{\text{f}}}{v_{\text{f}}} \right) ^2}\mid_{0}^{\infty}+\frac{e^{\frac{\lambda v_{\text{0}}-\lambda v_{\text{f}}-2\lambda _0v_{\text{f}}}{v_{\text{f}}}\tau _i}}{\frac{\lambda v_{\text{0}}-\lambda v_{\text{f}}-2\lambda _0v_{\text{f}}}{v_{\text{f}}}}\mid_{0}^{\infty} \right)\\\end{array} \right) \\
&=\left( v_{\text{f}}-v_{\text{0}} \right) \left( \frac{w^2}{\lambda ^2L^2}e^{-\lambda \frac{L}{w}}-\frac{w^2}{\lambda ^2L^2}+\frac{w}{\lambda L}+\frac{1}{2} \right) \cdot \frac{1}{L}\left( \begin{array}{c}	\frac{v_{\text{0}}-v_{\text{f}}}{\lambda}\cdot \left( \frac{1}{\lambda +2\lambda _0}-\frac{\lambda}{\left( \lambda +\lambda _0 \right) ^2} \right) +\left( \frac{L\lambda -v_{\text{0}}}{\lambda}-\frac{L}{2} \right) \cdot \frac{1}{2\lambda _0}\\	+\frac{{v_{\text{f}}}^2}{L\lambda ^2}\cdot \left( 1-e^{-\frac{\lambda L}{v_{\text{f}}}} \right) \cdot \left( \frac{\lambda v_{\text{f}}\left( v_{\text{0}}-v_{\text{f}} \right)}{\left( \lambda v_{\text{0}}-\lambda v_{\text{f}}-\lambda _0v_{\text{f}} \right) ^2}-\frac{v_{\text{f}}}{\lambda v_{\text{0}}-\lambda v_{\text{f}}-2\lambda _0v_{\text{f}}} \right)\\\end{array} \right) \\
\end{aligned}
\end{equation}

The complete derivation process of Equation \eqref{eq:term 6_x}: 

\begin{equation}
\label{eq:term 6_x_appendix}
\begin{aligned}
&\int_0^{\infty}{\lambda _0e^{-\lambda _0\tau _i}\cdot \left( v_{\text{f}}-v_{\text{0}} \right) \cdot}P_6\cdot \left( \tau _i-\frac{\int_0^{\tau _i}{z^2{\lambda _0}^2e^{-\lambda _0z}}dz}{\int_0^{\tau _i}{{\lambda _0}^2ze^{-\lambda _0z}}dz} \right) d\tau _i \\
&=\left( v_{\text{f}}-v_{\text{0}} \right) \left( \frac{w^2}{\lambda ^2L^2}e^{-\lambda \frac{L}{w}}-\frac{w^2}{\lambda ^2L^2}+\frac{w}{\lambda L}+\frac{1}{2} \right)\left( \left( \lambda _0\tau _i-2 \right) e^{-\lambda _0\tau _i}+\left( \lambda _0\tau _i+2 \right) e^{-2\lambda _0\tau _i} \right) \\
&\; \; \; \cdot \left(1 - \frac{1}{L}\left( \frac{{{v}_{0}}-{{v}_\text{f}}}{\lambda }{{e}^{-\lambda \tau_{i} }}+\frac{L\lambda -{{v}_{0}}}{\lambda }-\frac{L}{2}+\frac{{{v}_\text{f}}^{2}}{L{{\lambda }^{2}}}{{e}^{-\lambda \frac{-{{v}_{0}}\tau_{i} +{{v}_\text{f}}\tau_{i} }{{{v}_\text{f}}}}}-\frac{{{v}_\text{f}}^{2}}{L{{\lambda }^{2}}}{{e}^{-\lambda \frac{L-{{v}_{0}}\tau_{i} +{{v}_\text{f}}\tau_{i} }{{{v}_\text{f}}}}} \right) \right) \\
&=\left( v_{\text{f}}-v_{\text{0}} \right) \left( \frac{w^2}{\lambda ^2L^2}e^{-\lambda \frac{L}{w}}-\frac{w^2}{\lambda ^2L^2}+\frac{w}{\lambda L}+\frac{1}{2} \right) \\
&\; \; \; \cdot \left( \frac{1}{4\lambda _0}-\frac{1}{L}\left( \begin{array}{c}	\frac{v_{\text{0}}-v_{\text{f}}}{\lambda}\cdot \left( \frac{5\lambda _0+2\lambda}{\left( \lambda +2\lambda _0 \right) ^2}-\frac{\lambda _0+2\lambda}{\left( \lambda +\lambda _0 \right) ^2} \right) +\left( \frac{L\lambda -v_{\text{0}}}{\lambda}-\frac{L}{2} \right) \cdot \frac{1}{4\lambda _0}\\	+\frac{{v_{\text{f}}}^2}{L\lambda ^2}\cdot \left( 1-e^{-\frac{\lambda L}{v_{\text{f}}}} \right) \cdot \left( \frac{\lambda _0{v_{\text{f}}}^2+2v_{\text{f}}\left( \lambda v_{\text{0}}-\lambda v_{\text{f}}-\lambda _0v_{\text{f}} \right)}{\left( \lambda v_{\text{0}}-\lambda v_{\text{f}}-\lambda _0v_{\text{f}} \right) ^2}-\frac{2v_{\text{f}}\left( \lambda v_{\text{0}}-\lambda v_{\text{f}}-2\lambda _0v_{\text{f}} \right) -\lambda _0{v_{\text{f}}}^2}{\left( \lambda v_{\text{0}}-\lambda v_{\text{f}}-2\lambda _0v_{\text{f}} \right) ^2} \right)\\\end{array} \right) \right) \\
\end{aligned}
\end{equation}

The complete derivation process of Equation \eqref{eq:term 7_x}: 

\begin{equation}
\label{eq:term 7_x_appendix}
\begin{aligned}
&\int_{0}^{\infty} \lambda_{0} e^{-\lambda_{0} \tau_i} \cdot (v_{\text{f}} - v_{\text{0}}) P_7 \cdot \tau_i d\tau_{i}\\
&=\lambda_{0} \left( v_{\text{f}}-v_{\text{0}} \right) \cdot (\frac{{{w}^{2}}}{{{\lambda }^{2}}{{L}^{2}}}{{e}^{-\lambda \frac{L}{w}}}-\frac{{{w}^{2}}}{{{\lambda }^{2}}{{L}^{2}}}+\frac{w}{\lambda L}+\frac{1}{2}) \\
&\; \; \; \cdot \left( \left( -\frac{e^{-\lambda _0\tau _i}\left( \lambda _0\tau _i+1 \right)}{{\lambda _0}^2}\mid_{0}^{\infty} \right) -\frac{1}{L}\left( \begin{array}{c}	\frac{v_{\text{0}}-v_{\text{f}}}{\lambda}\cdot -\frac{e^{-\left( \lambda +\lambda _0 \right) \tau _i}\left( \lambda \tau _i+\lambda _0\tau _i+1 \right)}{\left( \lambda +\lambda _0 \right) ^2}\mid_{0}^{\infty}+\left( \frac{L\lambda -v_{\text{0}}}{\lambda}-\frac{L}{2} \right) \left( -\frac{e^{-\lambda _0\tau _i}\left( \lambda _0\tau _i+1 \right)}{{\lambda _0}^2}\mid_{0}^{\infty} \right)\\	+\frac{{v_{\text{f}}}^2}{L\lambda ^2}\cdot \left( 1-e^{-\frac{\lambda L}{v_{\text{f}}}} \right) \cdot \frac{v_{\text{f}}\left( \left( \lambda v_{\text{0}}-\lambda v_{\text{f}}-\lambda _0v_{\text{f}} \right) \tau _i-v_{\text{f}} \right)}{\left( \lambda v_{\text{0}}-\lambda v_{\text{f}}-\lambda _0v_{\text{f}} \right) ^2}\mid_{0}^{\infty}\\\end{array} \right) \right) \\
&= \lambda_{0} \left( v_{\text{f}}-v_{\text{0}} \right) \cdot (\frac{{{w}^{2}}}{{{\lambda }^{2}}{{L}^{2}}}{{e}^{-\lambda \frac{L}{w}}}-\frac{{{w}^{2}}}{{{\lambda }^{2}}{{L}^{2}}}+\frac{w}{\lambda L}+\frac{1}{2}) \\
&\; \; \; \cdot \left( \frac{1}{{\lambda _0}^2}-\frac{1}{L}\left( \frac{v_{\text{0}}-v_{\text{f}}}{\lambda}\cdot \frac{1}{\left( \lambda +\lambda _0 \right) ^2}+\left( \frac{L\lambda -v_{\text{0}}}{\lambda}-\frac{L}{2} \right) \frac{1}{{\lambda _0}^2}+\frac{{v_{\text{f}}}^2}{L\lambda ^2}\left( 1-e^{-\frac{L\lambda}{v_{\text{f}}}} \right) \left( \frac{{v_{\text{f}}}^2}{\left( \lambda v_{\text{0}}-\lambda v_{\text{f}}-\lambda _0v_{\text{f}} \right) ^2} \right) \right) \right) \\
\end{aligned}
\end{equation}

\end{appendix} 





\clearpage
\bibliographystyle{myplainnat}
\bibliography{main}
\end{document}